\documentclass{aa}
\AddToHook{begindocument/before}{\usepackage[colorlinks=true, linkcolor=blue, citecolor=blue]{hyperref}}
\usepackage{graphicx}
\usepackage[varg]{txfonts}
\usepackage{rotating}
\usepackage{changepage}
\usepackage{enumitem}
\usepackage{amsmath,amssymb}
\usepackage{booktabs}
\usepackage{xspace}
\usepackage{textcomp}
\usepackage{xcolor}	
\usepackage[euler]{textgreek}
\usepackage{adjustbox}
\usepackage{caption}
\usepackage{subcaption}
\usepackage{dblfloatfix}
\usepackage[nolist]{acronym}
\usepackage{tabularx}

\newcommand{\ci}{[\ion{C}{I}]\,(2--1)}
\newcommand{\cii}{[\ion{C}{II}]\,158\,\textmu m}
\newcommand{\kms}{\ensuremath{\rm{km}\, \rm{s}^{-1}}}

\newcommand{\Msun}{\ensuremath{\rm{M}_{\odot}}}

\graphicspath{{.}}

\begin{document}

	\begin{acronym}
	\acro{ism}[ISM]{interstellar medium}
	\acro{sed}[SED]{Spectral Energy Distribution}
	\acro{sfr}[SFR]{star formation rate}
	\acro{vla}[VLA]{Karl G. Jansky Very Large Array}
	\acro{alma}[ALMA]{Atacama Large Millimeter/submillimeter Array}
	\acro{noema}[NOEMA]{NOrthern Extended Millimeter Array}
	\acro{vlt}[VLT]{Very Large Telescope}
	\acro{casa}[{\sc{CASA}}]{Common Astronomy Software Application \citep{2022PASP..134k4501C}} 
	\acro{fwhm}[FWHM]{full width at half maximum}
	\acro{IR}[IR]{infrared}
	\acro{FIR}[FIR]{far-infared}
	\acro{UV}[UV]{ultraviolet}
	\acro{rms}[rms]{root mean square}
	\acro{sn}[S/N]{signal-to-noise ratio}
	\end{acronym}

	\title{The cold molecular gas in $z\gtrsim 6$ quasar host galaxies}
   
   		\titlerunning{The Cold, Molecular Gas of $z\gtrsim 6$ QSOs}

   	\author{Melanie Kaasinen\inst{1}, Bram Venemans\inst{2}, Kevin Harrington\inst{3}, Leindert A. Boogaard\inst{4}, Romain A. Meyer\inst{4,5}, Eduardo Ba\~{n}ados\inst{4}, Roberto Decarli\inst{6}, Fabian Walter\inst{4},  Marcel Neeleman\inst{7}, Gabriela Calistro Rivera\inst{1},  Elisabete da Cunha\inst{8,9}} 
    
    	\authorrunning{M. Kaasinen, B. Venemans, K. Harrington, et al.}    

   	\institute{European Southern Observatory, Karl-Schwarzschild-Str. 2, D-85748, Garching, Germany\\
        \email{melanie.kaasinen@eso.org}
          \and
          Leiden Observatory, Leiden University, PO Box 9513, 2300 RA Leiden, The Netherlands
          \and
          European Southern Observatory, Alonso de C\'{o}rdova 3107, Vitacura, Casilla 19001, Santiago de Chile, Chile
          \and
          Max-Planck-Institut f\"{u}r Astronomie, K\"{o}nigstuhl 17, 69117 Heidelberg, Germany
          \and
          Department of Astronomy, University of Geneva, Chemin Pegasi 51, 1290 Versoix, Switzerland
          \and
          INAF -- Osservatorio di Astrofisica e Scienza dello Spazio di Bologna, via Gobetti 93/3, 40129 Bologna, Italy
          \and
          National Radio Astronomy Observatory, 520 Edgemont Road, Charlottesville, VA, 22903, USA
          \and
          International Centre for Radio Astronomy Research, University of Western Australia, 35 Stirling Hwy, Crawley 26WA 6009, Australia 
          \and
          ARC Centre of Excellence for All Sky Astrophysics in 3 Dimensions (ASTRO 3D), Australia
             }

        \date{received 02 November 2023; accepted 02 February 2024}

    \abstract
   	{Probing the molecular gas reservoirs of $z\gtrsim6$ quasar (QSO) host galaxies is fundamental to understanding the coevolution of star formation and black hole growth in these extreme systems. Yet, there is still an inhomogeneous coverage of molecular gas tracers for $z\gtrsim6$ QSO hosts.}
   	{To measure the average excitation and mass of the molecular gas reservoirs in the brightest $z>6.5$ QSO hosts, we combined new observations of CO(2--1) emission with existing observations of CO(6--5), CO(7--6), \ci, \cii, and dust-continuum emission. }
   	{We reduced and analysed observations of CO(2--1), taken with the Karl G. Jansky Very Large Array, in three $z=6.5-6.9$ QSO hosts---the highest redshift observations of CO(2--1) to date. By combining these with the nine $z=5.7-6.4$ QSO hosts for which CO(2--1) emission has already been observed, we studied the spread in molecular gas masses and CO excitation of $z\gtrsim6$ QSOs.}
   	{Two of our three QSOs, P036+03 and J0305--3150, were not detected in CO(2--1), implying more highly excited CO than in the well-studied $z=6.4$ QSO J1148+5251. However, we detected CO(2--1) emission at $5.1\sigma$ for our highest-redshift target, J2348--3054, yielding a molecular gas mass of $(1.2\pm0.2)\times 10^{10}\, \mathrm{M}_\odot$, assuming $\alpha_\mathrm{CO} = 0.8$\,(K km s$^{-1}$ pc$^2$)$^{-1}$ and $r_\mathrm{2,1}=1$. This molecular gas mass is equivalent to the lower limit on the dynamical mass measured previously from resolved \cii\ observations, implying that there is little mass in stars or neutral gas within the [\ion{C}{II}]-emitting region and that a low CO-to-H$_2$ conversion factor is applicable. On average, these $z\gtrsim6$ QSO hosts have far higher CO(6--5)-, CO(7--6)-, and \cii\ versus CO(2--1) line ratios than the local gas-rich and IR-luminous galaxies that host active galactic nuclei, but with a large range of values, implying some variation in their interstellar medium conditions. We derived a mean CO(6--5)-to-CO(1--0) line luminosity ratio of $r_\mathrm{6,1}=0.9\pm0.2$.}
   	{Our new CO(2--1) observations show that even at 780 Myr after the Big Bang, QSO host galaxies can already have molecular gas masses of $10^{10}$ M$_\odot$, consistent with a picture in which these $z\gtrsim6$ QSOs reside in massive starbursts that are coevolving with the accreting supermassive black holes. Their high gas versus dynamical masses and extremely high line excitation imply the presence of extremely dense and warm molecular gas reservoirs illuminated by strong interstellar radiation fields.}


  	\maketitle

	\section{Introduction}
		\label{sec:intro}

		First discovered in optical surveys in the early 2000s, luminous $z\gtrsim6$ quasars (QSOs) have challenged our understanding of how supermassive black holes (SMBHs) grow and coevolve with their host galaxies \citep[see][and references therein]{2023ARA&A..61..373F}. Studies of the optical and near-infrared (NIR) spectra of these QSOs reveal that they host SMBHs with masses of $\gtrsim10^9\,\mathrm{M}_\odot$ \citep[e.g.][]{2007AJ....134.1150J,2007ApJ...669...32K,2014ApJ...790..145D,2015Natur.518..512W}, implying the presence of enormous black hole seeds at $z>15$ and/or early episodes of super-Eddington accretion \citep[][]{2016PASA...33...51L,2017MNRAS.468.5020I,2021IAUS..356..261T}. To understand how these SMBHs grow and impact star formation, it is crucial to constrain the mass and properties of their gas reservoirs. Doing so will also help reveal the evolutionary path of the gas-poor elliptical galaxies that host today's most massive SMBHs \citep[$M_\mathrm{BH}\gtrsim 10^{10}\,\mathrm{M}_\odot$,][]{2013ARA&A..51..511K}.  

		To discern the type of galaxies in which $z\gtrsim6$ QSOs reside, they have been observed at radio and (sub)millimetre wavelengths, where the emission from the host galaxies' \ac{ism} dominates that of the QSO. Observations of their dust-continuum emission have revealed that many of the hosts are bright in the far-infrared (FIR), with high FIR luminosities of $L_\mathrm{FIR} \sim 10^{12} - 10^{13}\,\mathrm{L}_\odot$, which have been interpreted as signatures of significant dust heating associated with high star formation rates \citep[e.g.][]{2003A&A...406L..55B,2006ApJ...642..694B,2007ApJ...671L..13R,2017Natur.545..457D}. The best studied of these appear to exhibit moderately high dust temperatures of $30-60$\,K \citep[e.g.][]{2013ApJ...772..103L,2023MNRAS.523.3119W} and dust masses of $\gtrsim 10^9\,\mathrm{M}_\odot$ \citep[e.g.][]{2017ApJ...851L...8V,2019ApJ...876...99S,2022ApJ...927..152M}, implying large molecular gas masses of $\gtrsim 10^{10}\,\mathrm{M}_\odot$ (assuming similar or higher gas-to-dust mass ratios than in the Milky Way) and efficient supernova dust production \citep[e.g.][]{2010A&A...522A..15M,Gall_2011,2012MNRAS.424L..34K}. Observations of the dominant ISM coolant, \cii, also show this emission to be bright, again indicating high SFRs \citep[][]{2009Natur.457..699W,2012ApJ...751L..25V,2013ApJ...773...44W,2015ApJ...805L...8B,2016ApJ...816...37V}. Combined with the underlying continuum emission, these results yield [\ion{C}{II}]/FIR luminosity ratios similar to high-redshift starbursts and local Ultra-Luminous Infrared Galaxies (ULIRGS). Moreover, the recent wealth of resolved studies of [\ion{C}{II}] \citep[e.g.][]{Willott_2017,2019ApJ...880....2W,2020ApJ...904..130V,2020ApJ...904..131N,2021ApJ...911..141N}, most recently pushing down to scales of $\sim 200-300$\,pc \citep{2022ApJ...927...21W,2023ApJ...956..127M}, have revealed the presence of some rotating gas disks, in which most of the gas is concentrated in highly compact, central regions. 

		For some $z\gtrsim6$ QSO hosts, observations of the warm and dense molecular gas component traced by CO(6--5) and/or CO(7--6) emission have also revealed the presence of massive molecular gas reservoirs of $\sim10^{10}\,\mathrm{M}_\odot$ \citep{2003A&A...409L..47B,2007ApJ...666L...9C,2010ApJ...714..699W,2011AJ....142..101W,2013ApJ...773...44W,2017ApJ...845..154V}. Yet, converting these observations to a molecular gas mass requires robust constraints on the CO excitation ladders. To test how much molecular gas is in the cold/diffuse state and hence correctly calibrate the CO excitation for these $z\gtrsim6$ QSO hosts, it is critical to also observe their low-$J$ ($J=1,2$) CO emission. But low-$J$ CO emission is intrinsically much fainter than the mid-$J$ CO and [\ion{C}{II}] emission. To date, nine QSOs at $z=5.7-6.4$ have observations of CO(2--1) emission \citep{2010ApJ...714..699W,2011ApJ...739L..34W,2015MNRAS.451.1713S,2016ApJ...830...53W,2019ApJ...876...99S}. In combination with their previously detected CO(5--4), CO(6--5), and/or CO(7--6) emission, these observations have revealed the presence of highly-excited CO, with what seems to be a dearth of diffuse cold gas. 

		To continue systematically exploring the molecular gas reservoirs of $z\gtrsim6$ QSO host galaxies, we used the \ac{vla} to observe the CO(2-1) emission of three QSOs with existing observations of CO(6--5), CO(7--6), [\ion{C}{I}] and [\ion{C}{II}] emission \citep{2015ApJ...805L...8B,2016ApJ...816...37V,2017ApJ...851L...8V}. By combining these lines, we robustly measured their molecular mass, anchored the CO excitation, and compared their ISM properties to those of highly star-forming galaxies and QSO hosts at lower redshift.

		This paper is organised as follows. In Section \ref{sec:data}, we present the VLA observations of our three QSO hosts and present the literature sample. In Section \ref{sec:analysis}, we infer physical properties from these observations. We place these properties in the context of other galaxy samples in Section \ref{sec:results}, and discuss the physical implications in Section \ref{sec:discussion}. In Section \ref{sec:conclusions}, we present a short summary. Throughout this work, we adopt a Lambda--cold dark matter cosmology, with $H_0 = 70$\, km s$^{-1}$ Mpc$^{-1}$, $\Omega_M = 0.3$, and $\Omega_\lambda = 0.7 $. 	
	


	\section{Observations and data reduction} 
		\label{sec:data}

		\subsection{Sample} 
			\label{sub:sample}

			In this work, we reduce the CO(2--1) observations of P036+03 (at $z=6.54$), J0305--3150 (at $z=6.61$), and J2348--3054 (at $z=6.90$). P036+03 was originally identified as part of the Panoramic Survey Telescope \& Rapid Response System 1 (Pan-STARRS1) \citep{2015ApJ...801L..11V} whereas the other two QSOs were identified from the VISTA Kilo-Degree Infrared Galaxy Survey (VIKING) \citep{2013ApJ...779...24V}. Their coordinates and previously derived properties are provided in Table~\ref{tab:source_prop}.  

			\begin{table*} 
			\begin{small} 
			\begin{center}
			\caption{QSO properties and data. \label{tab:source_prop}} 
			\begin{tabular}{@{}>{\raggedright}m{0.5 \columnwidth}>{\centering}m{0.4\columnwidth}>{\centering}m{0.4\columnwidth}>{\centering\arraybackslash}m{0.4\columnwidth}@{}}
			\toprule 
					 & P036+03 & J0305--3150 & J2348--3054 \\ 
			\midrule
			\textbf{Coordinates and BH properties} & & & \\
			\midrule  
			RA (J2000) & 02:26:01.875 & 03:05:16.91 & 23:48:33.34 \\ 
			DEC (J2000) & 03:02:59.40 & -31:50:55.9 & -30:54:10.24 \\ 
			z$_\mathrm{MgII}$  & $6.527\pm0.002$ & $6.605\pm0.002$ &  $6.889\pm0.007$ \\ 
			$M_\mathrm{BH}$ ($\times 10^9\,\mathrm{M}_\odot$) & $2.7 \pm 0.5^a$ & $ 0.8 \pm 0.2^b$ & $3.2^{+1.2}_{-0.9}$$^{,b}$ \\ 
			$L_\mathrm{bol}$ ($\times 10^{13}$ L$_\odot$) & $6.47\pm0.04^c$  & $2.45\pm0.03^c$ & $1.93\pm0.05^c$ \\ 
			\midrule
			\textbf{Previous sub-/mm observations} & & & \\
			\midrule
			z$_\mathrm{[C\,\textsc{ii}]}$  & $6.541\pm0.002$ & $6.6145\pm0.0001$ &  $6.9018\pm0.0007$ \\ 
			\cii\ flux (Jy \kms) & $3.16\pm0.09\,^d$ & $5.25\pm0.30\,^e$ & $1.53\pm0.16\,^{f,d}$ \\
			FWHM$_\mathrm{[C\,\textsc{ii}]}$ (km s$^{-1}$)  & $237\pm10\,^d$ & $268\pm20\,^e$ & $457\pm50\,^d$ \\
			Underlying ($\sim 1$\,mm) continuum flux density (mJy) & $2.55\pm0.05\,^d$ & $5.34\pm0.13\,^d$ & $2.28\pm0.07\,^f$ \\
			\midrule
			z$_\mathrm{CO, [CI]}$  & $6.5410\pm0.0003$ & $6.6139\pm0.0005$ &  $6.9007\pm0.0005$ \\  
			CO(6-5) flux (Jy \kms) & $0.35\pm0.03\,^g$ & $0.63\pm0.04\,^e$ & $0.28\pm0.05\,^f$ \\
			\ci\ flux (Jy \kms) & $0.21\pm0.04\,^g$ & $0.25\pm0.03\,^e$ & $0.16\pm0.06\,^f$ \\
			CO(7-6) flux (Jy \kms) & $0.40\pm0.04\,^g$ & $0.55\pm0.04\,^e$ & $0.26\pm0.06\,^f$ \\
			Underlying ($\sim 3$\,mm) continuum flux density (mJy) & $0.13\pm0.02\,^g$ & $0.27\pm0.02\,^e$ & $0.12\pm0.01\,^f$ \\
			\midrule
			CO(10-9) flux (Jy \kms) & $0.38\pm0.08\,^*$ & ... & ... \\
			\midrule
			\textbf{Previously measured sizes} & & & \\
			\midrule
			deconv. \cii\ size (kpc) $^d$  & $2.4\pm1.6$ & $2.8\times2.6$ &  $1.0\times0.5$ \\
			$2\, r_\mathrm{half,[C\,\textsc{ii}]}$ (kpc) $^{h,i}$  & $2.3\pm0.1$ & $3.6\pm0.1$ &  $0.88\pm0.03$ \\
			\midrule
			\textbf{VLA observations} & & & \\
			\midrule
			Beam FWHM ($"$)  & $1.00 \times 0.77$ & $1.86 \times 0.74$ & $1.49 \times 0.68$\\ 
			Beam FWHM (kpc)  & $5.5 \times 4.2$ & $10.2 \times 4.0$ & $9.6 \times 3.8$\\ 
			rms over channels of 1.2\,FWHM$_\mathrm{[CII]}$ (mJy beam$^{-1}$) & 29.3 & 33.4 & 12.3 \\ 
			CO(2-1) flux (Jy \kms) & \textbf{$<0.05^{4\sigma}$} & \textbf{$<0.06^{4\sigma}$} & $0.04\pm0.01$ \\
			\midrule
			\textbf{Derived dust and molecular gas masses} &  & & \\
			\midrule
			$M_\mathrm{mol, CO(2-1)} $ (M$_\odot$) & $<1.2\times10^{10}$ & $<1.6\times10^{10}$ &  $(1.2\pm0.2)\times10^{10}$ \\
			$M_\mathrm{mol, [C I](2-1)} $ (M$_\odot$) & $(1.9\pm0.9)\times10^{10}$ & $(2.4\pm1.0)\times10^{10}$ & $(1.6\pm0.9)\times10^{10}$ \\
			$M_\mathrm{dust} $ (M$_\odot$) & $(1.2\pm0.7)\times10^{8}$ & $(2.3\pm0.7)\times10^{9}$ & $(1.1\pm0.6)\times10^{8}$ \\
			\bottomrule 
			\end{tabular} 
			\end{center} 
			\begin{adjustwidth}{25pt}{25pt}
			\vspace{-0.2cm}
			\textbf{References: }
			$^a$ \cite{2023A&A...676A..71M}, 
			$^b$ \cite{2022ApJ...941..106F}, 
			$^c$ \cite{Schindler_2020},
			$^d$ \cite{2020ApJ...904..130V},
			$^e$ \cite{2022ApJ...930...27L},
			$^f$ \cite{2017ApJ...851L...8V},
			$^g$ \cite{2022A&A...662A..60D},
			$^h$ \cite{2021ApJ...911..141N},
			$^i$ \cite{2022ApJ...927...21W},
			$^*$from a Gaussian fit to the spectrum extracted from the NOEMA observations of programmes S15DA and W15FD (also used in \citealt{2022A&A...662A..60D})
			\end{adjustwidth}
			\end{small}
			\end{table*}

			The ISM emission of the three QSOs was first observed via the \cii\ and FIR emission, obtained as part of two programmes. P036+03 was observed with the NOEMA interferometer \citep[Programme E14AG][]{2015ApJ...805L...8B}, whereas the two, higher redshift QSOs from VIKING, were observed with ALMA \citep[Programme 2012.1.00882.S][]{2016ApJ...816...37V}. For J0305--3150 and J2348--3054, observations of the CO(6--5), CO(7--6), and \ci\ emission were conducted with ALMA \citep[Programme 2013.1.00273.S][]{2017ApJ...845..154V} and for P036+03 observations of CO(6--5), CO(7--6), \ci, and CO(10--9) emission were conducted with NOEMA \citep[Programmes S15DA and W15FD][]{2022A&A...662A..60D}.


		\subsection{Observations} 
			\label{sub:observations}

			The CO(2--1) observations analysed here were taken as part of the VLA programmes, 16A-160 and 17A-336 (PI: Venemans). For programme 16A-160, 11 observational blocks were conducted between 2016/03/04 and 2017/05/17, resulting in 10\,h of observations of J2348--3054, 2\,h of J0305--3150 and no observations of P036+03. To complete the experiment, 20 further observational blocks were requested as part of the programme 17A-336, with 18\,h to be spent on J2348--3054, 12\,h on J0305--3150 and 10\,h on P036+03. This supplementary programme was carried out in full between 2017/05/20 and 2017/08/08. Across the two programmes, almost all observation blocks were carried out in the C configuration, with one instead conducted in the B configuration. All observations were carried out in the Ka Band, spanning 26.5--40 GHz.

			We performed bandpass and flux calibrations using observations of 3C48 (for J0305--3150 and J2348--3054) and 3C147 (for P036+03). Phase and amplitude calibrations were performed using observations of J0201+0343, J0348-2749, and J2359-3133 (for P036+03, J0305--3150 and J2348--3054 respectively).  The total on-source integration times for the three targets were 5.3\,h, 7.42\,h, and 14.84\,h (for P036+03, J0305--3150, and J2348--3054 respectively). The synthesised beam sizes and $1\sigma$ root-mean-square (rms) noise values are provided in Table~\ref{tab:source_prop}.


			\begin{figure*}
				\centering
				\includegraphics[width=0.55\textwidth, trim={0.7cm -1.5cm 0.4cm 0cm},clip]{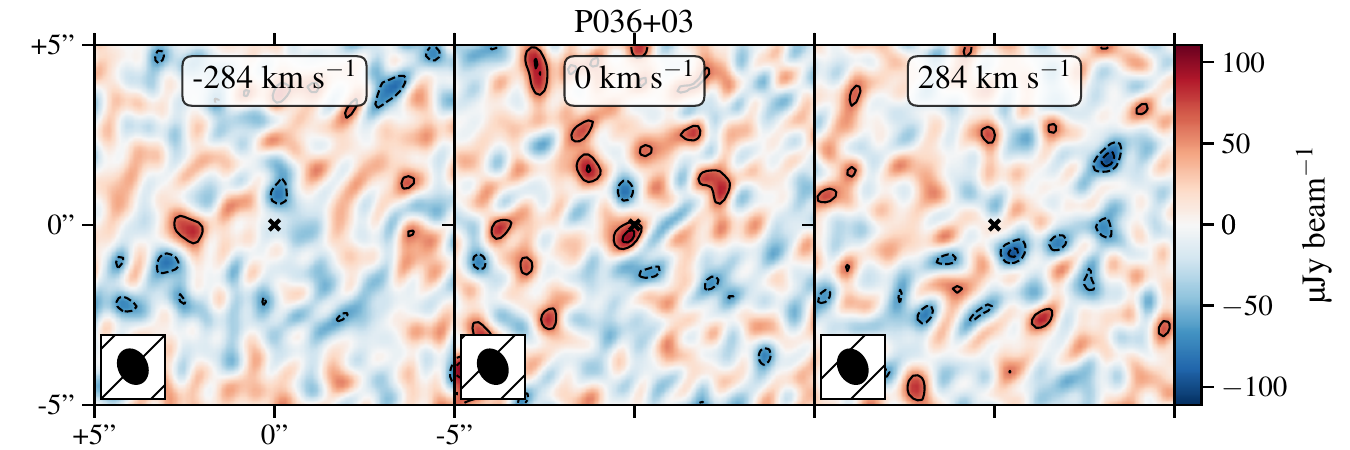}
				~
				\includegraphics[width=0.4\textwidth, trim={0.3cm 0cm 1cm 0cm},clip]{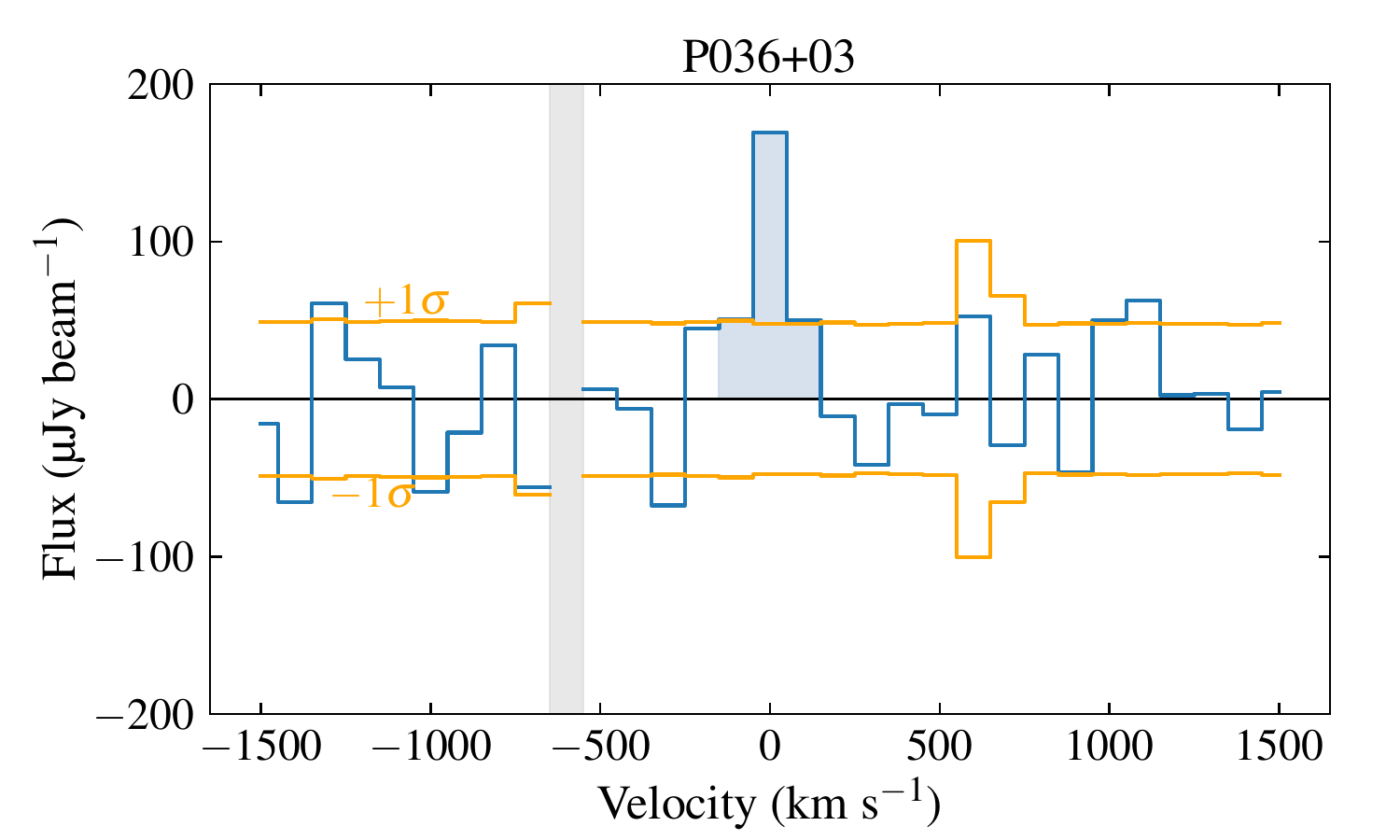}
				\\
				\includegraphics[width=0.55\textwidth, trim={0.7cm -1.5cm 0.4cm 0cm},clip]{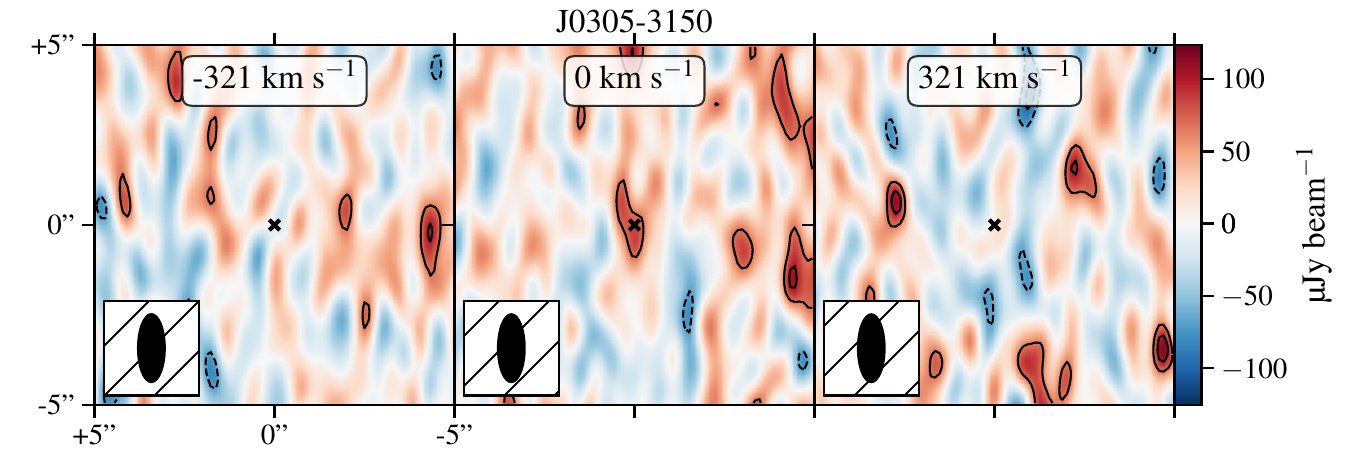}
				~
				\includegraphics[width=0.4\textwidth, trim={0.3cm 0cm 1cm 0cm},clip]{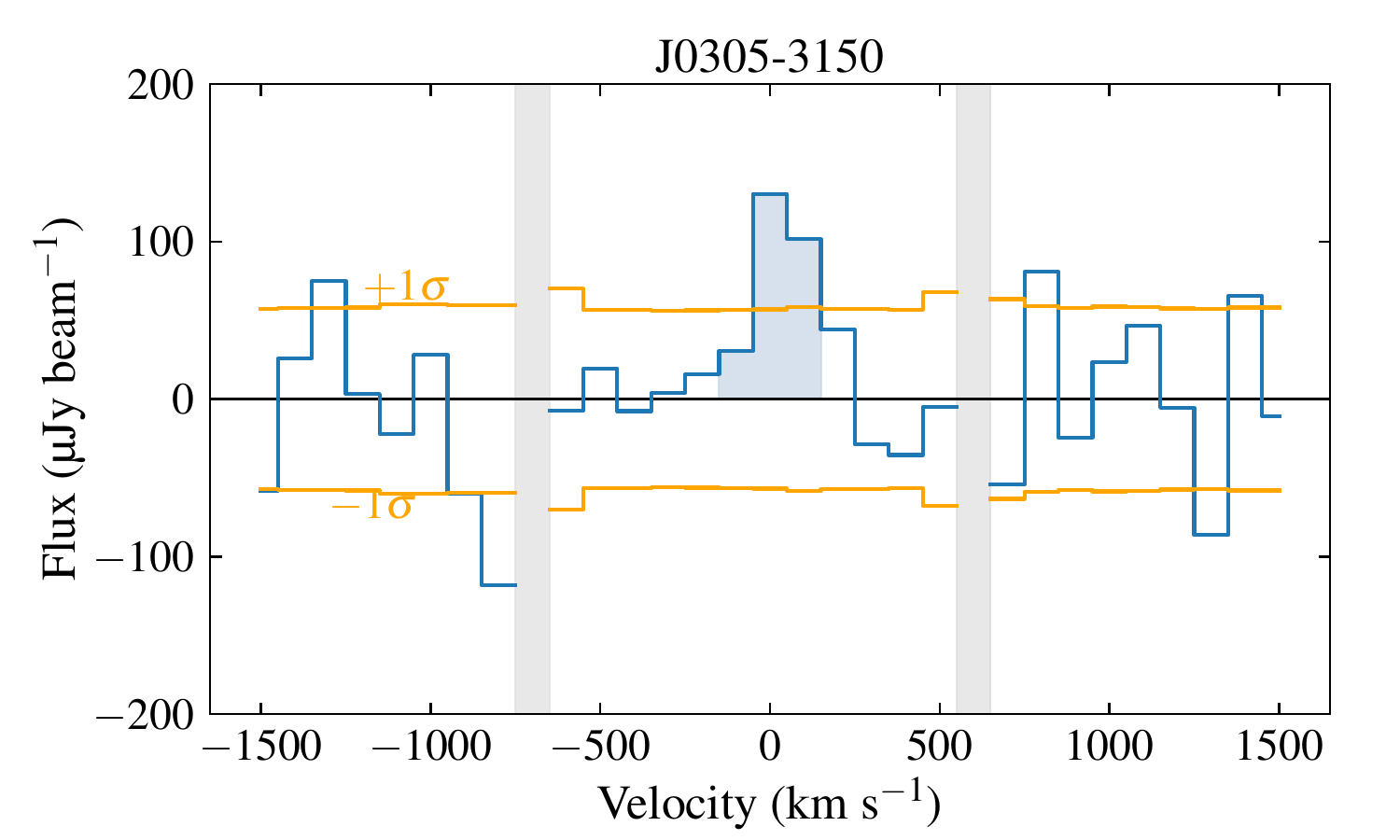}
				\\
				\includegraphics[width=0.55\textwidth, trim={0.7cm -1.5cm 0.4cm 0cm},clip]{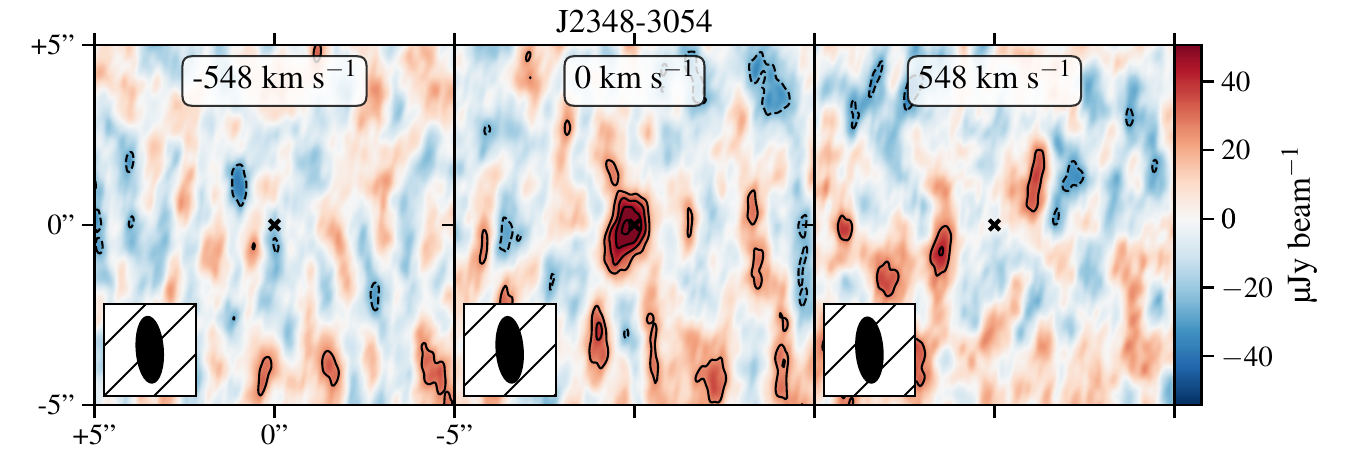}
				~
				\includegraphics[width=0.4\textwidth, trim={0.3cm 0cm 1cm 0cm},clip]{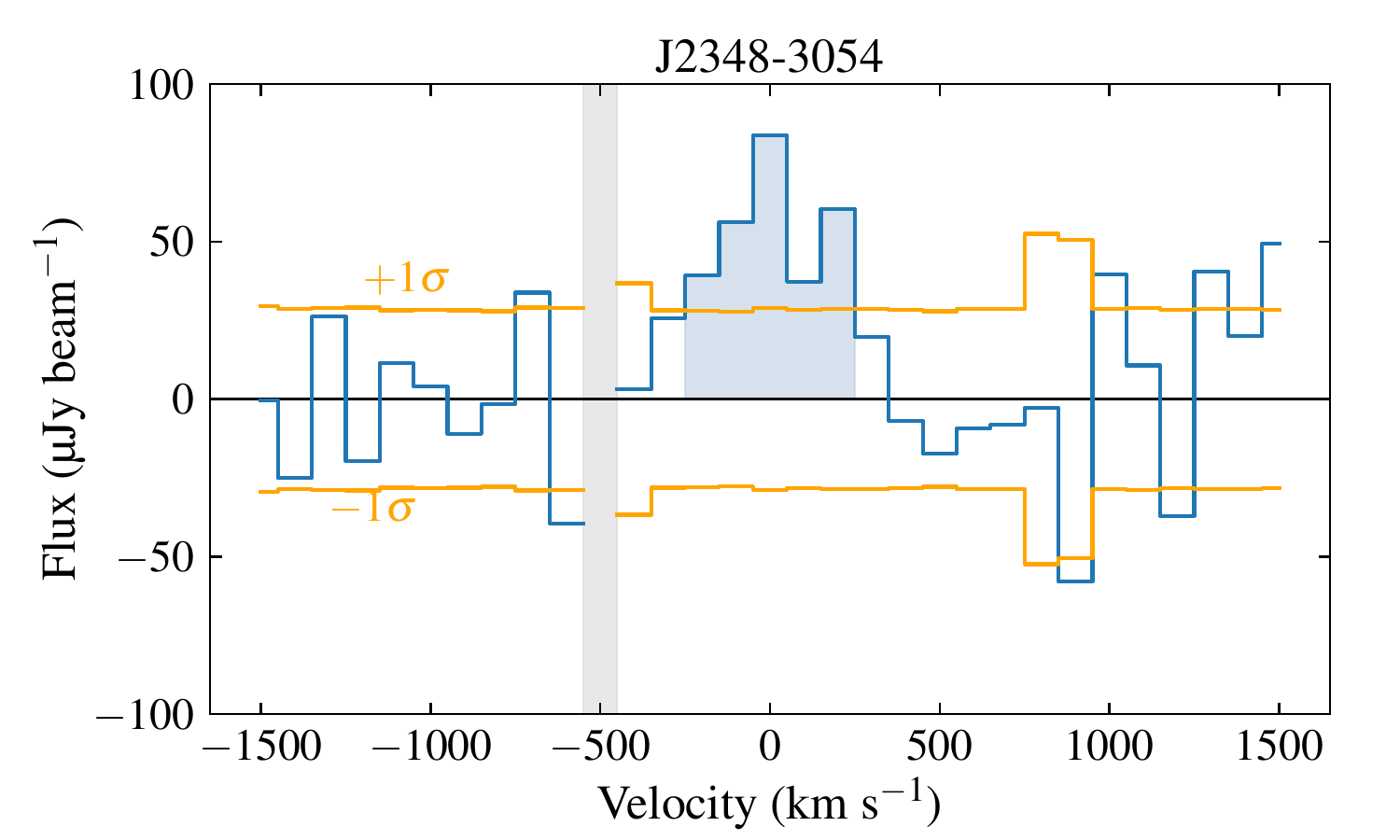}
				\\
				\caption{CO(2--1) observations of P036+03, J0305--3150, and J2348--3054 (from top to bottom, respectively). Left: Cleaned channel maps with the middle panel depicting the moment-0 map centred on the \cii\ redshift and integrated over $1.2\times$\,\ac{fwhm} of the \cii\ line, which should encompass 84\% of the CO(2--1) emission assuming there is no spatial or velocity offset. Right: Observed CO(2--1) spectra extracted from the pixel at the peak position of the moment 0 maps (blue line) and $1\sigma$ rms per channel (orange line). The spectra are shown for channels of 100 km s$^{-1}$ velocity width, the shaded region indicating the $1.2$\,FWHM region used to create the channel maps to the left. P036+03 and J0305--3150 are undetected (with a significant number of negative and positive peaks in the data cube exceeding the significance of the peaks in emission near the expected source centres), so we place $4\sigma$ upper limits on their CO(2--1) emission. J2348--3054 is detected at $5.1\sigma$ (central velocity-integrated map in bottom row). 
				\label{fig:channelmaps}}		
			\end{figure*} 

		\subsection{Data reduction} 
			\label{sub:data_reduction}

			To calibrate the data from each two-hour observation set, we used the \ac{casa} pipeline version 5.6.2 as described in \url{https://science.nrao.edu/facilities/vla/data-processing/pipeline#section-26} (that is, we included most of the standard calibration tasks and disabled the application of Hanning smoothing). After examining the calibrated data, we performed additional flagging, removing the RFI at 29.43 GHz remaining in the observation sets of J2348--3054 and the poor individual antenna data for short time windows in two observation sets (one for J0305--3150 and one set for J2348--3054). This leads to some of the gaps in data in the spectra of Fig.~\ref{fig:channelmaps}.

			To test whether the CO(2--1) emission is detected, we made three-channel cubes, by integrating each channel over $1.2\times$ the \ac{fwhm} of the previously detected \cii\ line emission (see Table~\ref{tab:source_prop}) and centring the frequency range on the CO(6--5)-derived redshift. Assuming that the CO(2--1) emission has the same linewidth as \cii\ and systemic velocity of the other CO transitions, this velocity range should encompass 84\% of the CO(2--1) emission. We applied a natural weighting scheme and cleaned down to $2\sigma$ within $3\farcs$ of the source centre. These cleaned maps are shown in Fig. \ref{fig:channelmaps}, with the beam properties and rms values provided at the bottom of Table~\ref{tab:source_prop}. Using the same central frequency, we also created image cubes of 100 km s$^{-1}$, cleaning down to $2\sigma$ and applying natural weighting. The spectra extracted from the single pixel  corresponding to the peak in the moment-0 maps of these cubes, are shown in the right panels of Fig.~\ref{fig:channelmaps}.  

		\subsection{Line fluxes and upper limits} 
			\label{sub:line_flux_and_upper_limits}

			We tested whether the CO(2--1) emission of the three QSO hosts was detected using the moment-0 maps described above. For our highest-redshift target, J2348--3054, we detect a $5.1\sigma$ peak in emission, consistent with the expected source position. We find a $3.3\,\sigma$ peak offset by $0\farcs43$ (2.4 kpc) from the source centre for P036+03; however, other peaks of emission exceeding this value are apparent in the integrated channel maps at the same frequency and in the channels on either side. Thus, we treat it as an upper limit. Likewise, we detect no CO(2--1) emission for J0305--3150, with a peak value near the source centre of $2.6 \,\sigma$. At a spectral resolution of 100 km s$^{-1}$, none of the three sources exceed $\mathrm{S/N}=3$ across the expected line width. Moreover, none of the three QSOs targeted show any evidence of continuum emission, nor is there any spectrally offset CO(2--1) emission. We also searched for evidence of the companions reported in \cite{2020ApJ...904..130V}, but find nothing exceeding $\sim2.5 \,\sigma$ at the expected positions and frequencies. 

			Based on the moment-0 maps, we treated the CO(2--1) emission from J2348--3054 as a significant detection and derived $4\sigma$ upper limits for the other two sources, consistent with our detection threshold. We derived the line flux measurement for J2348--3054 by first measuring the value of the single pixel at the peak position ($0\farcs3$ offset from the source centre quoted in Table ~\ref{tab:source_prop}) in the central channel of the bottom row in Fig.~\ref{fig:channelmaps}, multiplying by the channel width, and accounting for the extra 16\% of the line flux not included in $1.2\,\times$ FWHM. To derive the upper limits, we measured the root mean square (rms) of the moment-0 map, also multiplying this value by $1.2\,\times$ FWHM$_\mathrm{[CII]}$, accounting for the extra 16\%, and multiplying by our detection threshold of $4\sigma$. We also accounted for potentially extended emission underlying the noise, assuming that the CO(2--1) emission arises from the same size region as the \cii\ emission. We deem it unlikely that the CO emission would be significantly more extended than the \cii\ emission \citep[see also][]{2011ApJ...739L..32R}. To test how much flux we would miss by deriving an upper limit (line flux) from only the pixel rms (peak pixel value), we followed the approach of \cite{2020MNRAS.499.5136C}. That is, we simulated both a point source and extended (Gaussian) emission, using the \cii\ size ($2\,r_\mathrm{1/2,[C~II]}$, Table~\ref{tab:source_prop}), position angle, and inclination measured for each galaxy. We then performed mock VLA observations under the same configuration used for the real observations and compared the flux extracted from the peak pixel for both the point source and extended source. We performed noise-free tests to accurately recover the difference between the flux of the point versus extended source (which we cannot recover from our data). Based on this modelling, we found that for P036+03, we would miss 15\% of the flux if the CO(2--1) emission is as extended as the \cii\ emission. Similarly, for J0305-3150, we would miss 19\% of the flux. We added these potentially missing extended components when deriving the $4\sigma$ upper limits, based on the rms. For J2348-3054, the peak pixel extraction for the extended source is equivalent to that of the point source (only 0.004\% less), because the CO(2--1) beam ($7.8 \times 3.6$ kpc), is significantly larger than twice the measured \cii\ half-light radius ($\sim 0.8$\, kpc) \citep{2021ApJ...911..141N}. This is also consistent with the fact that the CO(2--1) emission from J2348--3054 is well fit by a (convolved) point source and the flux extracted from this point source matches that extracted from a larger circular aperture of $2\farcs5$. Thus, we performed no correction for any missing flux for J2348--3054.

			Given the low S/N of these observations, we derived the CO(2--1) line flux density and upper limits assuming that the CO(2--1) line width is the same as that measured for the much higher S/N \cii\ emission; that is, we use the highest S/N prior we have. This seems to be a reasonable assumption, as for most of the $z>5$ QSO host galaxies detected at $>5\sigma$ in both CO(2--1) and \cii, the reported FWHM values of the two lines have been consistent within uncertainties \citep[see Table~\ref{tab:lit_QSOs} for literature sample, values reported in][]{2013ApJ...773...44W,2016ApJ...830...53W,2019ApJ...876...99S,2022A&A...664A..39K,2022ApJ...927..152M}. Although we use the most recent measurements of the \cii\ line FWHM \citep{2020ApJ...904..130V,2022ApJ...930...27L} for the integrated CO(2--1) line flux density and upper limits, we also tested our results using the values derived in the original studies that reported the \cii\ detections \citep{2015ApJ...805L...8B,2016ApJ...816...37V}. The integrated CO(2--1) line flux densities (hereafter simply `line fluxes') and upper limits remain consistent, within errors. However, the S/N of the detection of J2348--3054 increases with the wider FWHM measured in \cite{2020ApJ...904..130V} versus \cite{2016ApJ...816...37V}.



	\section{Derived quantities} 
		\label{sec:analysis}

		\subsection{Line luminosities and dust properties} 
			\label{sub:luminosities}

			For the three galaxies studied here, and for the comparison samples, we computed the line luminosities as
			\begin{align}
				L_\mathrm{line}^\prime = 3.25\times 10^7 S_\mathrm{line} \Delta v \dfrac{D_L^2}{(1+z)^3 \nu_\mathrm{obs}^2} ~ \mathrm{K~km~s^{-1}~pc^2}
			\end{align}
			where $D_L$ is the luminosity distance in Mpc, $\nu_\mathrm{obs}$ is the observed central frequency of the line in GHz and $S_\mathrm{line} \Delta v$ is the line flux in Jy km s$^{-1}$. 

			At the redshift of these QSOs, the high temperature of the cosmic microwave background (CMB) may have a pronounced effect on the observed CO line fluxes, becoming a strong background against which lines are measured as well as potentially boosting the higher- versus lower-$J$ transitions \citep{2013ApJ...766...13D,2016RSOS....360025Z,2016ApJ...819..161T}. We therefore derived both CMB-corrected and uncorrected values for the CO line luminosities. As shown in \cite{2013ApJ...766...13D}, the ratio of the observed versus intrinsic line flux depends on the temperature and density of the gas. To perform the CMB corrections, we adopted their non-LTE, high-density ($n_\mathrm{H_2}=10^{4.2}$\,cm$^{-3}$) and high-temperature ($T_\mathrm{kin}=40$\,K) scenario. In their model, they assume that the kinetic temperature of the gas is coupled to that of the dust: $T_\mathrm{k} = T_\mathrm{d}$. Thus, we partly motivate the choice of this scenario by the high inferred dust temperatures ($\sim$40\,K), as well as the compact gas reservoirs and high observed line ratios, which imply high densities and temperatures (see Sec.~\ref{sec:discussion}). Adopting this correction, the fraction of CO(2--1) line flux observed against the CMB, $f_\mathrm{CMB}$, for P036+03, J0305-3150, and J2348-3054 is 0.56, 0.55, and 0.53 (respectively), whereas the fraction of CO(6--5) observed against the CMB is 0.67, 0.65, and 0.64 (respectively). This additional correction results in higher intrinsic line fluxes, and hence luminosities, than without the CMB correction.

			We also refit the dust SEDs, ensuring that we take into account the effects of the CMB by adopting a similar approach to \cite{2021ApJ...919...30D} and \cite{2022ApJ...927..152M} (see Appendix~\ref{sec:dust_sed_fitting}). We fit the dust SEDs with a modified blackbody, assuming optically thin dust emission at $\lambda>40$ \textmu m.  Our method accounts for the effects of the CMB, correcting for both the heating by and contrast against the CMB (as prescribed by \citealt{2013ApJ...766...13D}) and includes the most up-to-date continuum data. In this way, we obtained new values for the FIR luminosities, dust masses, dust temperatures, and emissivity indices of all $z\gtrsim6$ QSO hosts, including our three and the literature sample (Table~\ref{tab:dust_sed}). We note that in most cases, the dust temperatures and emissivity indices remain unconstrained due to the limited number of continuum data points. For these, the errors on the dust masses and FIR luminosities ($L_\mathrm{FIR}$) are higher, taking into account the full prior range on $T_\mathrm{dust}$ and $\beta$.  We provide two sets of $L_\mathrm{FIR}$ values to ease the literature comparison, integrating between 40-400 \textmu m and 42.5-122.5 \textmu m (e.g. as in \citealt{Liu_2015} versus \citealt{2011A&A...530L...8D}). 

			If the QSO host galaxies have high dust columns, the dust may actually be optically thick at $\lambda>40$ \textmu m, contrary to our assumption. Given the large dust masses of $10^8-10^9$ M$_\odot$ inferred here (albeit for the optically thin assumption) and small galaxy sizes of $r_\mathrm{half}\sim 1$\, kpc (as in the measurements for the highest-quality \cii\ observations of J2348+3054, \citealt{2022ApJ...927...21W}) this is certainly plausible. As shown by \cite{2021ApJ...919...30D}, accounting for the wavelength up to which dust remains optically thick ($\lambda_\mathrm{thick}=60-140$\,\textmu m for their sample) has the effect of shifting the best-fit dust temperatures of their sample 10\,K higher, on average, and dust masses to a few times lower values. However, the derived FIR luminosities and $\beta$ remain practically the same. Thus, our $L_\mathrm{FIR}$ should not change significantly in the case of optically thick dust emission but our dust temperatures and masses may be biased to slightly lower and higher values, respectively.

			\begin{figure}
				\centering
				\includegraphics[width=0.5\textwidth, trim={0.1cm 0.8cm 0.1cm 1cm},clip]{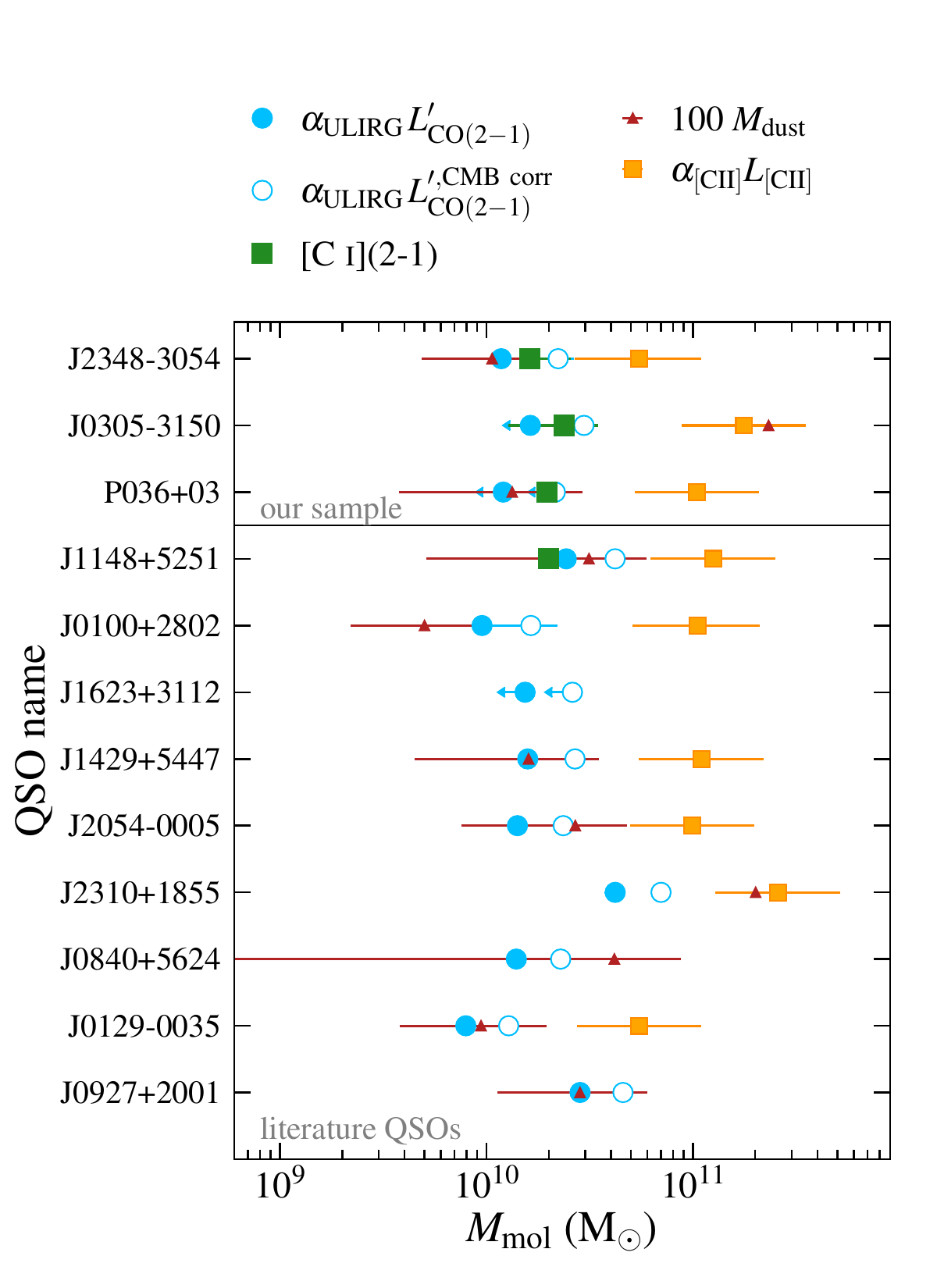}
				\caption{Comparison of molecular gas masses inferred from different sets of observations for our three QSO hosts (top three) and the literature sample of $z\gtrsim 6$ QSO hosts (bottom nine). For the CO-based values, we have assumed a ULIRG CO-to-H$_2$ conversion factor of $\alpha_\mathrm{CO}=0.8$\,M$_\odot$ (K km s$^{-1}$ pc$^2$)$^{-1}$.  We show two masses for CO(2--1), one derived without any CMB correction being applied (filled blue circles) and one derived by applying a correction assuming the high-density ($n_\mathrm{H_2}=10^{4.2}$\,cm$^{-3}$) and high-temperature ($T_\mathrm{kin}=40$\,K), non-LTE scenario presented in \cite{2013ApJ...766...13D} (blue outlined circles). We also show molecular gas masses derived from \ci\ emission (green squares) assuming the [\ion{C}{i}] is optically thin, in LTE and has the same abundance as the sample of lensed starbursts studied in \cite{2021ApJ...908...95H}, namely [\ion{C}{I}]/H$_2\,= (6.82\pm3.04)\times10^{-5}$. We also compare to molecular gas masses derived from the dust masses fit here, assuming the local gas-to-dust ratio of 100 (red triangles), as well as molecular gas masses derived from the \cii\ line luminosities, assuming the mean [\ion{C}{II}]-to-H$_2$ conversion factor ($\alpha_\mathrm{[CII]} = 31\,\Msun/L_{\odot}$) derived by \cite{2018MNRAS.481.1976Z}. \label{fig:Mmol_comp}}
			\end{figure}

			\begin{figure}
				\centering
				\includegraphics[width=0.5\textwidth, trim={0.8cm 0.5cm 0.2cm 0.4cm},clip]{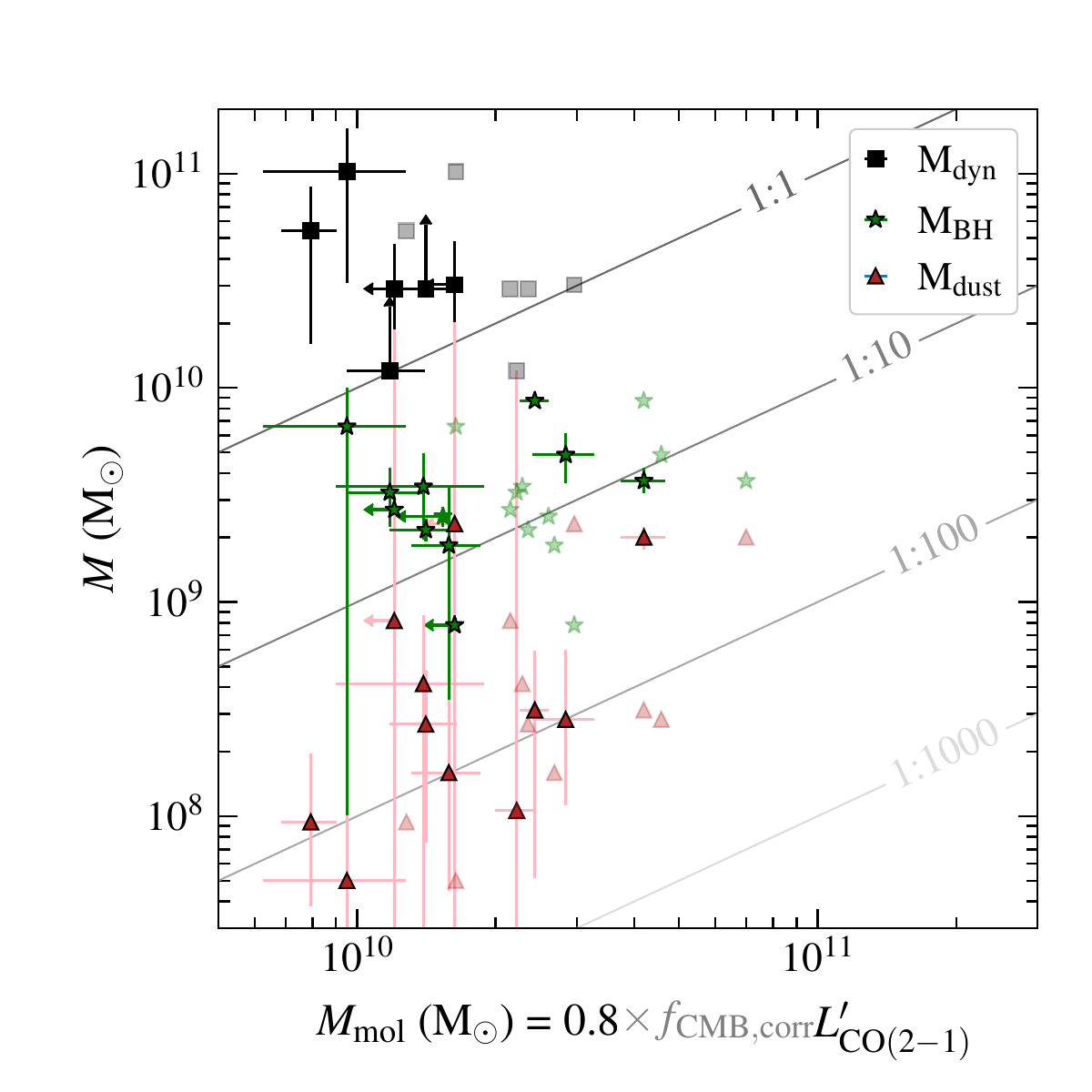}
				\caption{Dynamical masses (black squares), dust masses (red triangles), and black hole masses (green stars) compared to the CO(2--1)-derived molecular gas masses assuming a ULIRG $\alpha_\mathrm{CO}$, with a CMB correction (light shaded symbols) and without a CMB correction (dark shaded symbols). For the CO(2--1)-derived values, we assume that the low-$J$ CO emission is thermalised ($L_\mathrm{CO(1-0)}^\prime = L_\mathrm{CO(2-1)}^\prime$). Most molecular gas masses are high, spanning $0.1-1\times$ the dynamical mass. \label{fig:mass_comp}}	
			\end{figure}

		\subsection{Molecular gas masses} 
			\label{sub:molecular_gas_masses}

			We derived the molecular gas masses from the CO(2--1) emission assuming that the low-$J$ emission is thermalised, $L_\mathrm{CO(2-1)}^\prime = L_\mathrm{CO(1-0)}^\prime$, as in low-redshift QSOs \citep[e.g.][]{2023arXiv231010235M}. To convert $L_\mathrm{CO(1-0)}^\prime$ to a molecular gas mass, we assumed an intrinsic ULIRG-like $\alpha_\mathrm{CO}=0.8$\,M$_\odot$ (K km s$^{-1}$ pc$^2$)$^{-1}$, as derived in \cite{Downes_1998} (for ULIRG centres) based on resolved CO(2--1) or CO(1--0) observations of 10 local ULIRGs. These low values were found to be necessary in order for the molecular gas masses not to exceed the inferred dynamical masses. Similarly, early comparisons of dynamical mass estimates and CO-derived molecular gas masses in $z\gtrsim6$ QSO hosts motivated the choice of such low $\alpha_\mathrm{CO}$ also in these systems \citep[e.g.][]{2003Natur.424..406W,2004ApJ...615L..17W}. Following the majority of the literature on $z\gtrsim6$ QSOs \citep[see the review of][]{2013ARA&A..51..105C}, we adopted the same value but it is important to note that this may need to be adjusted in future, once robust constraints on $\alpha_\mathrm{CO}$ can be determined for each QSO. For example, by modelling the multi-line and continuum emission, \cite{2021ApJ...908...95H} show that for systems in which the CO gas column density is mostly traced by higher-$J$ lines, $\alpha_\mathrm{CO}$ is shifted to higher values. We also inferred the molecular gas masses based on the CMB-corrected CO(2--1) emission (see Sec.~\ref{sub:luminosities}). This additional correction results in a higher intrinsic molecular gas masses than without the CMB correction (although for lower assumed temperatures and densities these corrections would be much larger). In effect, this additional CMB correction leads to an increased $\alpha_\mathrm{CO}$, that is $\alpha_\mathrm{CO, tot.} = \alpha_\mathrm{CO, intr.} / f_\mathrm{CMB}$. Thus, Milky-Way-like values of $\alpha_\mathrm{CO}\sim 4\,\mathrm{M}_\odot$(K km s$^{-1}$ pc$^2$)$^{-1}$ are feasible for host galaxies with lower average molecular gas temperatures or densities than in the model we have assumed here.

			Dust emission can also be used as a molecular gas tracer, as long as the dust temperature, emissivity index, and dust-to-gas ratios can be well-constrained or safely assumed \citep[e.g.][]{Groves_2015,2016ApJ...820...83S,2018MNRAS.478L..83L}. In this work, we derived the molecular gas masses from the dust masses fit here (see Appendix~\ref{sec:dust_sed_fitting}). To derive molecular gas masses, we scaled the dust masses by the local gas-to-dust ratio of 100 \citep[e.g.][]{2007ApJ...663..866D}, which has been found to be consistent with some $z\gtrsim6$ QSO hosts \citep[e.g.][]{2016ApJ...830...53W,2022A&A...662A..60D,Feruglio_2023}.

			In recent years, atomic carbon, [\ion{C}{I}], has also garnered significant attention as a potentially reliable and efficient molecular gas tracer for $z>2$ star-forming galaxies \citep[e.g.][]{2011ApJ...730...18W,2013MNRAS.435.1493A,2017MNRAS.466.2825B,2017PhDT........21Y,2018A&A...615A.142A,2018ApJ...869...27V,2019A&A...624A..23N}. At these high redshifts, the two [\ion{C}{I}] transitions are shifted to millimetre wavelengths and can be detected in less time than CO(2--1). Unlike for $^{12}$CO, [\ion{C}{I}] emission has mostly been found to be optically thin, \citep[e.g.][]{2002ApJS..139..467I,2003A&A...409L..41W,2021ApJ...908...95H}, potentially making it a more straightforward tracer. Moreover, [\ion{C}{I}] emission is widespread within clouds, stemming from both the cloud interiors and surfaces \citep{2002ApJS..139..467I,2008A&A...477..547K,2015ApJ...803...37B,2017ApJ...839...90B,2015MNRAS.448.1607G,2019MNRAS.486.4622C}. [\ion{C}{I}] has therefore been proposed to be a more reliable molecular gas tracer than low-$J$ CO emission in diffuse and/or metal-poor gas \citep[e.g.][]{2004MNRAS.351..147P,2016MNRAS.456.3596G}. 

			To derive molecular gas masses from the observed \ci\ emission, we assumed the [\ion{C}{I}] emission is optically thin, and in LTE \citep[following][]{2005A&A...438..533W}. In this case, the mass of neutral carbon is given by,
			\begin{align}
				\dfrac{M_\mathrm{CI}}{\mathrm{M}_\odot} = \dfrac{4.556}{10^4} \dfrac{Q_\mathrm{ex}}{5} \exp(62.5/T_\mathrm{ex}) \dfrac{L_\mathrm{[CI]2-1}^\prime}{\mathrm{K~km~s^{-1}}} \, 
			\end{align}
			where $Q_\mathrm{ex} = 1+3\exp(-23.6/T_\mathrm{ex}) + 5\exp(-62.5/T_\mathrm{ex})$ is the partition function and $T_\mathrm{ex}$ is the excitation temperature. We assumed the atomic carbon is in thermal equilibrium with the dust, with the average dust temperature typically quoted in the literature, $T_\mathrm{ex} = 47$\, K \citep[][]{2006ApJ...642..694B,2014ApJ...785..154L}. If we instead assume $T_\mathrm{ex} =$ 30 K, the [\ion{C}{I}]-derived molecular gas masses would be 1.5$\times$ the values presented here, whereas if we assume $T_\mathrm{ex} =$ 100 K the molecular gas masses would be 0.8$\times$ our values, therefore making little difference to the derived values. 

			To convert the atomic carbon mass to a molecular gas mass, we applied the mean atomic carbon abundance of $\left<\mathrm{[\ion{C}{I}]/H_2}\right>=(6.82\pm3.04)\times10^{-5}$, measured by \cite{2021ApJ...908...95H} for their sample of 18, $z=2-3$ lensed starburst galaxies with [\ion{C}{I}] observations. This value is consistent with the implied high value of $7\times 10^{-5}$ found by \cite{2017MNRAS.466.2825B} for their sample of 13 lensed $z\sim 4$ dusty star-forming galaxies, when assuming the same low $\alpha_\mathrm{CO}$ as in this work. However, these values are higher, on average, than those measured for less IR-luminous, main-sequence star-forming galaxies at $z=1.2-2.5$. For example, \cite{2020ApJ...902..109B} derive a value of [\ion{C}{I}]/H$_2 = (1.9 \pm 0.4) \times 10^{-5}$ based on the [\ion{C}{I}] detections of six main-sequence star-forming galaxies, and \cite{2018ApJ...869...27V} find a value of [\ion{C}{I}]/H$_2 = (1.6 \pm 0.8) \times 10^{-5}$ for their sample of $z=1.2$ main-sequence galaxies. These lower values may be attributed to the difference in physical conditions (e.g. cosmic ray, X-ray, and/or mechanical heating effects) and/or the calibration methods.\footnote{\cite{2018ApJ...869...27V} and \cite{2020ApJ...902..109B} calibrated the atomic carbon abundance against the CO-derived molecular gas masses, for which they assume a Milky-Way-like $\alpha_\mathrm{CO}$, which is higher than the ULIRG value assumed for similar calibrations of higher-redshift sub-millimetre/IR-bright galaxies.} We take the value of \cite{2021ApJ...908...95H} here since their sample are comprised of starbursts, and thus are more similar to the $z\gtrsim6$ QSO hosts studied to date \citep[e.g.][]{2011AJ....142..101W,2016ApJ...830...53W,2019ApJ...876...99S}. Moreover, they do not assume an $\alpha_\mathrm{CO}$, but instead model the $\alpha_\mathrm{CO}$ and [\ion{C}{I}] abundance simultaneously.  

			\cii\ has been proposed as an even more efficient molecular gas tracer for high-redshift and metal-poor galaxies \citep[e.g.][]{2018MNRAS.481.1976Z,2020A&A...643A.141M}. However, as a multi-phase gas tracer, \cii\ emission is an indirect probe of the bulk of the cold, molecular gas and has not yet been systematically tested against more common molecular gas tracers at $z>6$. To infer the molecular gas masses from [\ion{C}{II}], we applied the mean conversion factor from the calibration of \cite{2018MNRAS.481.1976Z}, $\alpha_\mathrm{[C II]}=31\,\mathrm{M}_\odot/\mathrm{L}_\odot$, folding in the 0.3\,dex standard deviation around this value as the uncertainty. \cite{2018MNRAS.481.1976Z} derive this value based on a comparison of molecular gas masses versus \cii\ emission. For most of the $z=1-5$ main-sequence and starburst galaxies underlying their empirical calibration, the molecular gas masses were derived from CO observations. However, for the local starbursts underlying this calibration, the molecular gas masses were instead mostly derived using relations between the depletion timescale ($t_\mathrm{depl}=M_\mathrm{mol}/SFR$) and specific star formation rate, sSFR (see Sec.~\ref{sub:molecular_gas_masses}).

		\section{Results} 
			\label{sec:results}

			\subsection{Molecular gas mass comparison} 
				\label{sub:molecular_gas_masses}

				Based on the molecular gas mass comparison shown in Fig.~\ref{fig:Mmol_comp}, we find that the CO(2--1)-based molecular gas masses of the three, $z=6.6-6.9$ QSO hosts studied here (top three) are consistent with the $z=5.7-6.4$ QSO hosts with existing CO(2--1) measurements \citep[bottom nine in Fig.~\ref{fig:Mmol_comp},][]{2010ApJ...714..699W,2011ApJ...739L..34W,2015MNRAS.451.1713S,2016ApJ...830...53W,2019ApJ...876...99S}, with $M_\mathrm{mol}\sim(0.8-4.5)\times 10^{10}\,\mathrm{M}_\odot$. Even our newly CO(2--1)-detected QSO host, J2348--3054 at $z=6.9$, already has a molecular gas mass of $\sim 1\times 10^{10}\,\mathrm{M}_\odot$. The fact that these extremely IR-bright $z\gtrsim 6$ QSO hosts all have similarly massive molecular gas reservoirs implies that there is something about these systems that enables massive molecular gas reservoirs to form very rapidly. It also implies that the radiation from the rapidly accreting SMBHs is not dissociating significant amounts of CO. 

				In Fig.~\ref{fig:Mmol_comp}, we compare the CO(2--1)-derived masses to those inferred from \ci, \cii\ and dust-continuum emission. We find that the molecular gas masses inferred from the dust mass measurements are mostly consistent with those inferred from CO(2--1), implying that, on average, the gas-to-dust ratios are consistent with, or only somewhat higher than, the assumed Milky Way value of 100 (also shown in Fig.~\ref{fig:mass_comp}). The only two exceptions, J0305--3150 and J2310+1855, are systems for which the best-fit dust temperatures are low ($\lesssim 30$\,K). It may be that for these galaxies, the gas-to-dust ratios are actually lower (implying more plentiful dust), or that the best-fit solution is inaccurate due to optical depth effects or, poor sampling of the dust SED for J0305--3150. 

				For our our three $z>6.5$ QSO hosts, and J1148+5251 \citep{2009ApJ...703.1338R,2015MNRAS.451.1713S}, we find that the \ci-derived molecular gas masses are consistent with the CO(2--1)-derived values, when assuming both the ULIRG $\alpha_\mathrm{CO}$ value and the high atomic carbon abundance measured in \cite{2021ApJ...908...95H}. If we instead applied a Milky-Way-like $\alpha_\mathrm{CO}$, the [\ion{C}{I}]-derived molecular gas masses would be $\sim5\times$ lower in comparison. In that case, the lower carbon abundances measured by \cite{2020ApJ...902..109B} and \cite{2018ApJ...869...27V} would be appropriate. If the assumed $\alpha_\mathrm{CO}$ is correct, this comparison indicates that the gas in these QSO hosts may be highly carbon enriched and/or subject to several sources of strong heating. We discuss the implications, including additional effects that complicate this comparison, in Sec.~\ref{sec:discussion}. 

				In contrast to the other tracers, we find that the \cii-based molecular gas masses are systematically higher than those derived from the CMB-uncorrected CO(2--1), by a factor of $6.8\pm1.9$ (mean and standard deviation for the whole sample of $z\gtrsim6$ QSO hosts). For J2348--3054, the [\ion{C}{ii}]-derived value is 4.6$\times$ that of the CO(2--1) derived value. Assuming the other values are accurate, this implies that the mean conversion factor derived for local and high-redshift galaxies, $\alpha_\mathrm{[CII]} = 31\,\Msun/L_{\odot}$, is too high for these $z\gtrsim6$ QSO host galaxies. This may be because $\alpha_\mathrm{[CII]}$ is still uncertain for such highly star-forming galaxies. At high sSFR, the calibration from \cite{2018MNRAS.481.1976Z} mostly relies on local starbursts taken from \cite{2017ApJ...846...32D}, which were without CO observations. To derive the molecular gas masses of this sample (and hence calibrate these against the \cii\ luminosity), \cite{2018MNRAS.481.1976Z} rely on two literature scaling relations between the depletion time and sSFR---those of \cite{Sargent_2014} and \cite{2017ApJ...837..150S}---comparing the calibration derived from the average of the two, and from \cite{2017ApJ...837..150S} only (see Fig.~9 of \citealt{2018MNRAS.481.1976Z}). When also using the \cite{Sargent_2014} relation (which is based on a metallicity-dependent $\alpha_\mathrm{CO}$) they find $\alpha_\mathrm{[CII]}$ values of $\leq10$ at the highest sSFRs ($\mathrm{sSFR/sSFR_{MS}}>10$). Thus, they also derive some values consistent with what we find for the $z\gtrsim 6$ QSO host galaxies, depending on the assumed scaling relation.

      \subsection{Mass budget} 
        \label{sub:molecular_versus_other_mass}

				In Fig.~\ref{fig:mass_comp}, we compare the molecular gas masses derived here against the other main mass components of the QSO host galaxies---the dust, black hole, and total (dynamical) mass. We use the dust masses derived here (Sec.~\ref{sub:luminosities} and Appendix~\ref{sec:dust_sed_fitting}). For the black hole masses of our three QSOs, we take the measured values from \cite{2023A&A...676A..71M} and \cite{2022ApJ...941..106F}, which were derived from \ion{Mg}{II} following the method from \cite{2011ApJS..194...45S}. Likewise for the literature sample, we take the \ion{Mg}{II}-based values from \cite{2019ApJ...873...35S}, \cite{2022ApJ...941..106F}, and \cite{2023A&A...676A..71M}. For the dynamical mass estimates, we take the values derived by \cite{2021ApJ...911..141N}, based on $0\farcs25$ observations of \cii\ emission. These dynamical masses were estimated from the best-fit \cii\ half-light radius, rotation velocity, and velocity dispersion (where these values were fit using the parametric kinematic fitting code, \texttt{qubefit}; \citealt{qubefit}).  

				Assuming a ULIRG-like $\alpha_\mathrm{CO}$, the CO(2--1)-derived gas masses (opaque filled symbols with errorbars in Fig.~\ref{fig:mass_comp}) are 3-11 times greater than the black hole masses, $20-190\times$ greater than the dust masses, and $0.1-1\times$ the dynamical masses. In order for the CO(2--1)-derived molecular gas masses (and upper limits) of the three QSOs studied here not to exceed the estimated dynamical masses, the applied CO-to-H$_2$ conversion factors need to be low, similar to the ULIRG value we have adopted. This also means that only a minimal CMB correction (as for the non-LTE, high temperature and density case assumed here) is required. In particular for J2348--3054, the molecular gas mass is almost equivalent (0.98$\times$) to the lower limit on the dynamical mass of $\geq1.2\times10^{10}\,\mathrm{M}_\odot$, whereas the black hole mass is $(3.3\pm1)\times 10^9\,\mathrm{M}_\odot$. If the dynamical masses are reliable, the CO(2--1)-derived molecular gas masses also imply that the molecular gas in most of these systems, especially in J2348--3054, dominates the mass budget, leaving barely any mass in neutral gas or stars within the region traced by \cii. 

				Our finding that the molecular gas comprises a large fraction of the total mass, at least in the central regions, is consistent with earlier findings for some of the literature sample and for J2348--3054 based on dynamical modelling of \cii\ versus CO(6--5)- or \cii-derived gas masses \citep{2021ApJ...917...99Y,2021ApJ...914...36I,2021ApJ...911..141N,2022ApJ...927...21W}. However, it appears to be in tension with the mean stellar masses of $\geq 10^{10}\,\mathrm{M}_\odot$ recently inferred for six $z\gtrsim 6$ QSOs by modelling the spectral energy distribution of their JWST photometry \citep{2023arXiv230904614Y}. The only QSO host with both a dynamical mass \citep[from][]{2021ApJ...911..141N} and stellar mass (from their study) is J0100+2802, for which the dynamical mass is $1.2_{-0.7}^{+0.6}\times 10^{10}$, the CO(2--1)-derived molecular gas mass is $(0.95\pm0.32)\times 10^{10}$ and the stellar mass derived by \citep{2023arXiv230904614Y} is $<30\times10^{10}$. For this source, which is in both our samples, the upper limit could still yield a consistent mass budget. Better photometry are needed for the full sample of QSO hosts studied here to accurately characterise their stellar masses and compare these to the other baryonic components.

				The above argument that the molecular gas masses dominate the mass budget, relies on a big `if'; in reality, the measured dynamical masses are highly uncertain and could be systematically underestimated if the measured sizes and/or circular velocities are too small, or if the assumed geometry is incorrect \citep[see discussion in][]{2021ApJ...911..141N}. If the gas is not in a planar stucture or if the dynamics are not supported by gravity (but instead by large outflows or turbulence), then the models assumed in fitting the dynamics are inappropriate.  The dynamical masses may also be underestimated if these $z\gtrsim6$ QSO host systems are not in dynamical equilibrium. Indeed, all evidence points to these systems being in a state of rapid evolution, implying that the assumptions used in the modelling are not appropriate for these systems and leaving room for higher $\alpha_\mathrm{CO}$ factors.  	

				An additional complication for the dynamical versus gas mass comparison, is that current \cii\ observations may be missing the much fainter, extended component, as shown \cite{2020ApJ...904..131N} based on the stacked \cii\ emission of 20 $z\gtrsim6$ QSO hosts (including those studied here). The much larger extent of this faint component would mean that the measured rotation curves do not probe far enough into the galaxy `disk' to sample the maximum rotation velocity, nor would the measured sizes then represent the whole disk. This is why dynamical masses are quoted within certain radii. In the case of the QSO hosts studied here, we quote values inferred from within twice the \cii\ scale length (2$r_\mathrm{d,[CII]}$). In order for the molecular gas traced by CO(2--1) to make up a small fraction of the total mass within this region, it would have to be significantly more extended than the \cii\ emission. Although we cannot test this scenario at the resolution of our data, it seems unlikely based on low-redshift galaxies \citep[e.g.][]{Blok_2016,Bigiel_2020}. Accurately testing how much of the total mass is in the form of molecular gas, will require deep, high-resolution observations of the host galaxy (and its environment), as well as better sampling of the CO excitation ladders, enabling robust constraints on $\alpha_\mathrm{CO}$. 
			

      	\subsection{Depletion timescales} 
        	\label{sub:depletion_timescales}

        	Combining the CO(2--1)-derived molecular gas masses and IR-inferred SFRs, we find that these $z\gtrsim6$ QSO hosts have extremely short gas depletion timescales. Our highest-redshift target, J2348--3054, has a high molecular gas mass of $\sim1\times10^{10}\,\mathrm{M}_\odot$. Assuming that 25-60\% of its IR luminosity is attributable to star formation \citep[as for the QSO sample of][]{2013ApJ...772..103L} and following the SFR-IR calibration of \cite{Hao_2011}, this equates to a SFR of $500-1300 \,\mathrm{M}_\odot\,\mathrm{yr}^{-1}$. Combined with the CO(2--1)-derived molecular gas mass, this implies a short depletion timescale ($t_\mathrm{depl} = M_\mathrm{mol}/\mathrm{SFR}$) of only $8-20$ Myr. This value is consistent with the full literature sample, which span SFRs of $400-7000\,\mathrm{M}_\odot\,\mathrm{yr}^{-1}$ and depletion timescales of $4-50$ Myr. These depletion timescales are significantly shorter than for most ULIRGS and high-redshift star-forming galaxies, which have $t_\mathrm{depl} \sim 10^7-10^8$\,Myr \citep[e.g,][]{2005ARA&A..43..677S,2020ARA&A..58..157T}, but are comparable to a handful of hyperluminous starbursts at $z>2$ \citep[e.g.][]{2012A&A...538L...4C,2016MNRAS.457.4406A,2020A&A...635A..27C}. These results imply that IR-luminous $z\gtrsim6$ QSO hosts are undergoing an extreme starburst phase. However, better sampling of the dust-continuum emission is still required to disentangle the AGN and SFR contributions in these systems.


			\begin{figure*}
				\centering
				\includegraphics[width=\textwidth, trim={0cm 0.1cm 0cm 0cm},clip]{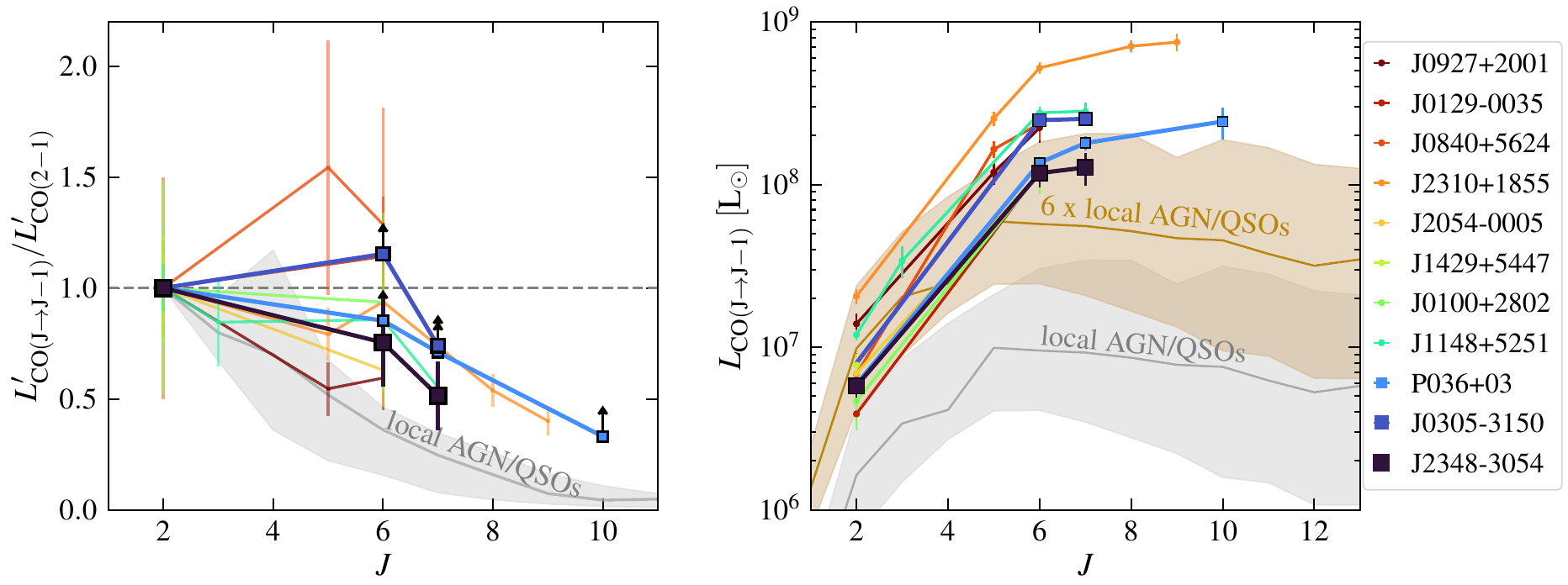}
				\caption{CO excitation ladders for the three $z>6.5$ QSO host galaxies studied here (filled squares), compared to other $z\gtrsim 6$ QSO hosts with CO(2--1) measurements (semi-transparent lines connected by small filled circles) and local AGN and QSO hosts from \cite{2015ApJ...801...72R} and \cite{2016ApJ...829...93K} (grey line and shading indicating the median and 16th to 84th percentiles). The colour-coding indicates the redshift, with the legend at the right providing the name of the corresponding QSO.  Left panel: CO line luminosity ratios relative to CO(2--1). Right panel: CO line luminosities in units of L$_\odot$. Most $z\gtrsim 6$ QSO host galaxies show significantly (factor of $\sim$3$\times$) higher mid-$J$ versus CO(2--1) emission than the local AGN and QSO hosts (near/exceeding the thermalised value). The mid--$J$ CO line emission for most $z\gtrsim 6$ QSO hosts exceeds that of the local AGN and QSO hosts by a greater factor than the CO(2--1) emission, implying that a greater fraction of these more massive molecular gas reservoirs is in more highly excited states. \label{fig:CO_ladders}}	
			\end{figure*} 

			\begin{figure*}
				\centering
				\includegraphics[width=\textwidth, trim={2cm 0.1cm 2cm 1cm},clip]{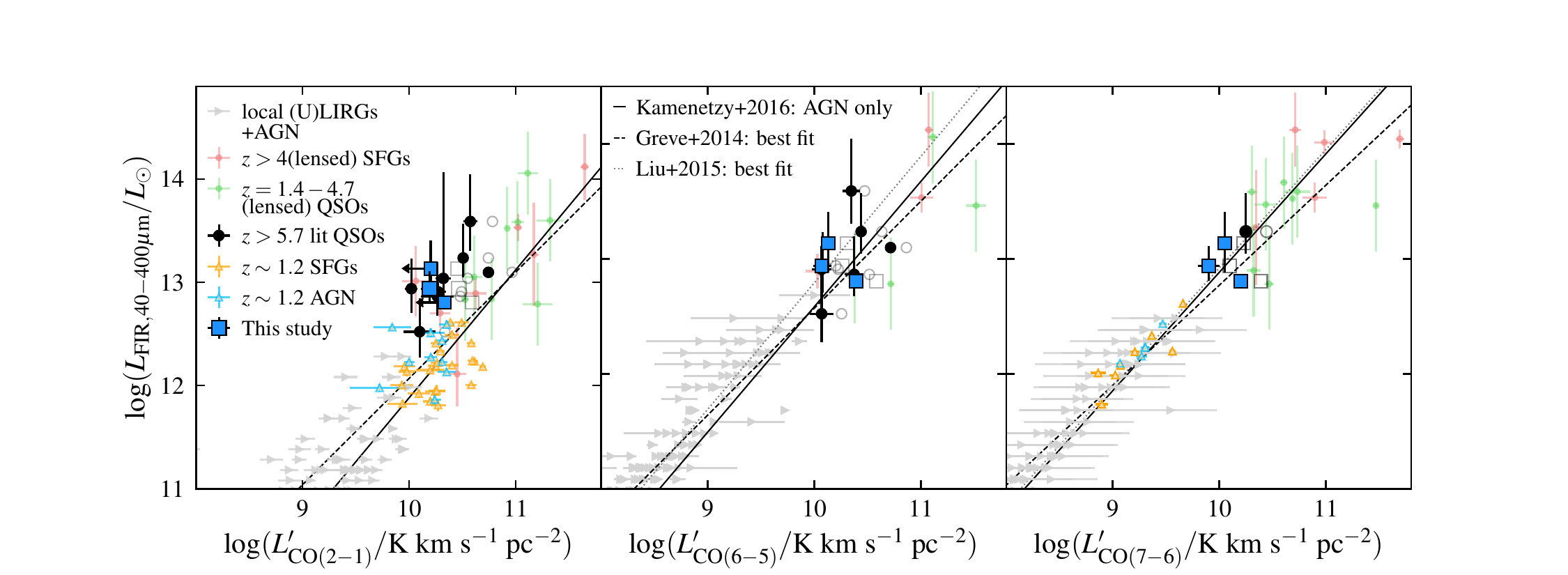}
				\caption{FIR luminosities versus CO line luminosities of our three QSOs (coloured squares), the literature sample of $z\gtrsim 6$ QSOs (filled, coloured circles), the $z>3$ SFGs in Table~\ref{tab:highz_SFGs} (orange circles), the $z\sim 1.2$ star-forming galaxies and AGN from \cite{2020ApJ...890...24V} (orange and blue triangles), and the local (U)LIRGS, AGN, and QSOs from \cite{2015ApJ...801...72R} and \cite{2016ApJ...829...93K} (grey, right-pointing triangles). The $z\gtrsim 6$ QSOs follow the literature relations for the CO(6--5) and CO(7-6) transitions but are shifted to 2-6$\times$ lower CO(2--1) line luminosities than predicted by these relations. \label{fig:Lco_Lfir}}	
			\end{figure*}

			\subsection{CO line excitation} 
				\label{sub:co_excitation}

				The fact that we do not detect the CO(2--1) emission of P036+03 and J0305--3150 implies that their average CO excitation is more extreme than that of the $z=6.4$ QSO host J1148+5251, which is the best-studied CO line excitation ladder of any $z\gtrsim6$ QSO studied to date. To investigate this further, we compare the CO line excitation ladders to those of the local AGN and QSOs investigated by \cite{2016ApJ...829...93K} and \cite{2015ApJ...801...72R} in Fig. \ref{fig:CO_ladders}. Unlike for the local AGN and QSOs, many of the $z\gtrsim 6$ QSO hosts have thermalised or even supra-thermal mid-to-low $J$ line ratios (left panel) \citep[as also shown in][]{2009ApJ...703.1338R,2010ApJ...714..699W,2011ApJ...739L..34W}. To test whether this is caused by significantly lower CO(2--1) or significantly higher mid-$J$ CO emission we also show the CO excitation ladder in units of $\mathrm{L_\odot}$ (right hand panel). From this comparison, it seems that the difference between the samples is much more pronounced at $J=6$ and $7$ than for $J=2$, implying that there is an even greater fraction of molecular gas in these, more highly excited states in the $z\gtrsim 6$ versus $z\sim0$ QSOs. We discuss this further in Sec.~\ref{sec:discussion}. 

				Based on our results, in combination with the literature $z>5.7$ QSO hosts, we caution that when inferring molecular gas masses from mid-$J$ observations of these extreme systems, it is important to apply a higher excitation factor than what is measured for local AGN and ULIRGs ($r_{6,1}=0.23$) or $z=2-3$ star-forming galaxies ($r_{6,1}=0.28$, \citealt{2020ApJ...902..109B}). For the full sample of $z\gtrsim6$ QSOs with observations of both the CO(6--5) and CO(2--1) transitions, we measure a mean and standard deviation on the CO(6--5)-to-CO(2--1) line luminosity ratio of $r_{6,2}=0.9\pm0.2$. Assuming that the first three levels are thermalised, this also equates to a CO(6--5)-to-CO(1-0) line luminosity ratio of $r_{6,1}=0.9\pm0.2$.  


			\subsection{FIR versus CO line luminosities} 
				\label{sub:fir_versus_co_line_luminosities}

				We test whether these $z\gtrsim6$ QSO hosts follow empirical relations between the CO line luminosities and FIR luminosities, and hence whether they are consistent with trends found for local (U)LIRGs and $z=1-4$ star-forming galaxies, in Fig. \ref{fig:Lco_Lfir}.  The FIR luminosities of the $z\gtrsim6$ QSO hosts are corrected for CMB effects as described in Sec.~\ref{sub:luminosities}. However, for the CO line luminosities, we show both the measured values and those calculated using a CMB correction consistent with the high-density and temperature, non-LTE case of \cite{2013ApJ...766...13D}. To compare FIR luminosities derived for different integration widths, we apply two scaling factors. For the sample of \cite{2020ApJ...890...24V}, we use the provided scaling $L_\mathrm{TIR, 8-1000 \mu m} = 1.2 L_\mathrm{FIR, 40-400 \mu m}$. We also assumed $L_\mathrm{TIR, 8-1000 \mu m} = 1.45 L_\mathrm{FIR, 42.5-122.5 \mu m}$ \citep{2011A&A...530L...8D}. We compare to three sets of $L_\mathrm{FIR}$ versus $L^\prime_\mathrm{CO}$ relations derived by \cite{Greve_2014}, \cite{Liu_2015} and \cite{2016ApJ...829...93K}. We adopt the FIR wavelength range convention of $L_\mathrm{FIR, 40-400 \mu m}$ also used in \cite{Liu_2015} and scale the \cite{Greve_2014} relations using the mean value of their local sample, $ \left< L_\mathrm{TIR, 8-1000 \mu m} / L_\mathrm{FIR, 30-500 \mu m} \right> = 1.77$. By converting in this manner, we are introducing an additional level of uncertainty. However, the conversion is crucial to avoid systematic offsets. 

				We find that the measured CO(2--1) line luminosities of the $z\gtrsim 6$ QSOs are offset, on average, to 2-6$\times$ lower values that predicted from the relations with $L_\mathrm{FIR}$ (although consistent with the scatter of lower-redshift sources), whereas the higher-$J$ transitions are consistent with these literature relations.  This minor systematic offset may be the result of various factors: 1) the FIR luminosities of QSOs may be the result of both the SFRs \emph{and} a significant contribution from the AGN, such that this comparison no longer reflects the correlation between the molecular gas mass and SFR, 2) the combined effects of the heating by, and contrast against the CMB serves to decrease the observed CO(2--1) line flux more than the $J=6$ or $7$ transitions, or 3) previous calibrations do not accurately constrain the slope of the relation as they did not sample the extremely high $L_\mathrm{FIR}$ regime well enough. 

				To test whether the dust is being significantly heated by the AGN in these QSOs (explanation 1), would require consistently well-sampled dust SEDs, with which the different dust components can be disentangled, as in \cite{2013ApJ...772..103L}. By modelling the contributions to the FIR luminosities of 11, $z>5$ QSO hosts, they find that the AGN contribute 40-75\% to the FIR luminosity. Subtracting the mean AGN contribution of their sample, shifts the $z\gtrsim6$ QSO hosts studied here close to the $L_\mathrm{CO(2-1)}^\prime$ versus $L_\mathrm{FIR}$ relation (systematically above but consistent with the scatter), and systematically below but within the scatter of the $L_\mathrm{CO(6-5)}^\prime$ and $L_\mathrm{CO(7-6)}^\prime$ versus $L_\mathrm{FIR}$ relations. In Fig.~\ref{fig:Mmol_comp}, we also show that most of the $z\sim1.2$ AGN from \cite{2020ApJ...890...24V} lie towards lower CO(2--1) line luminosities than predicted from the total FIR luminosities. However, when subtracting the AGN contribution (which they model in their SED fitting) this systematic offset also disappears. The adopted CMB correction (faint outlined squares and circles) also brings the values more in line with the CO(2--1) relation, while still keeping the high-$J$ lines consistent with these relations. However, it may well be that no correction is needed, due to the extremely high gas densities and temperatures in these QSO hosts. Isolating the CMB effects from the heating and turbulence generated through star formation and the central QSO will require dedicated simulations of these systems.



		\section{Discussion} 
			\label{sec:discussion}

				The high line ratios and implied low $\alpha_\mathrm{CO}$ of these $z\gtrsim 6$ QSO hosts indicate the presence of more extreme molecular ISM conditions than in the less-luminous, and mostly gas-poorer, local AGN and QSO hosts and ULIRGs we have compared them to. These are likely caused by a combination of both strong ISRFs (and high X-ray fluxes), and the presence of compact, dense gas reservoirs. Our qualitative argument for this can be summarised as follows.

				To keep a significant component of molecular gas (without dissociating it) this gas must be extremely dense and/or well shielded. Both appear to be the case, in that the $z\gtrsim6$ QSO hosts galaxies studied here already have a high implied gas-to-dust ratios and small measured sizes; with \cii\ half-light radi of 0.8 to 4.5 kpc \citep{2021ApJ...911..141N,2022ApJ...927...21W,2022ApJ...927..152M}. The extent of this \cii\ emission is significantly smaller (by at least a factor of five) than the scale of the emission in the local AGN and QSO hosts we are comparing to. Compounding this, the molecular gas masses inferred via the same $\alpha_\mathrm{CO}$ conversion factors are a few times larger on average (or of the same order as) the local comparison sample; the molecular gas masses of the $z\gtrsim6$ QSO hosts are $\sim$few\,$\times10^{10} \mathrm{M}_\odot$ versus the $\sim$few$\,\times10^{9} \mathrm{M}_\odot$ measured for gas-rich local AGN and QSO hosts \citep[e.g.][]{2012MNRAS.426.2601P,2015ApJ...801...72R,2016ApJ...829...93K}. Indeed, \cite{2019ApJ...876...99S} infer densities of a $10^5-10^6$\,cm$^{-3}$ for three of the literature QSO hosts included here. Not only do the $z\gtrsim 6$ QSO hosts have more molecular gas per unit volume than the local AGN and QSO hosts, they also have an order of magnitude higher (on average) FIR luminosities than the local AGN and QSO hosts. The implied SFRs of $400-7000$ M$_\odot$ yr$^{-1}$, which stem from a smaller volume than in the local galaxies, imply higher ISRF strengths (with $G_0 \gtrsim 10^3$ in Habing units for some of the $z\gtrsim6$ QSO hosts included here, according to the modelling performed by \citealt{2019ApJ...876...99S} and \citealt{2022A&A...662A..60D}). In combination, this is consistent with a picture in which the molecular clouds in these $z\gtrsim6$ QSO hosts are more closely spaced and/or much denser than in local AGN and QSO hosts and are illuminated by stronger radiation fields. 

				The high CO(7--6)- and CO(6--5)-to-CO(2--1) line ratios (consistent with thermalised and even supra-thermal CO) imply the presence of dense, warm, and highly turbulent gas. Similarly high line ratios have been found in the local photodissociation region (PDR) known as the Orion Bar. Modelling the far more extensive set of molecular and atomic emission lines (including isotopologues) observed for this region, several groups \citep{2015ApJ...812...75G,2016Natur.537..207G,2019A&A...622A..91G,2017A&A...598A...2A,2018A&A...617A..77P} have found evidence for clumps of extremely high-density ($n_H\sim10^5-10^6$\, cm$^{-3}$), and high-temperature ($T\sim150-300$s\,K) molecular gas. To produce such high line ratios, the gas temperatures may exceed the average dust temperature \citep[see also][]{2014PhR...539...49K,2021ApJ...908...95H}. Several sources of heating could elevate the kinetic temperatures of the gas. High star formation rate densities should give rise to more cosmic rays (compared to local ULIRGS/AGN), providing an additional, strong source of heating that increases the gas temperatures and causes the CO excitation ladders to peak at higher-$J$ \citep[e.g.][]{2012MNRAS.426.2601P,2023MNRAS.519..729B}. If large amounts of energy (e.g. from star formation and/or the QSO) are being injected into such a small volume of gas, then this gas is likely also more turbulent, providing another source of heating and implying a lower intrinsic $\alpha_\mathrm{CO}$ \citep[e.g.][]{2011MNRAS.418..664N,2012ApJ...751...10P}. In combination with the high densities, this increase in velocity dispersion would serve to counteract any increase in $\alpha_\mathrm{CO}$ caused by a higher ISRF or higher cosmic ray ionisation rate \citep[e.g.][]{2015MNRAS.452.2057C}. 

        		The high carbon abundances implied by our results are also consistent with the presence of highly turbulent gas. Mixing caused by strong turbulence may result in elevated [\ion{C}{I}]/H$_2$ in the cloud interiors, thereby giving rise to bright [\ion{C}{I}] emission \citep{1995ApJ...440..674X,2015MNRAS.448.1607G}. Moreover, strong contributions from cosmic rays, X-rays and/or shocks \citep{2014A&A...562A..96I,2015A&A...578A..95I} would enhance the brightness of [\ion{C}{i}] compared to low-$J$ CO emission \citep{2014A&A...562A..96I,2015A&A...578A..95I,2018MNRAS.478.1716P,2022MNRAS.510..725P}. These additional heating sources may drive atomic carbon out of the state of LTE asumed here. Indeed, \cite{2021ApJ...908...95H} and \cite{2022MNRAS.510..725P} show that the [\ion{C}{i}] excitation of high-redshift star-forming galaxies and local (U)LIRGS is best described by non-LTE radiative transfer models; with the LTE assumption easily accounting for a factor of a few too high $M_\mathrm{[CI]}$. In that case, a lower atomic carbon abundance than what we assumed may still be appropriate.  Testing the impact of turbulence, cosmic rays, and X-rays will require better-sampled CO excitation ladders, [\ion{C}{I}] transitions (including the addition of the [\ion{C}{I}](1--0) line) and dust spectral energy distributions than are currently available for these $z\gtrsim6$ QSO hosts. 

\begin{figure*}
				\centering
				\includegraphics[width=\textwidth, trim={0.6cm 0.5cm 0.3cm 1cm},clip]{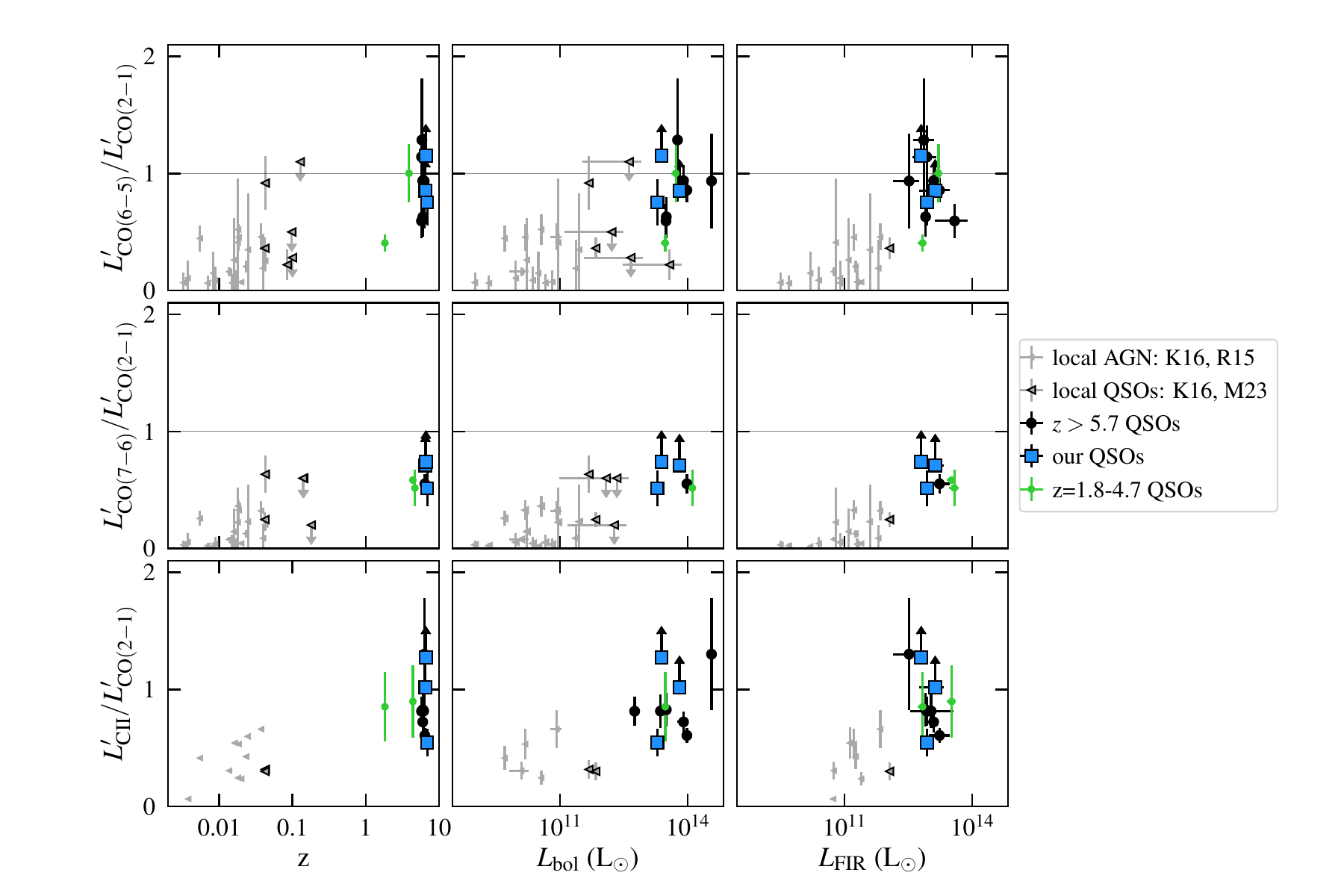}
				\caption{ Line luminosity ratios versus redshift (left column), $L_\mathrm{bol}$ (middle), and $L_\mathrm{FIR, 40-400\mu\,m}$ (right). The data for the three QSO hosts studied here are depicted via filled squares, whereas the previously studied $z\gtrsim6$ QSO hosts are depicted by filled circles. The colour-coding is the same as for Fig.~\ref{fig:CO_ladders}. We also show the local AGN and QSO hosts from \cite{2016ApJ...829...93K} and \cite{2023arXiv231010235M} (grey and black, respectively) and a few $z=1-4$ QSO hosts from \cite{2008ApJ...686L...9R,2009ApJ...703.1338R,2008A&A...491..173A}, corrected for lensing where appropriate (green circles). The bolometric luminosities of the different samples were inferred in different ways: scaling the 3000\AA\ luminosity following \cite{2011ApJS..194...45S} for the $z\gtrsim6$ QSO hosts, scaling the luminosity at 14–150 keV following \cite{2017ApJS..233...17R} for the local sample (BASS catalogue of \citealt{2022ApJS..261....2K}) and scaling the [\ion{O}{III}]$\lambda 5007$\AA\ line luminosity for the sample of \cite{2023arXiv231010235M}. For the $z=1-4$ QSO hosts we use values determined from the [\ion{O}{III}]$\lambda 5007$\AA\ line, 6\,\textmu m emission, or the full spectral fitting. The FIR luminosities of the $z<4$ samples were taken from the same studies, based on their SED fitting, whereas for the $z\gtrsim6$ QSOs, we use the values derived in this work. \label{fig:lumratio_versus}}		
			\end{figure*}

       			Given our data, it is unclear to what extent the extreme conditions within the molecular gas of these $z\gtrsim 6$ QSO host galaxies are driven by star formation versus the luminous accreting black holes. In Fig.~\ref{fig:lumratio_versus}, we show that the higher line ratios of the $z\gtrsim 6$ versus $z\sim 0$ QSO (and AGN) hosts correlate with proxies for both the AGN power and star formation activity. The left column illustrates that such luminous QSO hosts have not been observed in CO in the local Universe; instead the most powerful AGN in the local Universe are hosted in elliptical galaxies with little molecular gas. However, we compare to these $z\sim0$ systems to span a wider range of bolometric and FIR luminosities. We use the bolometric luminosity $L_\mathrm{bol}$ (middle column) as a proxy of the AGN power\footnote{representing the total luminosity produced by the accretion disk} and the FIR luminosity, $L_\mathrm{FIR}$ (right-most column) as a proxy for the SFR. The fact that these three line ratios correlate with the FIR luminosity is expected, e.g. high star formation rate densities lead to greater dust heating (and hence higher $L_\mathrm{FIR}$) and also higher CO excitation \citep[e.g.][]{2014MNRAS.442.1411N}. We also find a clear correlation with the bolometric luminosity, albeit with a large scatter. This scatter can be attributed to the different, indirect methods used to infer the $L_\mathrm{bol}$: scaling the 3000\AA\ luminosity following \cite{2011ApJS..194...45S}\footnote{$L_\mathrm{bol}=5.15\,\lambda L_\mathrm{\lambda,3000}$} for the $z\gtrsim6$ QSO hosts, scaling the luminosity at 14–150 keV following \cite{2017ApJS..233...17R} for the local sample (BASS catalogue of \citealt{2022ApJS..261....2K}) and scaling the [\ion{O}{III}]$\lambda 5007$\AA\ line luminosity\footnote{via $L_\mathrm{bol} = 3500 L_\mathrm{[O III]}$} for the sample of \cite{2023arXiv231010235M}. Thus, neither the tightness nor slope of these correlations allow us to disentangle the impact of star formation versus AGN activity on the molecular gas.   

        		The lack of high-$J$ CO emission and inhomogeneous X-ray coverage also do not allow us to disentangle the impact of star formation versus AGN activity.  These QSOs may be heavily obscured and have inhomogeneous X-ray coverage. Two of our sample (J0305--3150 and J2348--3054) have only upper limits on the X-ray fluxes from Chandra, with net counts at 0.5-7 keV of $\leq15$ and $\leq21$ \citep[][]{2021ApJ...908...53W}, whereas P036+03 and two of the literature sample, J0100+2802 and J1429+5447, are detected with XMM Newton \citep{2021MNRAS.504..576M,https://doi.org/10.48550/arxiv.2305.02347}. For this sample of five, the data reveal no correlation between the CO line ratios probed here and the X-ray luminosity (e.g. at 2-10 keV). Future X-ray observations and observations of high-$J$ CO transitions, up to CO(11--10), will enable tests against predicted trends with the X-ray flux \citep[e.g.][]{2019MNRAS.490.4502V}.

        	Overall, we find little evidence for any loss of signal due to the contrast against the CMB. The lack of a large systematic offset from empirical relations between the CO line versus CMB-corrected FIR luminosities indicates that little correction for the CMB contrast is needed for the CO lines. Likewise, the low $\alpha_\mathrm{CO}$ value needed to ensure the molecular gas masses do not exceed the dynamical masses also indicates that minimal correction for the CMB contrast is needed. However, the CMB can have multiple effects, which we cannot disentangle from variations in the ISRF and/or ISM density: it is a stronger background against which the lines are observed, but it is also an additional source of heating, thereby changing the temperature and density structure of the molecular ISM. As shown by \cite{2013ApJ...766...13D}, heating by the CMB can serve to make the CO excitation ladders peak at higher $J$, similar to effects of heating through strong ISRFs, cosmic rays, or X-rays. To disentangle these effects it is crucial to observe CO, [\ion{C}{I}], and dust-continuum emission from samples of QSOs across different epochs, with the same high IR luminosities and molecular gas masses. 
				


		We check whether the compactness (proxy for density) impacts the molecular gas conditions as traced by the relative line strengths. To this end, high-resolution observations of the \cii\ and FIR emission are available for our set of QSOs (P036+03, J0305--3150 and J2348--3054) and four of the literature sample (J0129-0035, J2054-0005, J0100+2802, and J1148+5251) \citep{2020ApJ...904..130V, 2021ApJ...911..141N,2022ApJ...927..152M,2023arXiv230601644T}. Measuring sizes has so far proven challenging, with different assumed surface brightness profiles and modelling tools being applied. Nonetheless, by any metric, J2348--3054 is the most compact of the seven galaxies ($r_\mathrm{eff,[C II]}\sim 0.46$\,kpc). J0129--0035 and J2054--0005 are the next most compact ($r_\mathrm{eff, [C II]}\sim 0.8$\, kpc). We find marginal evidence that the compactness is related to the CO excitation. Whereas J2348--3054 and J0129--0035 exhibit some of the highest CO(6--5)-to-CO(2--1) line ratios, J1148+5251, the most extended of these resolved $z\gtrsim 6$ QSOs, has line ratios consistent with the sample mean (neither at the low nor high end). The fact that the two most compact sources seem to have the highest excitation is consistent with the picture of a denser and warmer ISM in galaxies with higher UV flux per unit volume (originating either from the central black hole or star formation). However, the sample size with both multi-line observations and resolved \cii\ emission remains too small to draw any strong conclusions. 

        We also attempt to test whether the presence of companions (and hence interactions) affects the relative line strengths or molecular gas masses. J2348--3054 and J2054-0005 have no confirmed companions, whereas J0305--3150 and J0100+2802 do \citep[][]{2019ApJ...874L..30V,2023arXiv230601644T,2023ApJ...951L...4W}. J1429+5447 is a known merger \citep[e.g.][]{2022A&A...664A..39K}. Of these, J0305--3150 has the highest mid-to-low-$J$ CO line ratios, considering the upper limit on CO(2--1). J0100+2802 and J1429+5447 are consistent with the mean excitation of the full sample. Within this small sample, we find no clear connection yet between the presence of companions and the molecular gas masses or excitation. However, future JWST observations, as in \cite{2023ApJ...951L...4W}, will help shed light on this issue.



	\section{Conclusions} 
		\label{sec:conclusions}

			In this paper, we present the VLA Ka Band observations of CO(2--1) in three QSO host galaxies at $6.5 < z <6.9$, thereby probing their cold molecular gas reservoirs. These three QSO hosts were already detected in \cii\ emission and the underlying FIR continuum \citep{2015ApJ...805L...8B,2016ApJ...816...37V} and in CO(6--5), CO(7-6) and \ci\ emission \citep{2017ApJ...851L...8V}. For the highest-redshift QSO host, J2348--3054 at $z=6.9$, we robustly detected the CO(2--1) emission, revealing a molecular gas reservoir of $\sim 1\times 10^{10}\,\mathrm{M}_\odot$ already 780 Myr after the Big Bang. However, for the other two QSO hosts, P036+03 and J0305--3150, the CO(2--1) observations are consistent with noise, yielding $4\sigma$ upper limits on the molecular gas masses of $(1.0-1.5) \times 10^{10}\,\mathrm{M}_\odot$. We detect no underlying continuum emission for these three targets, nor any CO or continuum emission from their known companions. 

			We combine these three QSO hosts with the full set of nine $z>5$ QSO host galaxies that already have CO(2--1) observations (Table~\ref{tab:lit_QSOs}), thereby performing a broader study of trends in the molecular gas masses and CO excitation. To this end, we consistently re-derive the FIR luminosities and dust masses, taking CMB effects into account. Based on the CO(2--1) observations, and the assumptions of a ULIRG $\alpha_\mathrm{CO}=0.8$\,M$_\odot$ (K km s$^{-1}$ pc$^2$)$^{-1}$ (see \citealt{2013ARA&A..51..105C}) and thermalised low-$J$ transitions ($r_\mathrm{2,1}=1$), the molecular gas masses of the three QSO hosts studied here are consistent with the literature sample, which spans $(0.8 - 5) \times 10^{10}\, \mathrm{M}_\odot$. These values are mostly consistent with those inferred from the dust masses, assuming a gas-to-dust ratio of 100. However, the molecular gas masses estimated from the \cii\ line luminosity using the mean $\alpha_\mathrm{[C~II]}$ of \cite{2018MNRAS.481.1976Z} are a factor of $6.8\pm1.9$ higher than the CO(2--1)-based values, reflecting a potential decrease in $\alpha_\mathrm{[C~II]}$ with sSFR. In order for these CO(2--1)-derived values to also be consistent with those inferred from \ci, atomic carbon must be already be abundant ([\ion{C}{i}]/H$_2 \sim 9 \times 10^{-5}$) and/or be undergoing significant heating through cosmic rays, X-rays, turbulence, and/or shocks. 

      		The three QSO hosts studied here have high CO excitation, with CO(7--6)- and CO(6--5)- versus CO(2--1) line ratios consistent with or higher than most of the $z\gtrsim6$ QSO hosts studied previously \citep{2010ApJ...714..699W,2011ApJ...739L..34W,2013ApJ...773...44W,2015MNRAS.451.1713S}. For P036+03 and J0305--3150, the gas excitation is significantly higher than for the $z=6.4$ QSO J1148+5251 \citep{2009ApJ...703.1338R,2015MNRAS.451.1713S}, implying more extreme physical conditions in these systems. Despite the small range of molecular gas masses, the combined set of $z\gtrsim6$ QSO hosts spans a wide range in CO excitation. Nonetheless, the line ratios studied here are systematically higher than for local AGN and QSO host galaxies with observations of the same transitions; the mean CO(6--5)-to-CO(2--1) line luminosity ratio is $r_\mathrm{6,2} = 0.9 \pm 0.2$ (2.5$\times$ higher than in local AGN and QSOs) and the mean \cii-to-CO(2--1) line luminosity ratios is $1.0 \pm 0.4$ (2$\times$ higher than in local AGN and QSOs). Assuming that the CO(2--1) emission is thermalised, the mean CO(6--5)-to-CO(1--0) line luminosity ratio is also $r_\mathrm{6,1} = 0.9 \pm 0.2$. These higher line ratios should be considered when inferring the molecular gas masses of $z>5$ QSO hosts from only their mid-$J$ CO lines.

      		Comparing the CO(2--1) observations to the existing dynamical mass measurements, we find that the mass budget of the three QSOs studied here is likely dominated by the molecular gas. In particular, J2348--3054 has a molecular gas mass consistent with the lower limit on the dynamical mass, leaving little mass in neutral gas or stars within the \cii-emitting region. If the dynamical masses based on the resolved [\ion{C}{II}] observations are correct, these CO(2--1) observations imply that the CO-to-H$_2$ conversion factor must be low---similar to the ULIRG value---to yield molecular gas masses that are lower than the dynamical masses. Thus, little correction for the contrast against the CMB is needed for these bright, compact systems. Combined with the high CO excitation, these results imply the presence of warm and dense molecular gas reservoirs. In combination with the literature sample, we find that the CO(2--1)-derived molecular gas masses are $20-190\times$ higher than the dust masses, implying the presence of significant, enriched reservoirs of dense molecular gas and dust. Moreover, the CO(2--1)-derived molecular gas masses are $3-11\times$ higher than the black hole masses.

      		Our new CO(2--1) observations support a picture in which these IR-luminous $z\gtrsim6$ QSO host galaxies have rapidly assembled copious amounts of molecular gas (either through accretion or mergers). Combining the previously detected CO(6--5), CO(7--6), and \ci\ emission with the new CO(2--1) constraints, indicates that all three QSO hosts studied here harbour significant amounts of warm and dense, metal-enriched gas---indicating that these are already highly evolved galaxies. In particular, the newly CO(2--1)-detected $z=6.9$ QSO host J2348--3054, has properties consistent with a massive starburst that is rapidly coevolving with the luminous central AGN (as found previously for some of the literature sample). 

      		With these highest-redshift CO(2--1) observations, we confirm that it is possible to detect the cold, molecular gas traced by CO(2--1) emission at $z>6.5$ within $\sim15$\, hours of VLA time. These observations are also now feasible with ALMA Band 1 (although it remains less efficient than the VLA at $z>5$). Observing CO(3--2) is also now feasible up $z=8.8$ with ALMA's Band 1, whereas [\ion{C}{i}](1--0) can soon be observed with ALMA's Band 2, up to $z=6.3$. Complementing the existing plentiful \cii\ observations of $z\gtrsim6$ QSO hosts with observations of the cold, diffuse molecular gas is therefore achievable, albeit time-intensive. Next-generation facilities, such as the Next-Generation VLA \citep{ngVLA_book}, will enable such observations to be conducted even more efficiently. The next step is to systematically model the average conditions within these massive molecular gas reservoirs---constraining the turbulence within the molecular gas, the carbon abundance and $\alpha_\mathrm{CO}$. To achieve this, it is now crucial to obtain QSO samples with a homogeneous coverage of CO (up to $J>8$), [\ion{C}{i}], and dust emission. The compilation of $z\gtrsim6$ QSO hosts studied here will serve as a crucial basis for this work.




	\begin{acknowledgements}
		This paper makes use of the following VLA data: 16A-160 and 17A-336. The National Radio Astronomy Observatory is a facility of the National Science Foundation operated under cooperative agreement by Associated Universities, Inc. This work made use of the public code package \texttt{Interferopy} \citep{2021ascl.soft12005B}\footnote{\url{https://interferopy.readthedocs.io/en/latest/}} as well as \texttt{CASA} \citep{2022PASP..134k4501C}. L.A.B., R.A.M., M.N., and F.W. acknowledge support from the ERC Advanced Grant 740246 (Cosmic\_Gas). R.A.M. acknowledges support from the Swiss National Science Foundation (SNSF) through project grant 200020\_207349. EdC gratefully acknowledges the Australian Research Council as the recipient of a Future Fellowship (project FT150100079) and the ARC Centre of Excellence for All Sky Astrophysics in 3 Dimensions (ASTRO 3D; project CE170100013). The authors wish to thank F. Valentino for useful discussions that improved the quality of this work. We also thank the anonymous referee, whose feedback improved our interpretation. 
	\end{acknowledgements}  
	\vspace{5mm}

	\bibliographystyle{aa}
	\bibliography{mybib}

\begin{thebibliography}{185}
\expandafter\ifx\csname natexlab\endcsname\relax\def\natexlab#1{#1}\fi

\bibitem[{{Alaghband-Zadeh} {et~al.}(2013){Alaghband-Zadeh}, {Chapman},
  {Swinbank}, {Smail}, {Danielson}, {Decarli}, {Ivison}, {Meijerink}, {Weiss},
  \& {van der Werf}}]{2013MNRAS.435.1493A}
{Alaghband-Zadeh}, S., {Chapman}, S.~C., {Swinbank}, A.~M., {et~al.} 2013,
  \mnras, 435, 1493

\bibitem[{{Andreani} {et~al.}(2018){Andreani}, {Retana-Montenegro}, {Zhang},
  {Papadopoulos}, {Yang}, \& {Vegetti}}]{2018A&A...615A.142A}
{Andreani}, P., {Retana-Montenegro}, E., {Zhang}, Z.-Y., {et~al.} 2018, \aap,
  615, A142

\bibitem[{{Andree-Labsch} {et~al.}(2017){Andree-Labsch}, {Ossenkopf-Okada}, \&
  {R{\"o}llig}}]{2017A&A...598A...2A}
{Andree-Labsch}, S., {Ossenkopf-Okada}, V., \& {R{\"o}llig}, M. 2017, \aap,
  598, A2

\bibitem[{{Ao} {et~al.}(2008){Ao}, {Dinh-V-Trung}, {Lim}, {Yang}, \&
  {Matsushita}}]{2008AJ....136.1118A}
{Ao}, Y., {Dinh-V-Trung}, {Lim}, J., {Yang}, J., \& {Matsushita}, S. 2008, \aj,
  136, 1118

\bibitem[{{Aravena} {et~al.}(2008){Aravena}, {Bertoldi}, {Schinnerer}, {Weiss},
  {Jahnke}, {Carilli}, {Frayer}, {Henkel}, {Brusa}, {Menten}, {Salvato}, \&
  {Smolcic}}]{2008A&A...491..173A}
{Aravena}, M., {Bertoldi}, F., {Schinnerer}, E., {et~al.} 2008, \aap, 491, 173

\bibitem[{{Aravena} {et~al.}(2016){Aravena}, {Spilker}, {Bethermin},
  {Bothwell}, {Chapman}, {de Breuck}, {Furstenau}, {G{\'o}nzalez-L{\'o}pez},
  {Greve}, {Litke}, {Ma}, {Malkan}, {Marrone}, {Murphy}, {Stark}, {Strandet},
  {Vieira}, {Weiss}, {Welikala}, {Wong}, \& {Collier}}]{2016MNRAS.457.4406A}
{Aravena}, M., {Spilker}, J.~S., {Bethermin}, M., {et~al.} 2016, \mnras, 457,
  4406

\bibitem[{{Ba{\~n}ados} {et~al.}(2015){Ba{\~n}ados}, {Decarli}, {Walter},
  {Venemans}, {Farina}, \& {Fan}}]{2015ApJ...805L...8B}
{Ba{\~n}ados}, E., {Decarli}, R., {Walter}, F., {et~al.} 2015, \apjl, 805, L8

\bibitem[{Barvainis \& Ivison(2002)}]{Barvainis_2002}
Barvainis, R. \& Ivison, R. 2002, The Astrophysical Journal, 571, 712

\bibitem[{{Beelen} {et~al.}(2006){Beelen}, {Cox}, {Benford}, {Dowell},
  {Kov{\'a}cs}, {Bertoldi}, {Omont}, \& {Carilli}}]{2006ApJ...642..694B}
{Beelen}, A., {Cox}, P., {Benford}, D.~J., {et~al.} 2006, \apj, 642, 694

\bibitem[{{Beelen} {et~al.}(2004){Beelen}, {Cox}, {Pety}, {Carilli},
  {Bertoldi}, {Momjian}, {Omont}, {Petitjean}, \&
  {Petric}}]{2004A&A...423..441B}
{Beelen}, A., {Cox}, P., {Pety}, J., {et~al.} 2004, \aap, 423, 441

\bibitem[{{Bertoldi} {et~al.}(2003{\natexlab{a}}){Bertoldi}, {Carilli}, {Cox},
  {Fan}, {Strauss}, {Beelen}, {Omont}, \& {Zylka}}]{2003A&A...406L..55B}
{Bertoldi}, F., {Carilli}, C.~L., {Cox}, P., {et~al.} 2003{\natexlab{a}}, \aap,
  406, L55

\bibitem[{{Bertoldi} {et~al.}(2003{\natexlab{b}}){Bertoldi}, {Cox}, {Neri},
  {Carilli}, {Walter}, {Omont}, {Beelen}, {Henkel}, {Fan}, {Strauss}, \&
  {Menten}}]{2003A&A...409L..47B}
{Bertoldi}, F., {Cox}, P., {Neri}, R., {et~al.} 2003{\natexlab{b}}, \aap, 409,
  L47

\bibitem[{Bigiel {et~al.}(2020)Bigiel, de~Looze, Krabbe, Cormier, Barnes,
  Fischer, Bolatto, Bryant, Colditz, Geis, Herrera-Camus, Iserlohe, Klein,
  Leroy, Linz, Looney, Madden, Poglitsch, Stutzki, \& Vacca}]{Bigiel_2020}
Bigiel, F., de~Looze, I., Krabbe, A., {et~al.} 2020, The Astrophysical Journal,
  903, 30

\bibitem[{{Bisbas} {et~al.}(2015){Bisbas}, {Papadopoulos}, \&
  {Viti}}]{2015ApJ...803...37B}
{Bisbas}, T.~G., {Papadopoulos}, P.~P., \& {Viti}, S. 2015, \apj, 803, 37

\bibitem[{{Bisbas} {et~al.}(2023){Bisbas}, {van Dishoeck}, {Hu}, \&
  {Schruba}}]{2023MNRAS.519..729B}
{Bisbas}, T.~G., {van Dishoeck}, E.~F., {Hu}, C.-Y., \& {Schruba}, A. 2023,
  \mnras, 519, 729

\bibitem[{{Bisbas} {et~al.}(2017){Bisbas}, {van Dishoeck}, {Papadopoulos},
  {Sz{\H{u}}cs}, {Bialy}, \& {Zhang}}]{2017ApJ...839...90B}
{Bisbas}, T.~G., {van Dishoeck}, E.~F., {Papadopoulos}, P.~P., {et~al.} 2017,
  \apj, 839, 90

\bibitem[{{Boogaard} {et~al.}(2021){Boogaard}, {Meyer}, \&
  {Novak}}]{2021ascl.soft12005B}
{Boogaard}, L., {Meyer}, R.~A., \& {Novak}, M. 2021, {Interferopy: Analyzing
  datacubes from radio-to-submm observations}, Astrophysics Source Code
  Library, record ascl:2112.005

\bibitem[{{Boogaard} {et~al.}(2020){Boogaard}, {van der Werf}, {Weiss},
  {Popping}, {Decarli}, {Walter}, {Aravena}, {Bouwens}, {Riechers},
  {Gonz{\'a}lez-L{\'o}pez}, {Smail}, {Carilli}, {Kaasinen}, {Daddi}, {Cox},
  {D{\'\i}az-Santos}, {Inami}, {Cortes}, \& {Wagg}}]{2020ApJ...902..109B}
{Boogaard}, L.~A., {van der Werf}, P., {Weiss}, A., {et~al.} 2020, \apj, 902,
  109

\bibitem[{{Bothwell} {et~al.}(2017){Bothwell}, {Aguirre}, {Aravena},
  {Bethermin}, {Bisbas}, {Chapman}, {De Breuck}, {Gonzalez}, {Greve},
  {Hezaveh}, {Ma}, {Malkan}, {Marrone}, {Murphy}, {Spilker}, {Strandet},
  {Vieira}, \& {Wei{\ss}}}]{2017MNRAS.466.2825B}
{Bothwell}, M.~S., {Aguirre}, J.~E., {Aravena}, M., {et~al.} 2017, \mnras, 466,
  2825

\bibitem[{{Bothwell} {et~al.}(2013){Bothwell}, {Smail}, {Chapman}, {Genzel},
  {Ivison}, {Tacconi}, {Alaghband -Zadeh}, {Bertoldi}, {Blain}, {Casey}, {Cox},
  {Greve}, {Lutz}, {Neri}, {Omont}, \& {Swinbank}}]{2013MNRAS.429.3047B}
{Bothwell}, M.~S., {Smail}, I., {Chapman}, S.~C., {et~al.} 2013, \mnras, 429,
  3047

\bibitem[{Bradford {et~al.}(2009)Bradford, Aguirre, Aikin, Bock, Earle, Glenn,
  Inami, Maloney, Matsuhara, Naylor, Nguyen, \& Zmuidzinas}]{Bradford_2009}
Bradford, C.~M., Aguirre, J.~E., Aikin, R., {et~al.} 2009, The Astrophysical
  Journal, 705, 112

\bibitem[{{Capak} {et~al.}(2008){Capak}, {Carilli}, {Lee}, {Aldcroft},
  {Aussel}, {Schinnerer}, {Wilson}, {Yun}, {Blain}, {Giavalisco}, {Ilbert},
  {Kartaltepe}, {Lee}, {McCracken}, {Mobasher}, {Salvato}, {Sasaki}, {Scott},
  {Sheth}, {Shioya}, {Thompson}, {Elvis}, {Sanders}, {Scoville}, \&
  {Tanaguchi}}]{2008ApJ...681L..53C}
{Capak}, P., {Carilli}, C.~L., {Lee}, N., {et~al.} 2008, \apjl, 681, L53

\bibitem[{{Carilli} {et~al.}(2011){Carilli}, {Hodge}, {Walter}, {Riechers},
  {Daddi}, {Dannerbauer}, \& {Morrison}}]{2011ApJ...739L..33C}
{Carilli}, C.~L., {Hodge}, J., {Walter}, F., {et~al.} 2011, \apjl, 739, L33

\bibitem[{{Carilli} {et~al.}(2002){Carilli}, {Kohno}, {Kawabe}, {Ohta},
  {Henkel}, {Menten}, {Yun}, {Petric}, \& {Tutui}}]{2002AJ....123.1838C}
{Carilli}, C.~L., {Kohno}, K., {Kawabe}, R., {et~al.} 2002, \aj, 123, 1838

\bibitem[{{Carilli} {et~al.}(2007){Carilli}, {Neri}, {Wang}, {Cox}, {Bertoldi},
  {Walter}, {Fan}, {Menten}, {Wagg}, {Maiolino}, {Omont}, {Strauss},
  {Riechers}, {Lo}, {Bolatto}, \& {Scoville}}]{2007ApJ...666L...9C}
{Carilli}, C.~L., {Neri}, R., {Wang}, R., {et~al.} 2007, \apjl, 666, L9

\bibitem[{Carilli {et~al.}(2005)Carilli, Solomon, Bout, Walter, Beelen, Cox,
  Bertoldi, Menten, Isaak, Chandler, \& Omont}]{Carilli_2005}
Carilli, C.~L., Solomon, P., Bout, P.~V., {et~al.} 2005, The Astrophysical
  Journal, 618, 586

\bibitem[{{Carilli} \& {Walter}(2013)}]{2013ARA&A..51..105C}
{Carilli}, C.~L. \& {Walter}, F. 2013, \araa, 51, 105

\bibitem[{{Carniani} {et~al.}(2020){Carniani}, {Ferrara}, {Maiolino},
  {Castellano}, {Gallerani}, {Fontana}, {Kohandel}, {Lupi}, {Pallottini},
  {Pentericci}, {Vallini}, \& {Vanzella}}]{2020MNRAS.499.5136C}
{Carniani}, S., {Ferrara}, A., {Maiolino}, R., {et~al.} 2020, \mnras, 499, 5136

\bibitem[{{Carniani} {et~al.}(2019){Carniani}, {Gallerani}, {Vallini},
  {Pallottini}, {Tazzari}, {Ferrara}, {Maiolino}, {Cicone}, {Feruglio}, {Neri},
  {D'Odorico}, {Wang}, \& {Li}}]{2019MNRAS.489.3939C}
{Carniani}, S., {Gallerani}, S., {Vallini}, L., {et~al.} 2019, \mnras, 489,
  3939

\bibitem[{{Carniani} {et~al.}(2013){Carniani}, {Marconi}, {Biggs}, {Cresci},
  {Cupani}, {D'Odorico}, {Humphreys}, {Maiolino}, {Mannucci}, {Molaro},
  {Nagao}, {Testi}, \& {Zwaan}}]{2013A&A...559A..29C}
{Carniani}, S., {Marconi}, A., {Biggs}, A., {et~al.} 2013, \aap, 559, A29

\bibitem[{{CASA Team} {et~al.}(2022){CASA Team}, {Bean}, {Bhatnagar}, {Castro},
  {Donovan Meyer}, {Emonts}, {Garcia}, {Garwood}, {Golap}, {Gonzalez Villalba},
  {Harris}, {Hayashi}, {Hoskins}, {Hsieh}, {Jagannathan}, {Kawasaki},
  {Keimpema}, {Kettenis}, {Lopez}, {Marvil}, {Masters}, {McNichols},
  {Mehringer}, {Miel}, {Moellenbrock}, {Montesino}, {Nakazato}, {Ott}, {Petry},
  {Pokorny}, {Raba}, {Rau}, {Schiebel}, {Schweighart}, {Sekhar}, {Shimada},
  {Small}, {Steeb}, {Sugimoto}, {Suoranta}, {Tsutsumi}, {van Bemmel},
  {Verkouter}, {Wells}, {Xiong}, {Szomoru}, {Griffith}, {Glendenning}, \&
  {Kern}}]{2022PASP..134k4501C}
{CASA Team}, {Bean}, B., {Bhatnagar}, S., {et~al.} 2022, \pasp, 134, 114501

\bibitem[{{Ciesla} {et~al.}(2020){Ciesla}, {B{\'e}thermin}, {Daddi}, {Richard},
  {Diaz-Santos}, {Sargent}, {Elbaz}, {Boquien}, {Wang}, {Schreiber}, {Yang},
  {Zabl}, {Fraser}, {Aravena}, {Assef}, {Baker}, {Beelen}, {Boselli},
  {Bournaud}, {Burgarella}, {Charmandaris}, {C{\^o}t{\'e}}, {Epinat},
  {Ferrarese}, {Gobat}, \& {Ilbert}}]{2020A&A...635A..27C}
{Ciesla}, L., {B{\'e}thermin}, M., {Daddi}, E., {et~al.} 2020, \aap, 635, A27

\bibitem[{{Clark} \& {Glover}(2015)}]{2015MNRAS.452.2057C}
{Clark}, P.~C. \& {Glover}, S. C.~O. 2015, \mnras, 452, 2057

\bibitem[{{Clark} {et~al.}(2019){Clark}, {Glover}, {Ragan}, \&
  {Duarte-Cabral}}]{2019MNRAS.486.4622C}
{Clark}, P.~C., {Glover}, S. C.~O., {Ragan}, S.~E., \& {Duarte-Cabral}, A.
  2019, \mnras, 486, 4622

\bibitem[{{Clements} {et~al.}(1993){Clements}, {Andreani}, \&
  {Chase}}]{1993MNRAS.261..299C}
{Clements}, D.~L., {Andreani}, P., \& {Chase}, S.~T. 1993, \mnras, 261, 299

\bibitem[{{Combes} {et~al.}(2012){Combes}, {Rex}, {Rawle}, {Egami}, {Boone},
  {Smail}, {Richard}, {Ivison}, {Gurwell}, {Casey}, {Omont}, {Berciano Alba},
  {Dessauges-Zavadsky}, {Edge}, {Fazio}, {Kneib}, {Okabe}, {Pell{\'o}},
  {P{\'e}rez-Gonz{\'a}lez}, {Schaerer}, {Smith}, {Swinbank}, \& {van der
  Werf}}]{2012A&A...538L...4C}
{Combes}, F., {Rex}, M., {Rawle}, T.~D., {et~al.} 2012, \aap, 538, L4

\bibitem[{{Coppin} {et~al.}(2010){Coppin}, {Chapman}, {Smail}, {Swinbank},
  {Walter}, {Wardlow}, {Weiss}, {Alexander}, {Brandt}, {Dannerbauer}, {De
  Breuck}, {Dickinson}, {Dunlop}, {Edge}, {Emonts}, {Greve}, {Huynh}, {Ivison},
  {Knudsen}, {Menten}, {Schinnerer}, \& {van der Werf}}]{2010MNRAS.407L.103C}
{Coppin}, K.~E.~K., {Chapman}, S.~C., {Smail}, I., {et~al.} 2010, \mnras, 407,
  L103

\bibitem[{Cox {et~al.}(2011)Cox, Krips, Neri, Omont, G{\"u}sten, Menten,
  Wyrowski, Wei{\ss}, Beelen, Gurwell, Dannerbauer, Ivison, Negrello, Aretxaga,
  Hughes, Auld, Baes, Blundell, Buttiglione, Cava, Cooray, Dariush, Dunne, Dye,
  Eales, Frayer, Fritz, Gavazzi, Hopwood, Ibar, Jarvis, Maddox, Micha{\l}owski,
  Pascale, Pohlen, Rigby, Smith, Swinbank, Temi, Valtchanov, van~der Werf, \&
  de~Zotti}]{Cox_2011}
Cox, P., Krips, M., Neri, R., {et~al.} 2011, The Astrophysical Journal, 740, 63

\bibitem[{{da Cunha} {et~al.}(2013{\natexlab{a}}){da Cunha}, {Groves},
  {Walter}, {Decarli}, {Weiss}, {Bertoldi}, {Carilli}, {Daddi}, {Elbaz},
  {Ivison}, {Maiolino}, {Riechers}, {Rix}, {Sargent}, \&
  {Smail}}]{2013ApJ...766...13D}
{da Cunha}, E., {Groves}, B., {Walter}, F., {et~al.} 2013{\natexlab{a}}, \apj,
  766, 13

\bibitem[{{da Cunha} {et~al.}(2021){da Cunha}, {Hodge}, {Casey}, {Algera},
  {Kaasinen}, {Smail}, {Walter}, {Brandt}, {Dannerbauer}, {Decarli}, {Groves},
  {Knudsen}, {Swinbank}, {Weiss}, {van der Werf}, \&
  {Zavala}}]{2021ApJ...919...30D}
{da Cunha}, E., {Hodge}, J.~A., {Casey}, C.~M., {et~al.} 2021, \apj, 919, 30

\bibitem[{{da Cunha} {et~al.}(2013{\natexlab{b}}){da Cunha}, {Walter},
  {Decarli}, {Bertoldi}, {Carilli}, {Daddi}, {Elbaz}, {Ivison}, {Maiolino},
  {Riechers}, {Rix}, {Sargent}, {Smail}, \& {Weiss}}]{2013ApJ...765....9D}
{da Cunha}, E., {Walter}, F., {Decarli}, R., {et~al.} 2013{\natexlab{b}}, \apj,
  765, 9

\bibitem[{de~Blok {et~al.}(2016)de~Blok, Walter, Smith, Herrera-Camus, Bolatto,
  Requena-Torres, Crocker, Croxall, Kennicutt, Koda, Armus, Boquien, Dale,
  Kreckel, \& Meidt}]{Blok_2016}
de~Blok, W. J.~G., Walter, F., Smith, J.-D.~T., {et~al.} 2016, The Astronomical
  Journal, 152, 51

\bibitem[{{De Breuck} {et~al.}(2011){De Breuck}, {Maiolino}, {Caselli},
  {Coppin}, {Hailey-Dunsheath}, \& {Nagao}}]{2011A&A...530L...8D}
{De Breuck}, C., {Maiolino}, R., {Caselli}, P., {et~al.} 2011, \aap, 530, L8

\bibitem[{{De Rosa} {et~al.}(2014){De Rosa}, {Venemans}, {Decarli}, {Gennaro},
  {Simcoe}, {Dietrich}, {Peterson}, {Walter}, {Frank}, {McMahon}, {Hewett},
  {Mortlock}, \& {Simpson}}]{2014ApJ...790..145D}
{De Rosa}, G., {Venemans}, B.~P., {Decarli}, R., {et~al.} 2014, \apj, 790, 145

\bibitem[{{Decarli} {et~al.}(2022){Decarli}, {Pensabene}, {Venemans}, {Walter},
  {Ba{\~n}ados}, {Bertoldi}, {Carilli}, {Cox}, {Fan}, {Farina}, {Ferkinhoff},
  {Groves}, {Li}, {Mazzucchelli}, {Neri}, {Riechers}, {Uzgil}, {Wang}, {Wang},
  {Weiss}, {Winters}, \& {Yang}}]{2022A&A...662A..60D}
{Decarli}, R., {Pensabene}, A., {Venemans}, B., {et~al.} 2022, \aap, 662, A60

\bibitem[{{Decarli} {et~al.}(2017){Decarli}, {Walter}, {Venemans},
  {Ba{\~n}ados}, {Bertoldi}, {Carilli}, {Fan}, {Farina}, {Mazzucchelli},
  {Riechers}, {Rix}, {Strauss}, {Wang}, \& {Yang}}]{2017Natur.545..457D}
{Decarli}, R., {Walter}, F., {Venemans}, B.~P., {et~al.} 2017, \nat, 545, 457

\bibitem[{{D{\'\i}az-Santos} {et~al.}(2017){D{\'\i}az-Santos}, {Armus},
  {Charmandaris}, {Lu}, {Stierwalt}, {Stacey}, {Malhotra}, {van der Werf},
  {Howell}, {Privon}, {Mazzarella}, {Goldsmith}, {Murphy}, {Barcos-Mu{\~n}oz},
  {Linden}, {Inami}, {Larson}, {Evans}, {Appleton}, {Iwasawa}, {Lord},
  {Sanders}, \& {Surace}}]{2017ApJ...846...32D}
{D{\'\i}az-Santos}, T., {Armus}, L., {Charmandaris}, V., {et~al.} 2017, \apj,
  846, 32

\bibitem[{Downes \& Solomon(1998)}]{Downes_1998}
Downes, D. \& Solomon, P.~M. 1998, The Astrophysical Journal, 507, 615

\bibitem[{Downes \& Solomon(2003)}]{Downes_2003}
Downes, D. \& Solomon, P.~M. 2003, The Astrophysical Journal, 582, 37

\bibitem[{{Draine} {et~al.}(2007){Draine}, {Dale}, {Bendo}, {Gordon}, {Smith},
  {Armus}, {Engelbracht}, {Helou}, {Kennicutt}, {Li}, {Roussel}, {Walter},
  {Calzetti}, {Moustakas}, {Murphy}, {Rieke}, {Bot}, {Hollenbach}, {Sheth}, \&
  {Teplitz}}]{2007ApJ...663..866D}
{Draine}, B.~T., {Dale}, D.~A., {Bendo}, G., {et~al.} 2007, \apj, 663, 866

\bibitem[{{Dunne} {et~al.}(2003){Dunne}, {Eales}, \&
  {Edmunds}}]{2003MNRAS.341..589D}
{Dunne}, L., {Eales}, S.~A., \& {Edmunds}, M.~G. 2003, \mnras, 341, 589

\bibitem[{{Eric Murphy and the ngVLA Science Advisory
  Council}(2018)}]{ngVLA_book}
{Eric Murphy and the ngVLA Science Advisory Council}. 2018, Science with a Next
  Generation Very Large Array (Astronomical Society of the Pacific Conference
  Series)

\bibitem[{{Fan} {et~al.}(2023){Fan}, {Ba{\~n}ados}, \&
  {Simcoe}}]{2023ARA&A..61..373F}
{Fan}, X., {Ba{\~n}ados}, E., \& {Simcoe}, R.~A. 2023, \araa, 61, 373

\bibitem[{{Farina} {et~al.}(2022){Farina}, {Schindler}, {Walter},
  {Ba{\~n}ados}, {Davies}, {Decarli}, {Eilers}, {Fan}, {Hennawi},
  {Mazzucchelli}, {Meyer}, {Trakhtenbrot}, {Volonteri}, {Wang}, {Worseck},
  {Yang}, {Gutcke}, {Venemans}, {Bosman}, {Costa}, {De Rosa}, {Drake}, \&
  {Onoue}}]{2022ApJ...941..106F}
{Farina}, E.~P., {Schindler}, J.-T., {Walter}, F., {et~al.} 2022, \apj, 941,
  106

\bibitem[{{Ferkinhoff} {et~al.}(2011){Ferkinhoff}, {Brisbin}, {Nikola},
  {Parshley}, {Stacey}, {Phillips}, {Falgarone}, {Benford}, {Staguhn}, \&
  {Tucker}}]{2011ApJ...740L..29F}
{Ferkinhoff}, C., {Brisbin}, D., {Nikola}, T., {et~al.} 2011, \apjl, 740, L29

\bibitem[{Feruglio {et~al.}(2023)Feruglio, Maio, Tripodi, Winters, Zappacosta,
  Bischetti, Civano, Carniani, D'Odorico, Fiore, Gallerani, Ginolfi, Maiolino,
  Piconcelli, Valiante, \& Zanchettin}]{Feruglio_2023}
Feruglio, C., Maio, U., Tripodi, R., {et~al.} 2023, The Astrophysical Journal
  Letters, 954, L10

\bibitem[{{Foreman-Mackey} {et~al.}(2013){Foreman-Mackey}, {Hogg}, {Lang}, \&
  {Goodman}}]{2013PASP..125..306F}
{Foreman-Mackey}, D., {Hogg}, D.~W., {Lang}, D., \& {Goodman}, J. 2013, \pasp,
  125, 306

\bibitem[{Gall {et~al.}(2011)Gall, Andersen, \& Hjorth}]{Gall_2011}
Gall, C., Andersen, A.~C., \& Hjorth, J. 2011, Astronomy \& Astrophysics, 528,
  A14

\bibitem[{Gilli {et~al.}(2011)Gilli, Su, Norman, Vignali, Comastri, Tozzi,
  Rosati, Stiavelli, Brandt, Xue, Luo, Castellano, Fontana, Fiore, Mainieri, \&
  Ptak}]{Gilli_2011}
Gilli, R., Su, J., Norman, C., {et~al.} 2011, The Astrophysical Journal, 730,
  L28

\bibitem[{{Glover} \& {Clark}(2016)}]{2016MNRAS.456.3596G}
{Glover}, S. C.~O. \& {Clark}, P.~C. 2016, \mnras, 456, 3596

\bibitem[{{Glover} {et~al.}(2015){Glover}, {Clark}, {Micic}, \&
  {Molina}}]{2015MNRAS.448.1607G}
{Glover}, S. C.~O., {Clark}, P.~C., {Micic}, M., \& {Molina}, F. 2015, \mnras,
  448, 1607

\bibitem[{{Goicoechea} {et~al.}(2016){Goicoechea}, {Pety}, {Cuadrado},
  {Cernicharo}, {Chapillon}, {Fuente}, {Gerin}, {Joblin}, {Marcelino}, \&
  {Pilleri}}]{2016Natur.537..207G}
{Goicoechea}, J.~R., {Pety}, J., {Cuadrado}, S., {et~al.} 2016, \nat, 537, 207

\bibitem[{{Goicoechea} {et~al.}(2019){Goicoechea}, {Santa-Maria}, {Bron},
  {Teyssier}, {Marcelino}, {Cernicharo}, \& {Cuadrado}}]{2019A&A...622A..91G}
{Goicoechea}, J.~R., {Santa-Maria}, M.~G., {Bron}, E., {et~al.} 2019, \aap,
  622, A91

\bibitem[{{Goicoechea} {et~al.}(2015){Goicoechea}, {Teyssier}, {Etxaluze},
  {Goldsmith}, {Ossenkopf}, {Gerin}, {Bergin}, {Black}, {Cernicharo},
  {Cuadrado}, {Encrenaz}, {Falgarone}, {Fuente}, {Hacar}, {Lis}, {Marcelino},
  {Melnick}, {M{\"u}ller}, {Persson}, {Pety}, {R{\"o}llig}, {Schilke}, {Simon},
  {Snell}, \& {Stutzki}}]{2015ApJ...812...75G}
{Goicoechea}, J.~R., {Teyssier}, D., {Etxaluze}, M., {et~al.} 2015, \apj, 812,
  75

\bibitem[{Greve {et~al.}(2014)Greve, Leonidaki, Xilouris, Wei{\ss}, Zhang,
  van~der Werf, Aalto, Armus, D{\'{\i}}az-Santos, Evans, Fischer, Gao,
  Gonz{\'{a}}lez-Alfonso, Harris, Henkel, Meijerink, Naylor, Smith, Spaans,
  Stacey, Veilleux, \& Walter}]{Greve_2014}
Greve, T.~R., Leonidaki, I., Xilouris, E.~M., {et~al.} 2014, The Astrophysical
  Journal, 794, 142

\bibitem[{Groves {et~al.}(2015)Groves, Schinnerer, Leroy, Galametz, Walter,
  Bolatto, Hunt, Dale, Calzetti, Croxall, \& Jr.}]{Groves_2015}
Groves, B.~A., Schinnerer, E., Leroy, A., {et~al.} 2015, The Astrophysical
  Journal, 799, 96

\bibitem[{Hao {et~al.}(2011)Hao, Kennicutt, Johnson, Calzetti, Dale, \&
  Moustakas}]{Hao_2011}
Hao, C.-N., Kennicutt, R.~C., Johnson, B.~D., {et~al.} 2011, The Astrophysical
  Journal, 741, 124

\bibitem[{{Harrington} {et~al.}(2021){Harrington}, {Weiss}, {Yun}, {Magnelli},
  {Sharon}, {Leung}, {Vishwas}, {Wang}, {Frayer}, {Jim{\'e}nez-Andrade}, {Liu},
  {Garc{\'\i}a}, {Romano-D{\'\i}az}, {Frye}, {Jarugula}, {B{\u{a}}descu},
  {Berman}, {Dannerbauer}, {D{\'\i}az-S{\'a}nchez}, {Grassitelli},
  {Kamieneski}, {Kim}, {Kirkpatrick}, {Lowenthal}, {Messias}, {Puschnig},
  {Stacey}, {Torne}, \& {Bertoldi}}]{2021ApJ...908...95H}
{Harrington}, K.~C., {Weiss}, A., {Yun}, M.~S., {et~al.} 2021, \apj, 908, 95

\bibitem[{{Ikeda} {et~al.}(2002){Ikeda}, {Oka}, {Tatematsu}, {Sekimoto}, \&
  {Yamamoto}}]{2002ApJS..139..467I}
{Ikeda}, M., {Oka}, T., {Tatematsu}, K., {Sekimoto}, Y., \& {Yamamoto}, S.
  2002, \apjs, 139, 467

\bibitem[{{Inayoshi} {et~al.}(2017){Inayoshi}, {Hirai}, {Kinugawa}, \&
  {Hotokezaka}}]{2017MNRAS.468.5020I}
{Inayoshi}, K., {Hirai}, R., {Kinugawa}, T., \& {Hotokezaka}, K. 2017, \mnras,
  468, 5020

\bibitem[{{Israel} {et~al.}(2014){Israel}, {G{\"u}sten}, {Meijerink}, {Loenen},
  {Requena-Torres}, {Stutzki}, {van der Werf}, {Harris}, {Kramer},
  {Martin-Pintado}, \& {Weiss}}]{2014A&A...562A..96I}
{Israel}, F.~P., {G{\"u}sten}, R., {Meijerink}, R., {et~al.} 2014, \aap, 562,
  A96

\bibitem[{{Israel} {et~al.}(2015){Israel}, {Rosenberg}, \& {van der
  Werf}}]{2015A&A...578A..95I}
{Israel}, F.~P., {Rosenberg}, M.~J.~F., \& {van der Werf}, P. 2015, \aap, 578,
  A95

\bibitem[{{Izumi} {et~al.}(2021){Izumi}, {Matsuoka}, {Fujimoto}, {Onoue},
  {Strauss}, {Umehata}, {Imanishi}, {Kohno}, {Kawaguchi}, {Kawamuro}, {Baba},
  {Nagao}, {Toba}, {Inayoshi}, {Silverman}, {Inoue}, {Ikarashi}, {Iwasawa},
  {Kashikawa}, {Hashimoto}, {Nakanishi}, {Ueda}, {Schramm}, {Lee}, \&
  {Suh}}]{2021ApJ...914...36I}
{Izumi}, T., {Matsuoka}, Y., {Fujimoto}, S., {et~al.} 2021, \apj, 914, 36

\bibitem[{{Jiang} {et~al.}(2007){Jiang}, {Fan}, {Vestergaard}, {Kurk},
  {Walter}, {Kelly}, \& {Strauss}}]{2007AJ....134.1150J}
{Jiang}, L., {Fan}, X., {Vestergaard}, M., {et~al.} 2007, \aj, 134, 1150

\bibitem[{{Kamenetzky} {et~al.}(2016){Kamenetzky}, {Rangwala}, {Glenn},
  {Maloney}, \& {Conley}}]{2016ApJ...829...93K}
{Kamenetzky}, J., {Rangwala}, N., {Glenn}, J., {Maloney}, P.~R., \& {Conley},
  A. 2016, \apj, 829, 93

\bibitem[{{Khusanova} {et~al.}(2022){Khusanova}, {Ba{\~n}ados}, {Mazzucchelli},
  {Rojas-Ruiz}, {Momjian}, {Walter}, {Decarli}, {Venemans}, {Farina}, {Meyer},
  {Wang}, \& {Yang}}]{2022A&A...664A..39K}
{Khusanova}, Y., {Ba{\~n}ados}, E., {Mazzucchelli}, C., {et~al.} 2022, \aap,
  664, A39

\bibitem[{{Kormendy} \& {Ho}(2013)}]{2013ARA&A..51..511K}
{Kormendy}, J. \& {Ho}, L.~C. 2013, \araa, 51, 511

\bibitem[{{Koss} {et~al.}(2022){Koss}, {Ricci}, {Trakhtenbrot}, {Oh}, {den
  Brok}, {Mej{\'\i}a-Restrepo}, {Stern}, {Privon}, {Treister}, {Powell},
  {Mushotzky}, {Bauer}, {Ananna}, {Balokovi{\'c}}, {B{\"a}r}, {Becker},
  {Bessiere}, {Burtscher}, {Caglar}, {Congiu}, {Evans}, {Harrison}, {Heida},
  {Ichikawa}, {Kamraj}, {Lamperti}, {Pacucci}, {Ricci}, {Riffel}, {Rojas},
  {Schawinski}, {Temple}, {Urry}, {Veilleux}, \&
  {Williams}}]{2022ApJS..261....2K}
{Koss}, M.~J., {Ricci}, C., {Trakhtenbrot}, B., {et~al.} 2022, \apjs, 261, 2

\bibitem[{{Kramer} {et~al.}(2008){Kramer}, {Cubick}, {R{\"o}llig}, {Sun},
  {Yonekura}, {Aravena}, {Bensch}, {Bertoldi}, {Bronfman}, {Fujishita},
  {Fukui}, {Graf}, {Hitschfeld}, {Honingh}, {Ito}, {Jakob}, {Jacobs}, {Klein},
  {Koo}, {May}, {Miller}, {Miyamoto}, {Mizuno}, {Onishi}, {Park}, {Pineda},
  {Rabanus}, {Sasago}, {Schieder}, {Simon}, {Stutzki}, {Volgenau}, \&
  {Yamamoto}}]{2008A&A...477..547K}
{Kramer}, C., {Cubick}, M., {R{\"o}llig}, M., {et~al.} 2008, \aap, 477, 547

\bibitem[{{Krumholz}(2014)}]{2014PhR...539...49K}
{Krumholz}, M.~R. 2014, \physrep, 539, 49

\bibitem[{{Kuo} \& {Hirashita}(2012)}]{2012MNRAS.424L..34K}
{Kuo}, T.-M. \& {Hirashita}, H. 2012, \mnras, 424, L34

\bibitem[{{Kurk} {et~al.}(2007){Kurk}, {Walter}, {Fan}, {Jiang}, {Riechers},
  {Rix}, {Pentericci}, {Strauss}, {Carilli}, \& {Wagner}}]{2007ApJ...669...32K}
{Kurk}, J.~D., {Walter}, F., {Fan}, X., {et~al.} 2007, \apj, 669, 32

\bibitem[{{Latif} \& {Ferrara}(2016)}]{2016PASA...33...51L}
{Latif}, M.~A. \& {Ferrara}, A. 2016, \pasa, 33, e051

\bibitem[{{Leipski} {et~al.}(2013){Leipski}, {Meisenheimer}, {Walter}, {Besel},
  {Dannerbauer}, {Fan}, {Haas}, {Klaas}, {Krause}, \&
  {Rix}}]{2013ApJ...772..103L}
{Leipski}, C., {Meisenheimer}, K., {Walter}, F., {et~al.} 2013, \apj, 772, 103

\bibitem[{{Leipski} {et~al.}(2014){Leipski}, {Meisenheimer}, {Walter}, {Klaas},
  {Dannerbauer}, {De Rosa}, {Fan}, {Haas}, {Krause}, \&
  {Rix}}]{2014ApJ...785..154L}
{Leipski}, C., {Meisenheimer}, K., {Walter}, F., {et~al.} 2014, \apj, 785, 154

\bibitem[{Leipski {et~al.}(2014)Leipski, Meisenheimer, Walter, Klaas,
  Dannerbauer, Rosa, Fan, Haas, Krause, \& Rix}]{Leipski_2014}
Leipski, C., Meisenheimer, K., Walter, F., {et~al.} 2014, The Astrophysical
  Journal, 785, 154

\bibitem[{{Lestrade} {et~al.}(2011){Lestrade}, {Carilli}, {Thanjavur}, {Kneib},
  {Riechers}, {Bertoldi}, {Walter}, \& {Omont}}]{2011ApJ...739L..30L}
{Lestrade}, J.-F., {Carilli}, C.~L., {Thanjavur}, K., {et~al.} 2011, \apjl,
  739, L30

\bibitem[{{Lestrade} {et~al.}(2010){Lestrade}, {Combes}, {Salom{\'e}}, {Omont},
  {Bertoldi}, {Andr{\'e}}, \& {Schneider}}]{2010A&A...522L...4L}
{Lestrade}, J.~F., {Combes}, F., {Salom{\'e}}, P., {et~al.} 2010, \aap, 522, L4

\bibitem[{{Li} {et~al.}(2022){Li}, {Venemans}, {Walter}, {Decarli}, {Wang}, \&
  {Cai}}]{2022ApJ...930...27L}
{Li}, J., {Venemans}, B.~P., {Walter}, F., {et~al.} 2022, \apj, 930, 27

\bibitem[{{Liang} {et~al.}(2018){Liang}, {Feldmann}, {Faucher-Gigu{\`e}re},
  {Kere{\v{s}}}, {Hopkins}, {Hayward}, {Quataert}, \&
  {Scoville}}]{2018MNRAS.478L..83L}
{Liang}, L., {Feldmann}, R., {Faucher-Gigu{\`e}re}, C.-A., {et~al.} 2018,
  \mnras, 478, L83

\bibitem[{Liu {et~al.}(2015)Liu, Gao, Isaak, Daddi, Yang, Lu, \& van~der
  Werf}]{Liu_2015}
Liu, D., Gao, Y., Isaak, K., {et~al.} 2015, The Astrophysical Journal, 810, L14

\bibitem[{{Lu} {et~al.}(2018){Lu}, {Cao}, {D{\'\i}az-Santos}, {Zhao}, {Privon},
  {Cheng}, {Gao}, {Xu}, {Charmandaris}, {Rigopoulou}, {van der Werf}, {Huang},
  {Wang}, {Evans}, \& {Sanders}}]{2018ApJ...864...38L}
{Lu}, N., {Cao}, T., {D{\'\i}az-Santos}, T., {et~al.} 2018, \apj, 864, 38

\bibitem[{Lutz {et~al.}(2007)Lutz, Sturm, Tacconi, Valiante, Schweitzer,
  Netzer, Maiolino, Andreani, Shemmer, \& Veilleux}]{Lutz_2007}
Lutz, D., Sturm, E., Tacconi, L.~J., {et~al.} 2007, The Astrophysical Journal,
  661, L25

\bibitem[{{Madden} {et~al.}(2020){Madden}, {Cormier}, {Hony}, {Lebouteiller},
  {Abel}, {Galametz}, {De Looze}, {Chevance}, {Polles}, {Lee}, {Galliano},
  {Lambert-Huyghe}, {Hu}, \& {Ramambason}}]{2020A&A...643A.141M}
{Madden}, S.~C., {Cormier}, D., {Hony}, S., {et~al.} 2020, \aap, 643, A141

\bibitem[{{Mazzucchelli} {et~al.}(2023){Mazzucchelli}, {Bischetti},
  {D'Odorico}, {Feruglio}, {Schindler}, {Onoue}, {Ba{\~n}ados}, {Becker},
  {Bian}, {Carniani}, {Decarli}, {Eilers}, {Farina}, {Gallerani}, {Lai},
  {Meyer}, {Rojas-Ruiz}, {Satyavolu}, {Venemans}, {Wang}, {Yang}, \&
  {Zhu}}]{2023A&A...676A..71M}
{Mazzucchelli}, C., {Bischetti}, M., {D'Odorico}, V., {et~al.} 2023, \aap, 676,
  A71

\bibitem[{{Medvedev} {et~al.}(2021){Medvedev}, {Gilfanov}, {Sazonov},
  {Schartel}, \& {Sunyaev}}]{2021MNRAS.504..576M}
{Medvedev}, P., {Gilfanov}, M., {Sazonov}, S., {Schartel}, N., \& {Sunyaev}, R.
  2021, \mnras, 504, 576

\bibitem[{{Meyer} {et~al.}(2023){Meyer}, {Neeleman}, {Walter}, \&
  {Venemans}}]{2023ApJ...956..127M}
{Meyer}, R.~A., {Neeleman}, M., {Walter}, F., \& {Venemans}, B. 2023, \apj,
  956, 127

\bibitem[{{Meyer} {et~al.}(2022){Meyer}, {Walter}, {Cicone}, {Cox}, {Decarli},
  {Neri}, {Novak}, {Pensabene}, {Riechers}, \& {Weiss}}]{2022ApJ...927..152M}
{Meyer}, R.~A., {Walter}, F., {Cicone}, C., {et~al.} 2022, \apj, 927, 152

\bibitem[{{Micha{\l}owski} {et~al.}(2010){Micha{\l}owski}, {Murphy}, {Hjorth},
  {Watson}, {Gall}, \& {Dunlop}}]{2010A&A...522A..15M}
{Micha{\l}owski}, M.~J., {Murphy}, E.~J., {Hjorth}, J., {et~al.} 2010, \aap,
  522, A15

\bibitem[{{Molyneux} {et~al.}(2023){Molyneux}, {Calistro Rivera}, {De Breuck},
  {Harrison}, {Mainieri}, {Lundgren}, {Kakkad}, {Circosta}, {Girdhar}, {Costa},
  {Mullaney}, {Kharb}, {Arrigoni Battaia}, {Farina}, {Alexander}, {Ward},
  {Silpa}, \& {Smit}}]{2023arXiv231010235M}
{Molyneux}, S.~J., {Calistro Rivera}, G., {De Breuck}, C., {et~al.} 2023, arXiv
  e-prints, arXiv:2310.10235

\bibitem[{{Narayanan} {et~al.}(2011){Narayanan}, {Krumholz}, {Ostriker}, \&
  {Hernquist}}]{2011MNRAS.418..664N}
{Narayanan}, D., {Krumholz}, M., {Ostriker}, E.~C., \& {Hernquist}, L. 2011,
  \mnras, 418, 664

\bibitem[{{Narayanan} \& {Krumholz}(2014)}]{2014MNRAS.442.1411N}
{Narayanan}, D. \& {Krumholz}, M.~R. 2014, \mnras, 442, 1411

\bibitem[{Neeleman(2021)}]{qubefit}
Neeleman, M. 2021, Qubefit v1.0.1

\bibitem[{{Neeleman} {et~al.}(2021){Neeleman}, {Novak}, {Venemans}, {Walter},
  {Decarli}, {Kaasinen}, {Schindler}, {Ba{\~n}ados}, {Carilli}, {Drake}, {Fan},
  \& {Rix}}]{2021ApJ...911..141N}
{Neeleman}, M., {Novak}, M., {Venemans}, B.~P., {et~al.} 2021, \apj, 911, 141

\bibitem[{{Nesvadba} {et~al.}(2019){Nesvadba}, {Ca{\~n}ameras}, {Kneissl},
  {Koenig}, {Yang}, {Le Floc'h}, {Omont}, \& {Scott}}]{2019A&A...624A..23N}
{Nesvadba}, N.~P.~H., {Ca{\~n}ameras}, R., {Kneissl}, R., {et~al.} 2019, \aap,
  624, A23

\bibitem[{{Novak} {et~al.}(2020){Novak}, {Venemans}, {Walter}, {Neeleman},
  {Kaasinen}, {Liang}, {Feldmann}, {Ba{\~n}ados}, {Carilli}, {Decarli},
  {Drake}, {Fan}, {Farina}, {Mazzucchelli}, {Rix}, \&
  {Wang}}]{2020ApJ...904..131N}
{Novak}, M., {Venemans}, B.~P., {Walter}, F., {et~al.} 2020, \apj, 904, 131

\bibitem[{Omont {et~al.}(2013)Omont, Willott, Beelen, Bergeron, Orellana, \&
  Delorme}]{Omont_2013}
Omont, A., Willott, C.~J., Beelen, A., {et~al.} 2013, Astronomy \&
  Astrophysics, 552, A43

\bibitem[{{Papadopoulos} {et~al.}(2022){Papadopoulos}, {Dunne}, \&
  {Maddox}}]{2022MNRAS.510..725P}
{Papadopoulos}, P., {Dunne}, L., \& {Maddox}, S. 2022, \mnras, 510, 725

\bibitem[{{Papadopoulos} {et~al.}(2018){Papadopoulos}, {Bisbas}, \&
  {Zhang}}]{2018MNRAS.478.1716P}
{Papadopoulos}, P.~P., {Bisbas}, T.~G., \& {Zhang}, Z.-Y. 2018, \mnras, 478,
  1716

\bibitem[{{Papadopoulos} {et~al.}(2004){Papadopoulos}, {Thi}, \&
  {Viti}}]{2004MNRAS.351..147P}
{Papadopoulos}, P.~P., {Thi}, W.~F., \& {Viti}, S. 2004, \mnras, 351, 147

\bibitem[{{Papadopoulos} {et~al.}(2012{\natexlab{a}}){Papadopoulos}, {van der
  Werf}, {Xilouris}, {Isaak}, \& {Gao}}]{2012ApJ...751...10P}
{Papadopoulos}, P.~P., {van der Werf}, P., {Xilouris}, E., {Isaak}, K.~G., \&
  {Gao}, Y. 2012{\natexlab{a}}, \apj, 751, 10

\bibitem[{{Papadopoulos} {et~al.}(2012{\natexlab{b}}){Papadopoulos}, {van der
  Werf}, {Xilouris}, {Isaak}, {Gao}, \& {M{\"u}hle}}]{2012MNRAS.426.2601P}
{Papadopoulos}, P.~P., {van der Werf}, P.~P., {Xilouris}, E.~M., {et~al.}
  2012{\natexlab{b}}, \mnras, 426, 2601

\bibitem[{{Parikka} {et~al.}(2018){Parikka}, {Habart}, {Bernard-Salas},
  {K{\"o}hler}, \& {Abergel}}]{2018A&A...617A..77P}
{Parikka}, A., {Habart}, E., {Bernard-Salas}, J., {K{\"o}hler}, M., \&
  {Abergel}, A. 2018, \aap, 617, A77

\bibitem[{{Pavesi} {et~al.}(2018){Pavesi}, {Sharon}, {Riechers}, {Hodge},
  {Decarli}, {Walter}, {Carilli}, {Daddi}, {Smail}, {Dickinson}, {Ivison},
  {Sargent}, {da Cunha}, {Aravena}, {Darling}, {Smol{\v c}i{\'c}}, {Scoville},
  {Capak}, \& {Wagg}}]{2018ApJ...864...49P}
{Pavesi}, R., {Sharon}, C.~E., {Riechers}, D.~A., {et~al.} 2018, \apj, 864, 49

\bibitem[{{Petric} {et~al.}(2006){Petric}, {Carilli}, {Bertoldi}, {Beelen},
  {Cox}, \& {Omont}}]{2006AJ....132.1307P}
{Petric}, A.~O., {Carilli}, C.~L., {Bertoldi}, F., {et~al.} 2006, \aj, 132,
  1307

\bibitem[{{Planesas} {et~al.}(1999){Planesas}, {Martin-Pintado}, {Neri}, \&
  {Colina}}]{1999Sci...286.2493P}
{Planesas}, P., {Martin-Pintado}, J., {Neri}, R., \& {Colina}, L. 1999,
  Science, 286, 2493

\bibitem[{{Ricci} {et~al.}(2017){Ricci}, {Trakhtenbrot}, {Koss}, {Ueda}, {Del
  Vecchio}, {Treister}, {Schawinski}, {Paltani}, {Oh}, {Lamperti}, {Berney},
  {Gandhi}, {Ichikawa}, {Bauer}, {Ho}, {Asmus}, {Beckmann}, {Soldi},
  {Balokovi{\'c}}, {Gehrels}, \& {Markwardt}}]{2017ApJS..233...17R}
{Ricci}, C., {Trakhtenbrot}, B., {Koss}, M.~J., {et~al.} 2017, \apjs, 233, 17

\bibitem[{{Riechers} {et~al.}(2013){Riechers}, {Bradford}, {Clements},
  {Dowell}, {P{\'e}rez-Fournon}, {Ivison}, {Bridge}, {Conley}, {Fu}, {Vieira},
  {Wardlow}, {Calanog}, {Cooray}, {Hurley}, {Neri}, {Kamenetzky}, {Aguirre},
  {Altieri}, {Arumugam}, {Benford}, {B{\'e}thermin}, {Bock}, {Burgarella},
  {Cabrera-Lavers}, {Chapman}, {Cox}, {Dunlop}, {Earle}, {Farrah}, {Ferrero},
  {Franceschini}, {Gavazzi}, {Glenn}, {Solares}, {Gurwell}, {Halpern},
  {Hatziminaoglou}, {Hyde}, {Ibar}, {Kov{\'a}cs}, {Krips}, {Lupu}, {Maloney},
  {Martinez-Navajas}, {Matsuhara}, {Murphy}, {Naylor}, {Nguyen}, {Oliver},
  {Omont}, {Page}, {Petitpas}, {Rangwala}, {Roseboom}, {Scott}, {Smith},
  {Staguhn}, {Streblyanska}, {Thomson}, {Valtchanov}, {Viero}, {Wang},
  {Zemcov}, \& {Zmuidzinas}}]{2013Natur.496..329R}
{Riechers}, D.~A., {Bradford}, C.~M., {Clements}, D.~L., {et~al.} 2013, \nat,
  496, 329

\bibitem[{{Riechers} {et~al.}(2011{\natexlab{a}}){Riechers}, {Carilli},
  {Maddalena}, {Hodge}, {Harris}, {Baker}, {Walter}, {Wagg}, {Vanden Bout},
  {Wei{\ss}}, \& {Sharon}}]{2011ApJ...739L..32R}
{Riechers}, D.~A., {Carilli}, C.~L., {Maddalena}, R.~J., {et~al.}
  2011{\natexlab{a}}, \apjl, 739, L32

\bibitem[{{Riechers} {et~al.}(2011{\natexlab{b}}){Riechers}, {Hodge}, {Walter},
  {Carilli}, \& {Bertoldi}}]{2011ApJ...739L..31R}
{Riechers}, D.~A., {Hodge}, J., {Walter}, F., {Carilli}, C.~L., \& {Bertoldi},
  F. 2011{\natexlab{b}}, \apjl, 739, L31

\bibitem[{{Riechers} {et~al.}(2009){Riechers}, {Walter}, {Bertoldi}, {Carilli},
  {Aravena}, {Neri}, {Cox}, {Wei{\ss}}, \& {Menten}}]{2009ApJ...703.1338R}
{Riechers}, D.~A., {Walter}, F., {Bertoldi}, F., {et~al.} 2009, \apj, 703, 1338

\bibitem[{{Riechers} {et~al.}(2007){Riechers}, {Walter}, {Carilli}, \&
  {Bertoldi}}]{2007ApJ...671L..13R}
{Riechers}, D.~A., {Walter}, F., {Carilli}, C.~L., \& {Bertoldi}, F. 2007,
  \apjl, 671, L13

\bibitem[{{Riechers} {et~al.}(2008){Riechers}, {Walter}, {Carilli}, {Bertoldi},
  \& {Momjian}}]{2008ApJ...686L...9R}
{Riechers}, D.~A., {Walter}, F., {Carilli}, C.~L., {Bertoldi}, F., \&
  {Momjian}, E. 2008, \apjl, 686, L9

\bibitem[{{Riechers} {et~al.}(2006){Riechers}, {Walter}, {Carilli}, {Knudsen},
  {Lo}, {Benford}, {Staguhn}, {Hunter}, {Bertoldi}, {Henkel}, {Menten},
  {Weiss}, {Yun}, \& {Scoville}}]{2006ApJ...650..604R}
{Riechers}, D.~A., {Walter}, F., {Carilli}, C.~L., {et~al.} 2006, \apj, 650,
  604

\bibitem[{{Rosenberg} {et~al.}(2015){Rosenberg}, {van der Werf}, {Aalto},
  {Armus}, {Charmandaris}, {D{\'\i}az-Santos}, {Evans}, {Fischer}, {Gao},
  {Gonz{\'a}lez-Alfonso}, {Greve}, {Harris}, {Henkel}, {Israel}, {Isaak},
  {Kramer}, {Meijerink}, {Naylor}, {Sanders}, {Smith}, {Spaans}, {Spinoglio},
  {Stacey}, {Veenendaal}, {Veilleux}, {Walter}, {Wei{\ss}}, {Wiedner}, {van der
  Wiel}, \& {Xilouris}}]{2015ApJ...801...72R}
{Rosenberg}, M.~J.~F., {van der Werf}, P.~P., {Aalto}, S., {et~al.} 2015, \apj,
  801, 72

\bibitem[{Sargent {et~al.}(2014)Sargent, Daddi, B{\'e}thermin, Aussel, Magdis,
  Hwang, Juneau, Elbaz, \& da~Cunha}]{Sargent_2014}
Sargent, M.~T., Daddi, E., B{\'e}thermin, M., {et~al.} 2014, The Astrophysical
  Journal, 793, 19

\bibitem[{Schindler {et~al.}(2020)Schindler, Farina, Ba{\~{n}}ados, Eilers,
  Hennawi, Onoue, Venemans, Walter, Wang, Davies, Decarli, Rosa, Drake, Fan,
  Mazzucchelli, Rix, Worseck, \& Yang}]{Schindler_2020}
Schindler, J.-T., Farina, E.~P., Ba{\~{n}}ados, E., {et~al.} 2020, The
  Astrophysical Journal, 905, 51

\bibitem[{{Schinnerer} {et~al.}(2008){Schinnerer}, {Carilli}, {Capak},
  {Martinez-Sansigre}, {Scoville}, {Smol{\v{c}}i{\'c}}, {Taniguchi}, {Yun},
  {Bertoldi}, {Le Fevre}, \& {de Ravel}}]{2008ApJ...689L...5S}
{Schinnerer}, E., {Carilli}, C.~L., {Capak}, P., {et~al.} 2008, \apjl, 689, L5

\bibitem[{{Schumacher} {et~al.}(2012){Schumacher}, {Mart{\'\i}nez-Sansigre},
  {Lacy}, {Rawlings}, \& {Schinnerer}}]{2012MNRAS.423.2132S}
{Schumacher}, H., {Mart{\'\i}nez-Sansigre}, A., {Lacy}, M., {Rawlings}, S., \&
  {Schinnerer}, E. 2012, \mnras, 423, 2132

\bibitem[{{Scoville} {et~al.}(2017){Scoville}, {Lee}, {Vanden Bout},
  {Diaz-Santos}, {Sanders}, {Darvish}, {Bongiorno}, {Casey}, {Murchikova},
  {Koda}, {Capak}, {Vlahakis}, {Ilbert}, {Sheth}, {Morokuma-Matsui}, {Ivison},
  {Aussel}, {Laigle}, {McCracken}, {Armus}, {Pope}, {Toft}, \&
  {Masters}}]{2017ApJ...837..150S}
{Scoville}, N., {Lee}, N., {Vanden Bout}, P., {et~al.} 2017, \apj, 837, 150

\bibitem[{{Scoville} {et~al.}(2016){Scoville}, {Sheth}, {Aussel}, {Vanden
  Bout}, {Capak}, {Bongiorno}, {Casey}, {Murchikova}, {Koda},
  {{\'A}lvarez-M{\'a}rquez}, {Lee}, {Laigle}, {McCracken}, {Ilbert}, {Pope},
  {Sanders}, {Chu}, {Toft}, {Ivison}, \& {Manohar}}]{2016ApJ...820...83S}
{Scoville}, N., {Sheth}, K., {Aussel}, H., {et~al.} 2016, \apj, 820, 83

\bibitem[{{Shao} {et~al.}(2019){Shao}, {Wang}, {Carilli}, {Wagg}, {Walter},
  {Li}, {Fan}, {Jiang}, {Riechers}, {Bertoldi}, {Strauss}, {Cox}, {Omont}, \&
  {Menten}}]{2019ApJ...876...99S}
{Shao}, Y., {Wang}, R., {Carilli}, C.~L., {et~al.} 2019, \apj, 876, 99

\bibitem[{{Shen} {et~al.}(2011){Shen}, {Richards}, {Strauss}, {Hall},
  {Schneider}, {Snedden}, {Bizyaev}, {Brewington}, {Malanushenko},
  {Malanushenko}, {Oravetz}, {Pan}, \& {Simmons}}]{2011ApJS..194...45S}
{Shen}, Y., {Richards}, G.~T., {Strauss}, M.~A., {et~al.} 2011, \apjs, 194, 45

\bibitem[{{Shen} {et~al.}(2019){Shen}, {Wu}, {Jiang}, {Ba{\~n}ados}, {Fan},
  {Ho}, {Riechers}, {Strauss}, {Venemans}, {Vestergaard}, {Walter}, {Wang},
  {Willott}, {Wu}, \& {Yang}}]{2019ApJ...873...35S}
{Shen}, Y., {Wu}, J., {Jiang}, L., {et~al.} 2019, \apj, 873, 35

\bibitem[{{Solomon} \& {Vanden Bout}(2005)}]{2005ARA&A..43..677S}
{Solomon}, P.~M. \& {Vanden Bout}, P.~A. 2005, \araa, 43, 677

\bibitem[{{Stacey} {et~al.}(2010){Stacey}, {Hailey-Dunsheath}, {Ferkinhoff},
  {Nikola}, {Parshley}, {Benford}, {Staguhn}, \&
  {Fiolet}}]{2010ApJ...724..957S}
{Stacey}, G.~J., {Hailey-Dunsheath}, S., {Ferkinhoff}, C., {et~al.} 2010, \apj,
  724, 957

\bibitem[{{Stefan} {et~al.}(2015){Stefan}, {Carilli}, {Wagg}, {Walter},
  {Riechers}, {Bertoldi}, {Green}, {Fan}, {Menten}, \&
  {Wang}}]{2015MNRAS.451.1713S}
{Stefan}, I.~I., {Carilli}, C.~L., {Wagg}, J., {et~al.} 2015, \mnras, 451, 1713

\bibitem[{{Tacconi} {et~al.}(2020){Tacconi}, {Genzel}, \&
  {Sternberg}}]{2020ARA&A..58..157T}
{Tacconi}, L.~J., {Genzel}, R., \& {Sternberg}, A. 2020, \araa, 58, 157

\bibitem[{{Trakhtenbrot}(2021)}]{2021IAUS..356..261T}
{Trakhtenbrot}, B. 2021, in Nuclear Activity in Galaxies Across Cosmic Time,
  ed. M.~{Povi{\'c}}, P.~{Marziani}, J.~{Masegosa}, H.~{Netzer}, S.~H. {Negu},
  \& S.~B. {Tessema}, Vol. 356, 261--275

\bibitem[{{Tripodi} {et~al.}(2023){Tripodi}, {Scholtz}, {Maiolino}, {Fujimoto},
  {Carniani}, {Silverman}, {Feruglio}, {Ginolfi}, {Zappacosta}, {Costa},
  {Jones}, {Piconcelli}, {Bischetti}, \& {Fiore}}]{2023arXiv230601644T}
{Tripodi}, R., {Scholtz}, J., {Maiolino}, R., {et~al.} 2023, arXiv e-prints,
  arXiv:2306.01644

\bibitem[{{Tunnard} \& {Greve}(2016)}]{2016ApJ...819..161T}
{Tunnard}, R. \& {Greve}, T.~R. 2016, \apj, 819, 161

\bibitem[{{Valentino} {et~al.}(2018){Valentino}, {Magdis}, {Daddi}, {Liu},
  {Aravena}, {Bournaud}, {Cibinel}, {Cormier}, {Dickinson}, {Gao}, {Jin},
  {Juneau}, {Kartaltepe}, {Lee}, {Madden}, {Puglisi}, {Sanders}, \&
  {Silverman}}]{2018ApJ...869...27V}
{Valentino}, F., {Magdis}, G.~E., {Daddi}, E., {et~al.} 2018, \apj, 869, 27

\bibitem[{{Valentino} {et~al.}(2020){Valentino}, {Magdis}, {Daddi}, {Liu},
  {Aravena}, {Bournaud}, {Cortzen}, {Gao}, {Jin}, {Juneau}, {Kartaltepe},
  {Kokorev}, {Lee}, {Madden}, {Narayanan}, {Popping}, \&
  {Puglisi}}]{2020ApJ...890...24V}
{Valentino}, F., {Magdis}, G.~E., {Daddi}, E., {et~al.} 2020, \apj, 890, 24

\bibitem[{{Vallini} {et~al.}(2019){Vallini}, {Tielens}, {Pallottini},
  {Gallerani}, {Gruppioni}, {Carniani}, {Pozzi}, \&
  {Talia}}]{2019MNRAS.490.4502V}
{Vallini}, L., {Tielens}, A.~G.~G.~M., {Pallottini}, A., {et~al.} 2019, \mnras,
  490, 4502

\bibitem[{{Vanden Bout} {et~al.}(2004){Vanden Bout}, {Solomon}, \&
  {Maddalena}}]{2004ApJ...614L..97V}
{Vanden Bout}, P.~A., {Solomon}, P.~M., \& {Maddalena}, R.~J. 2004, \apjl, 614,
  L97

\bibitem[{{Vayner} {et~al.}(2021){Vayner}, {Zakamska}, {Wright}, {Armus},
  {Murray}, \& {Walth}}]{2021ApJ...923...59V}
{Vayner}, A., {Zakamska}, N., {Wright}, S.~A., {et~al.} 2021, \apj, 923, 59

\bibitem[{{Venemans} {et~al.}(2015){Venemans}, {Ba{\~n}ados}, {Decarli},
  {Farina}, {Walter}, {Chambers}, {Fan}, {Rix}, {Schlafly}, {McMahon},
  {Simcoe}, {Stern}, {Burgett}, {Draper}, {Flewelling}, {Hodapp}, {Kaiser},
  {Magnier}, {Metcalfe}, {Morgan}, {Price}, {Tonry}, {Waters}, {AlSayyad},
  {Banerji}, {Chen}, {Gonz{\'a}lez-Solares}, {Greiner}, {Mazzucchelli},
  {McGreer}, {Miller}, {Reed}, \& {Sullivan}}]{2015ApJ...801L..11V}
{Venemans}, B.~P., {Ba{\~n}ados}, E., {Decarli}, R., {et~al.} 2015, \apjl, 801,
  L11

\bibitem[{{Venemans} {et~al.}(2013){Venemans}, {Findlay}, {Sutherland}, {De
  Rosa}, {McMahon}, {Simcoe}, {Gonz{\'a}lez-Solares}, {Kuijken}, \&
  {Lewis}}]{2013ApJ...779...24V}
{Venemans}, B.~P., {Findlay}, J.~R., {Sutherland}, W.~J., {et~al.} 2013, \apj,
  779, 24

\bibitem[{{Venemans} {et~al.}(2012){Venemans}, {McMahon}, {Walter}, {Decarli},
  {Cox}, {Neri}, {Hewett}, {Mortlock}, {Simpson}, \&
  {Warren}}]{2012ApJ...751L..25V}
{Venemans}, B.~P., {McMahon}, R.~G., {Walter}, F., {et~al.} 2012, \apjl, 751,
  L25

\bibitem[{{Venemans} {et~al.}(2019){Venemans}, {Neeleman}, {Walter}, {Novak},
  {Decarli}, {Hennawi}, \& {Rix}}]{2019ApJ...874L..30V}
{Venemans}, B.~P., {Neeleman}, M., {Walter}, F., {et~al.} 2019, \apjl, 874, L30

\bibitem[{{Venemans} {et~al.}(2017{\natexlab{a}}){Venemans}, {Walter},
  {Decarli}, {Ba{\~n}ados}, {Carilli}, {Winters}, {Schuster}, {da Cunha},
  {Fan}, {Farina}, {Mazzucchelli}, {Rix}, \& {Weiss}}]{2017ApJ...851L...8V}
{Venemans}, B.~P., {Walter}, F., {Decarli}, R., {et~al.} 2017{\natexlab{a}},
  \apjl, 851, L8

\bibitem[{{Venemans} {et~al.}(2017{\natexlab{b}}){Venemans}, {Walter},
  {Decarli}, {Ferkinhoff}, {Wei{\ss}}, {Findlay}, {McMahon}, {Sutherland}, \&
  {Meijerink}}]{2017ApJ...845..154V}
{Venemans}, B.~P., {Walter}, F., {Decarli}, R., {et~al.} 2017{\natexlab{b}},
  \apj, 845, 154

\bibitem[{{Venemans} {et~al.}(2020){Venemans}, {Walter}, {Neeleman}, {Novak},
  {Otter}, {Decarli}, {Ba{\~n}ados}, {Drake}, {Farina}, {Kaasinen},
  {Mazzucchelli}, {Carilli}, {Fan}, {Rix}, \& {Wang}}]{2020ApJ...904..130V}
{Venemans}, B.~P., {Walter}, F., {Neeleman}, M., {et~al.} 2020, \apj, 904, 130

\bibitem[{{Venemans} {et~al.}(2016){Venemans}, {Walter}, {Zschaechner},
  {Decarli}, {De Rosa}, {Findlay}, {McMahon}, \&
  {Sutherland}}]{2016ApJ...816...37V}
{Venemans}, B.~P., {Walter}, F., {Zschaechner}, L., {et~al.} 2016, \apj, 816,
  37

\bibitem[{Venturini \& Solomon(2003)}]{Venturini_2003}
Venturini, S. \& Solomon, P.~M. 2003, The Astrophysical Journal, 590, 740

\bibitem[{{Wagg} {et~al.}(2012){Wagg}, {Wiklind}, {Carilli}, {Espada}, {Peck},
  {Riechers}, {Walter}, {Wootten}, {Aravena}, {Barkats}, {Cortes}, {Hills},
  {Hodge}, {Impellizzeri}, {Iono}, {Leroy}, {Mart{\'\i}n}, {Rawlings},
  {Maiolino}, {McMahon}, {Scott}, {Villard}, \&
  {Vlahakis}}]{2012ApJ...752L..30W}
{Wagg}, J., {Wiklind}, T., {Carilli}, C.~L., {et~al.} 2012, \apjl, 752, L30

\bibitem[{{Walter} {et~al.}(2003){Walter}, {Bertoldi}, {Carilli}, {Cox}, {Lo},
  {Neri}, {Fan}, {Omont}, {Strauss}, \& {Menten}}]{2003Natur.424..406W}
{Walter}, F., {Bertoldi}, F., {Carilli}, C., {et~al.} 2003, \nat, 424, 406

\bibitem[{{Walter} {et~al.}(2004){Walter}, {Carilli}, {Bertoldi}, {Menten},
  {Cox}, {Lo}, {Fan}, \& {Strauss}}]{2004ApJ...615L..17W}
{Walter}, F., {Carilli}, C., {Bertoldi}, F., {et~al.} 2004, \apjl, 615, L17

\bibitem[{{Walter} {et~al.}(2012){Walter}, {Decarli}, {Carilli}, {Bertoldi},
  {Cox}, {da Cunha}, {Daddi}, {Dickinson}, {Downes}, {Elbaz}, {Ellis}, {Hodge},
  {Neri}, {Riechers}, {Weiss}, {Bell}, {Dannerbauer}, {Krips}, {Krumholz},
  {Lentati}, {Maiolino}, {Menten}, {Rix}, {Robertson}, {Spinrad}, {Stark}, \&
  {Stern}}]{2012Natur.486..233W}
{Walter}, F., {Decarli}, R., {Carilli}, C., {et~al.} 2012, \nat, 486, 233

\bibitem[{{Walter} {et~al.}(2022){Walter}, {Neeleman}, {Decarli}, {Venemans},
  {Meyer}, {Weiss}, {Ba{\~n}ados}, {Bosman}, {Carilli}, {Fan}, {Riechers},
  {Rix}, \& {Thompson}}]{2022ApJ...927...21W}
{Walter}, F., {Neeleman}, M., {Decarli}, R., {et~al.} 2022, \apj, 927, 21

\bibitem[{{Walter} {et~al.}(2009){Walter}, {Riechers}, {Cox}, {Neri},
  {Carilli}, {Bertoldi}, {Weiss}, \& {Maiolino}}]{2009Natur.457..699W}
{Walter}, F., {Riechers}, D., {Cox}, P., {et~al.} 2009, \nat, 457, 699

\bibitem[{{Walter} {et~al.}(2011){Walter}, {Wei{\ss}}, {Downes}, {Decarli}, \&
  {Henkel}}]{2011ApJ...730...18W}
{Walter}, F., {Wei{\ss}}, A., {Downes}, D., {Decarli}, R., \& {Henkel}, C.
  2011, \apj, 730, 18

\bibitem[{{Wang} {et~al.}(2021){Wang}, {Fan}, {Yang}, {Mazzucchelli}, {Wu},
  {Li}, {Ba{\~n}ados}, {Farina}, {Nanni}, {Ai}, {Bian}, {Davies}, {Decarli},
  {Hennawi}, {Schindler}, {Venemans}, \& {Walter}}]{2021ApJ...908...53W}
{Wang}, F., {Fan}, X., {Yang}, J., {et~al.} 2021, \apj, 908, 53

\bibitem[{{Wang} {et~al.}(2019){Wang}, {Wang}, {Fan}, {Wu}, {Yang}, {Neri}, \&
  {Yue}}]{2019ApJ...880....2W}
{Wang}, F., {Wang}, R., {Fan}, X., {et~al.} 2019, \apj, 880, 2

\bibitem[{{Wang} {et~al.}(2023){Wang}, {Yang}, {Hennawi}, {Fan}, {Sun},
  {Champagne}, {Costa}, {Habouzit}, {Endsley}, {Li}, {Lin}, {Meyer},
  {Schindler}, {Wu}, {Ba{\~n}ados}, {Barth}, {Bhowmick}, {Bieri}, {Blecha},
  {Bosman}, {Cai}, {Colina}, {Connor}, {Davies}, {Decarli}, {De Rosa}, {Drake},
  {Egami}, {Eilers}, {Evans}, {Farina}, {Haiman}, {Jiang}, {Jin}, {Jun},
  {Kakiichi}, {Khusanova}, {Kulkarni}, {Li}, {Liu}, {Loiacono}, {Lupi},
  {Mazzucchelli}, {Onoue}, {Pudoka}, {Rojas-Ruiz}, {Shen}, {Strauss}, {Tee},
  {Trakhtenbrot}, {Trebitsch}, {Venemans}, {Volonteri}, {Walter}, {Xie}, {Yue},
  {Zhang}, {Zhang}, \& {Zou}}]{2023ApJ...951L...4W}
{Wang}, F., {Yang}, J., {Hennawi}, J.~F., {et~al.} 2023, \apjl, 951, L4

\bibitem[{{Wang} {et~al.}(2010){Wang}, {Carilli}, {Neri}, {Riechers}, {Wagg},
  {Walter}, {Bertoldi}, {Menten}, {Omont}, {Cox}, \&
  {Fan}}]{2010ApJ...714..699W}
{Wang}, R., {Carilli}, C.~L., {Neri}, R., {et~al.} 2010, \apj, 714, 699

\bibitem[{{Wang} {et~al.}(2011{\natexlab{a}}){Wang}, {Wagg}, {Carilli}, {Neri},
  {Walter}, {Omont}, {Riechers}, {Bertoldi}, {Menten}, {Cox}, {Strauss}, {Fan},
  \& {Jiang}}]{2011AJ....142..101W}
{Wang}, R., {Wagg}, J., {Carilli}, C.~L., {et~al.} 2011{\natexlab{a}}, \aj,
  142, 101

\bibitem[{{Wang} {et~al.}(2013){Wang}, {Wagg}, {Carilli}, {Walter}, {Lentati},
  {Fan}, {Riechers}, {Bertoldi}, {Narayanan}, {Strauss}, {Cox}, {Omont},
  {Menten}, {Knudsen}, {Neri}, \& {Jiang}}]{2013ApJ...773...44W}
{Wang}, R., {Wagg}, J., {Carilli}, C.~L., {et~al.} 2013, \apj, 773, 44

\bibitem[{{Wang} {et~al.}(2011{\natexlab{b}}){Wang}, {Wagg}, {Carilli},
  {Walter}, {Riechers}, {Willott}, {Bertoldi}, {Omont}, {Beelen}, {Cox},
  {Strauss}, {Bergeron}, {Forveille}, {Menten}, \& {Fan}}]{2011ApJ...739L..34W}
{Wang}, R., {Wagg}, J., {Carilli}, C.~L., {et~al.} 2011{\natexlab{b}}, \apjl,
  739, L34

\bibitem[{{Wang} {et~al.}(2016){Wang}, {Wu}, {Neri}, {Fan}, {Walter},
  {Carilli}, {Momjian}, {Bertoldi}, {Strauss}, {Li}, {Wang}, {Riechers},
  {Jiang}, {Omont}, {Wagg}, \& {Cox}}]{2016ApJ...830...53W}
{Wang}, R., {Wu}, X.-B., {Neri}, R., {et~al.} 2016, \apj, 830, 53

\bibitem[{{Wei{\ss}} {et~al.}(2007){Wei{\ss}}, {Downes}, {Neri}, {Walter},
  {Henkel}, {Wilner}, {Wagg}, \& {Wiklind}}]{2007A&A...467..955W}
{Wei{\ss}}, A., {Downes}, D., {Neri}, R., {et~al.} 2007, \aap, 467, 955

\bibitem[{{Wei{\ss}} {et~al.}(2003){Wei{\ss}}, {Henkel}, {Downes}, \&
  {Walter}}]{2003A&A...409L..41W}
{Wei{\ss}}, A., {Henkel}, C., {Downes}, D., \& {Walter}, F. 2003, \aap, 409,
  L41

\bibitem[{{Wei{\ss}} {et~al.}(2012){Wei{\ss}}, {Walter}, {Downes}, {Carrili},
  {Henkel}, {Menten}, \& {Cox}}]{2012ApJ...753..102W}
{Wei{\ss}}, A., {Walter}, F., {Downes}, D., {et~al.} 2012, \apj, 753, 102

\bibitem[{{Wei{\ss}} {et~al.}(2005){Wei{\ss}}, {Walter}, \&
  {Scoville}}]{2005A&A...438..533W}
{Wei{\ss}}, A., {Walter}, F., \& {Scoville}, N.~Z. 2005, \aap, 438, 533

\bibitem[{Willott {et~al.}(2017)Willott, Bergeron, \& Omont}]{Willott_2017}
Willott, C.~J., Bergeron, J., \& Omont, A. 2017, The Astrophysical Journal,
  850, 108

\bibitem[{{Willott} {et~al.}(2007){Willott}, {Mart{\'\i}nez-Sansigre}, \&
  {Rawlings}}]{2007AJ....133..564W}
{Willott}, C.~J., {Mart{\'\i}nez-Sansigre}, A., \& {Rawlings}, S. 2007, \aj,
  133, 564

\bibitem[{{Witstok} {et~al.}(2023){Witstok}, {Jones}, {Maiolino}, {Smit}, \&
  {Schneider}}]{2023MNRAS.523.3119W}
{Witstok}, J., {Jones}, G.~C., {Maiolino}, R., {Smit}, R., \& {Schneider}, R.
  2023, \mnras, 523, 3119

\bibitem[{{Wu} {et~al.}(2015){Wu}, {Wang}, {Fan}, {Yi}, {Zuo}, {Bian}, {Jiang},
  {McGreer}, {Wang}, {Yang}, {Yang}, {Thompson}, \&
  {Beletsky}}]{2015Natur.518..512W}
{Wu}, X.-B., {Wang}, F., {Fan}, X., {et~al.} 2015, \nat, 518, 512

\bibitem[{{Xie} {et~al.}(1995){Xie}, {Allen}, \&
  {Langer}}]{1995ApJ...440..674X}
{Xie}, T., {Allen}, M., \& {Langer}, W.~D. 1995, \apj, 440, 674

\bibitem[{{Yang}(2017)}]{2017PhDT........21Y}
{Yang}, C. 2017, PhD thesis, Institut d'Astrophysique Spatiale; CAS, Purple
  Mountain Observatory

\bibitem[{{Yue} {et~al.}(2023){Yue}, {Eilers}, {Simcoe}, {Mackenzie},
  {Matthee}, {Kashino}, {Bordoloi}, {Lilly}, \& {Naidu}}]{2023arXiv230904614Y}
{Yue}, M., {Eilers}, A.-C., {Simcoe}, R.~A., {et~al.} 2023, arXiv e-prints,
  arXiv:2309.04614

\bibitem[{{Yue} {et~al.}(2021){Yue}, {Yang}, {Fan}, {Wang}, {Spilker},
  {Georgiev}, {Keeton}, {Litke}, {Marrone}, {Walter}, {Wang}, {Wu}, {Venemans},
  \& {Zabludoff}}]{2021ApJ...917...99Y}
{Yue}, M., {Yang}, J., {Fan}, X., {et~al.} 2021, \apj, 917, 99

\bibitem[{{Zanella} {et~al.}(2018){Zanella}, {Daddi}, {Magdis}, {Diaz Santos},
  {Cormier}, {Liu}, {Cibinel}, {Gobat}, {Dickinson}, {Sargent}, {Popping},
  {Madden}, {Bethermin}, {Hughes}, {Valentino}, {Rujopakarn}, {Pannella},
  {Bournaud}, {Walter}, {Wang}, {Elbaz}, \& {Coogan}}]{2018MNRAS.481.1976Z}
{Zanella}, A., {Daddi}, E., {Magdis}, G., {et~al.} 2018, \mnras, 481, 1976

\bibitem[{Zappacosta {et~al.}(2023)Zappacosta, Piconcelli, Fiore, Saccheo,
  Valiante, Vignali, Vito, Volonteri, Bischetti, Comastri, Done, Elvis,
  Giallongo, La~Franca, Lanzuisi, Laurenti, Miniutti, Bongiorno, Brusa, Civano,
  Carniani, D'Odorico, Feruglio, Gallerani, Gilli, Grazian, Guainazzi,
  Marinucci, Menci, Middei, Nicastro, Puccetti, Tombesi, Tortosa, Testa,
  Vietri, Cristiani, Haardt, Maiolino, Schneider, Tripodi, Vallini, \&
  Vanzella}]{https://doi.org/10.48550/arxiv.2305.02347}
Zappacosta, L., Piconcelli, E., Fiore, F., {et~al.} 2023, HYPerluminous quasars
  at the Epoch of ReionizatION (HYPERION). A new regime for the X-ray nuclear
  properties of the first quasars

\bibitem[{{Zhang} {et~al.}(2016){Zhang}, {Papadopoulos}, {Ivison}, {Galametz},
  {Smith}, \& {Xilouris}}]{2016RSOS....360025Z}
{Zhang}, Z.-Y., {Papadopoulos}, P.~P., {Ivison}, R.~J., {et~al.} 2016, Royal
  Society Open Science, 3, 160025

\end{thebibliography}

	\begin{appendix}

	\section{Literature samples} 
		\label{sec:lit}

			To help interpret our results, we compiles a list of all $z>5$ QSO host galaxies with existing CO(2--1) observations from the literature (Table~\ref{tab:lit_QSOs}). We also put these galaxies in context with a sample of star-forming galaxies at $z\gtrsim4$ with CO(2--1), CO(6--5) and/or CO(7-6) detections (Table~\ref{tab:highz_SFGs}), a sample of $z=1-4$ QSO hosts (Table~\ref{tab:intz_qsos}) the $z\sim 1.2$ star-forming galaxies and AGN of \cite{2020ApJ...890...24V} and the low-redshift (U)LIRGS, especially those hosting AGN or QSOs, in \cite{2016ApJ...829...93K} and \cite{2015ApJ...801...72R}.

			\begin{sidewaystable}  %
			\small
			\begin{center}
			\caption{$z>5.7$ quasar host -- literature sample. \label{tab:lit_QSOs}} 
			\renewcommand{\arraystretch}{1.5}
			\begin{adjustbox}{scale=0.95,center}
			\begin{tabular}{@{}lccccccccccccc@{}}
			\toprule
			  \multicolumn{1}{c}{ID} &
			  \multicolumn{1}{c}{z} &
			  \multicolumn{1}{c}{$M_{BH}$} &
			  \multicolumn{1}{c}{$L_\mathrm{bol}$} &
			  \multicolumn{1}{c}{$M_{dyn}$}   &
			  \multicolumn{1}{c}{$S\Delta v_\mathrm{CO(2-1)}$}    &
			  \multicolumn{1}{c}{$S\Delta v_\mathrm{CO(3-2)}$}    &
			  \multicolumn{1}{c}{$S\Delta v_\mathrm{CO(5-4)}$}    &
			  \multicolumn{1}{c}{$S\Delta v_\mathrm{CO(6-5)}$}    &
			  \multicolumn{1}{c}{$S\Delta v_\mathrm{CO(7-6)}$}    &
			  \multicolumn{1}{c}{$S\Delta v_\mathrm{CO(8-7)}$}    &
			  \multicolumn{1}{c}{$S\Delta v_\mathrm{CO(9-8)}$}    &
			  \multicolumn{1}{c}{$S\Delta v_\mathrm{[C \textsc{i}(2-1)]}$}        &
			  \multicolumn{1}{c}{$S\Delta v_\mathrm{[C \textsc{ii}]}$}        \\
			  \multicolumn{1}{c}{} &
			  \multicolumn{1}{c}{} &
			   \multicolumn{1}{c}{($\times 10^9$ M$_\odot$)} &
			  \multicolumn{1}{c}{($\times 10^{13}$ L$_\odot$)} &
			 \multicolumn{1}{c}{($\times 10^{10}$ M$_\odot$)} &
			  \multicolumn{1}{c}{(Jy km s$^{-1}$)}    &
			  \multicolumn{1}{c}{(Jy km s$^{-1}$)}    &
			  \multicolumn{1}{c}{(Jy km s$^{-1}$)}    &
			  \multicolumn{1}{c}{(Jy km s$^{-1}$)}    &
			  \multicolumn{1}{c}{(Jy km s$^{-1}$)}    &
			  \multicolumn{1}{c}{(Jy km s$^{-1}$)}    &
			  \multicolumn{1}{c}{(Jy km s$^{-1}$)}    &
			  \multicolumn{1}{c}{(Jy km s$^{-1}$)}    &
			  \multicolumn{1}{c}{(Jy km s$^{-1}$)}        \\
			\midrule
			  J0927+2001 & 5.7722 &  $4.9_{-1.3}^{+1.3}$\,$^a$   &   $3.1 ^c$   &   -                         & $0.13\pm0.02~^j$  & -               & $0.44\pm0.07^o$ & $0.69\pm0.13^o$ & -              & -               & -               & -                    & -                \\
			  J0129-0035 & 5.7794 &  -                           &   $5.7 ^d$   &  $5.4_{-3.8}^{+3.3}$\,$^i$  & $0.036\pm0.005^k$ & -               & -               & $0.37\pm0.07^e$ & -              & -               & -               & -                    & $1.99\pm0.12^d$    \\
			  J0840+5624 & 5.8437 &  $3.5_{-1.5}^{+1.5} $\,$^a$  &   $5.8 ^e$   &  -                          & $0.06\pm0.02^j$   & -               & $0.60\pm0.07^o$ & $0.72\pm0.15^o$ & -              & -               & -               & -                    & -                \\
			  J2310+1855 & 6.003  &  $3.7_{-0.5}^{+0.6} $\,$^b$  &   $8.1 ^c$   &  -                          & $0.18\pm0.02^k$   & -               & $0.89\pm0.09^d$ & $1.52\pm0.13^d$ & -              & $1.55\pm0.13^q$ & $1.46\pm0.18^q$ & -                    & $8.83\pm0.44^d$    \\
			  J2054-0005 & 6.0379 &  $2.2_{-0.3}^{+0.3} $\,$^b$  &   $3.2 ^f$   &  $>2.9$\,$^i$               & $0.06\pm0.01^k$   & -               & -               & $0.34\pm0.07^o$ & -              & -               & -               & -                    & $3.37\pm0.12^d$    \\
			  J1429+5447 & 6.1831 &  $1.8_{-1.6}^{+1.6} $\,$^a$  &   $2.3 ^g$   &   -                         & $0.065\pm0.011^j$ & -               & -               & -               & -              & -               & -               & -                    & $3.6\phantom{0}\pm0.2\phantom{0}$\,$^r$ \\
			  J1623+3112 & 6.26   &  $2.5_{-0.3}^{+0.3} $\,$^a$  &   -          &   -                         & $ < 0.062^j$      & -               & -               & -               & -              & -               & -               & -                    & -                \\
			  J0100+2802 & 6.3    &  $6.6_{-0.7}^{+0.3} $\,$^c$  &   $3.7 ^f$   &  $10.2_{-7.1}^{+6.1}$\,$^i$ & $0.04\pm0.01^l$   & -               & -               & $0.32\pm0.084^l$& -              & -               & -               & -                    & $3.36\pm0.46^l$    \\
			  J1148+5251 & 6.4187 &  $8.7_{-0.4}^{+0.4} $\,$^a$  &   $9.7 ^h$   &   -                         & $0.095\pm0.007^m$ & $0.18\pm0.04^n$ & -               & $0.73\pm0.07^p$ & $0.64\pm0.08^p$& -               & -               & $0.22\pm0.05^p$      & $3.9\phantom{0}\pm0.3\phantom{0}$\,$^s$  \\
			\bottomrule
			\end{tabular}
			\end{adjustbox}
			\renewcommand{\arraystretch}{1}
			\end{center} 
			\vspace{-0.1cm}
			\textbf{References: }
			$^a$ \cite{2019ApJ...873...35S},
			$^b$ \cite{2022ApJ...941..106F}
			$^c$ \cite{2023A&A...676A..71M},
			$^d$ \cite{2013ApJ...773...44W},
			$^e$ \cite{2011AJ....142..101W},
			$^f$ \cite{Schindler_2020},
			$^g$ \cite{Omont_2013},
			$^h$ \cite{https://doi.org/10.48550/arxiv.2305.02347},
			$^i$ \cite{2021ApJ...911..141N}, 
			$^j$ \cite{2011ApJ...739L..34W}, 
			$^k$ \cite{2019ApJ...876...99S},
			$^l$ \cite{2016ApJ...830...53W},
			$^m$ \cite{2015MNRAS.451.1713S},
			$^n$ \cite{2004ApJ...615L..17W},
			$^o$ \cite{2010ApJ...714..699W},
			$^p$ \cite{2009ApJ...703.1338R},
			$^q$ \cite{2019MNRAS.489.3939C},
			$^r$ \cite{2022A&A...664A..39K},
			$^s$ \cite{2022ApJ...927..152M}
			\end{sidewaystable} 

			\begin{sidewaystable} 
			\begin{small} 
			\begin{center}
			\caption{$z\gtrsim4$ star-forming galaxies -- from literature. \label{tab:highz_SFGs}} 
			\renewcommand{\arraystretch}{1.5}
			\begin{adjustbox}{scale=0.9,center}
			\begin{tabular}{@{}lccccccccccc@{}}
			\toprule
			  \multicolumn{1}{l}{ID} &
			  \multicolumn{1}{c}{z} &
			  \multicolumn{1}{c}{$\log(L_\mathrm{FIR}/L_\odot)$} &
			  \multicolumn{1}{c}{FIR $\lambda$ range} &
			  \multicolumn{1}{c}{magnfication factor} &
			  \multicolumn{1}{c}{$S\Delta v_\mathrm{CO(2-1)}$} &
			  \multicolumn{1}{c}{$S\Delta v_\mathrm{CO(6-5)}$} &
			  \multicolumn{1}{c}{$S\Delta v_\mathrm{CO(7-6)}$} \\
			  \multicolumn{1}{c}{} &
			  \multicolumn{1}{c}{} &
			  \multicolumn{1}{c}{} &
			  \multicolumn{1}{c}{\textmu m} &
			  \multicolumn{1}{c}{\textmu} &
			  \multicolumn{1}{c}{(Jy km s$^{-1}$)}  &
			  \multicolumn{1}{c}{(Jy km s$^{-1}$)}  &
			  \multicolumn{1}{c}{(Jy km s$^{-1}$)}  \\
			\midrule
			  MM18423+5938                    & 3.929   & $14.2_{-0.4}^{+0.4}$ $^a$      & 14.0-608.0  & 1.0     & $2.74\pm0.23$ $^a$     & $6.6\pm0.6$ $^m$   & $3.9\pm0.5$ $^m$     \\
			  GN20 / SMMJ123711+622212        & 4.050   & $13.27_{-0.5}^{+0.5}$ $^b$     & 40-400      & 1.0     & $0.87\pm0.09$ $^j$     &                   & $1.6\pm0.2$ $^b$      \\
			  H-ATLASJ142413.9+023040/ID141   & 4.243   & $13.93_{-0.02}^{+0.02}$ $^c$   & 42.5-122.5  & 4.6     &                        &                   & $6.5\pm1.4$ $^c$      \\
			  J1000+0234                      & 4.542   & $12.93_{-0.26}^{+0.26}$ $^d$   & 42.5-122.5  & 1.0     & $0.057\pm0.017$ $^k$   &                   & -                   \\
			  J0332-2756                      & 4.755   & $12.62_{-0.06}^{+0.06}$ $^e$   & 42.5-122.5  & 1.0     & $0.09\pm0.02$ $^l$     &                   & -                    \\
			  SMMJ12365+621226, HDF850.1      & 5.183   & $12.81_{-0.07}^{+0.07}$ $^f$   & 42.5-122.5  & 1.4     & $0.17\pm0.04$ $^f$     & $0.39\pm0.1$ $^f$ & -                   \\
			  J0918+5142                      & 5.2429  & $13.96_{-0.0 }^{+0.0 }$ $^g$   & 42.5-122.5  & 11.0    & $7.1\pm1.0$ $^g$       & $27.0\pm2.8$ $^g$ & $24.0\pm2.1$ $^g$   \\
			  HZ10                            & 5.6543  & $12.03_{-0.23}^{+0.37}$ $^h$   & 42.5-122.5  & 1.0     & $0.1\pm0.02$ $^h$      &                   &                     \\
			  J170647.8+584623, HFLS3        & 6.3369  & $13.45_{-0.05}^{+0.05}$ $^i$    & 42.5-122.5  & 1.0     & $0.315\pm0.028$ $^i$   & $2.74\pm0.68$ $^i$& $2.9\pm0.77$$^i$     \\
			\bottomrule
			\end{tabular}
			\end{adjustbox}
			\end{center} 
			\renewcommand{\arraystretch}{1}
			\begin{adjustwidth}{70pt}{70pt}
			\vspace{-0.2cm}
			\textbf{References: }
			$^a$  \cite{2011ApJ...739L..30L},
			$^b$  \cite{2013MNRAS.429.3047B},
			$^c$  \cite{Cox_2011},
			$^d$  \cite{2008ApJ...681L..53C},
			$^e$  \cite{2011A&A...530L...8D},
			$^f$  \cite{2012Natur.486..233W},
			$^g$  \cite{2012A&A...538L...4C},
			$^h$  \cite{2018ApJ...864...49P},
			$^i$  \cite{2013Natur.496..329R},
			$^j$ \cite{2011ApJ...739L..33C},
			$^k$ \cite{2008ApJ...689L...5S},
			$^l$ \cite{2010MNRAS.407L.103C},
			$^m$ \cite{2010A&A...522L...4L}
			\end{adjustwidth}
			\end{small}
			\end{sidewaystable}

			\begin{sidewaystable} 
			\begin{small} 
			\begin{center}
			\caption{$z=1-5$ quasars -- literature sample. \label{tab:intz_qsos}} 
			\renewcommand{\arraystretch}{1.5}
			\begin{adjustbox}{scale=0.9,center}
			\begin{tabular}{@{}lccccccccccc@{}}
			\toprule
			  \multicolumn{1}{l}{ID} &
			  \multicolumn{1}{c}{z} &
			  \multicolumn{1}{c}{$L_\mathrm{FIR}/L_\odot$} &
			  \multicolumn{1}{c}{$L_\mathrm{bol}/L_\odot$} &
			  \multicolumn{1}{c}{\textmu} &
			  \multicolumn{1}{c}{$S\Delta v_\mathrm{CO(2-1)}$} &
			  \multicolumn{1}{c}{$S\Delta v_\mathrm{CO(6-5)}$} &
			  \multicolumn{1}{c}{$S\Delta v_\mathrm{CO(7-6)}$} &
			  \multicolumn{1}{c}{$S\Delta v_\mathrm{[C \textsc{i}(2-1)]}$}        &
			  \multicolumn{1}{c}{$S\Delta v_\mathrm{[C \textsc{ii}]}$}        \\
			  \multicolumn{1}{c}{} &
			  \multicolumn{1}{c}{} &
			  \multicolumn{1}{c}{} &
			  \multicolumn{1}{c}{} &
			  \multicolumn{1}{c}{} &
			  \multicolumn{1}{c}{(Jy km s$^{-1}$)}  &
			  \multicolumn{1}{c}{(Jy km s$^{-1}$)}  &
			  \multicolumn{1}{c}{(Jy km s$^{-1}$)}  &
			  \multicolumn{1}{c}{(Jy km s$^{-1}$)}  &
			  \multicolumn{1}{c}{(Jy km s$^{-1}$)}  \\
			\midrule
			Q0957+561                         & 1.414  & $5.63\times10^{12}$ $^1$    &  -                           & -           & $1.2\pm0.1$ $^{24}$   &                       &                       &               &             \\  
			3C318                             & 1.5771 & $9.34\times10^{12}$ $^2$    &  $2.1\times10^{12}$ $^{15}$  & -           & $1.19\pm0.22$ $^2$    &                       &                       &               &             \\
			COSBO11 / J100038.01+020822.4& 1.8275 & $5.70\times10^{12}$ $^3$    &  $3.0\times10^{13}$ $^{16}$  & 1 $^3$      & $1.33\pm0.08$ $^3$    & $4.86\pm0.8$ $^3$     &                       &               &  $77$ $^3$       \\
			B1938+666                         & 2.059  & $3.30\times10^{13}$ $^4$    &  -                           & -           & $3.8\pm1.1$ $^4$      &                       &                       &               &             \\ 
			IRASF10214+4724                   & 2.2855 & $5.00\times10^{12}$ $^5$    &  $1.4\times10^{14}$ $^{17}$  & -           &                       & $7.09\pm0.47$ $^{27}$ & $5.43\pm0.56$$^{27}$  &               &             \\
			Cloverleaf                        & 2.558  & $2.40\times10^{13}$ $^6$    &  $7.0\times10^{14}$ $^{18}$  & 11 $^{23}$  &                       & $37\pm8.1$ $^{28}$    & $45\pm2$ $^{28}$      &               &             \\
			SMMJ14011+0252                    & 2.56   & $6.55\times10^{12}$ $^7$    &  -                           & -           &                       &                       & $3.2\pm0.5$ $^7$      & $3.1\pm0.3$$^7$   &             \\
			VCVJ1409+5628 / J140955.5+562827  & 2.58   & $2.45\times10^{13}$ $^8$    &  $2.4\times10^{14}$ $^{19}$  & -           &                       &                       & $4.1\pm1.0$ $^{30}$   &               &             \\
			AMS12                             & 2.7668 & $3.16\times10^{13}$ $^9$    &  -                           & -           &                       &                       & $2.7\pm0.3$ $^9$      & $1.5\pm0.3$ $^9$   &             \\
			RXJ0911+0551                      & 2.796  & $3.81\times10^{13}$ $^{10}$ &  -                           & -           &                       &                       & $5.3\pm0.4$ $^{31}$   & $2.3\pm0.4$ $^{31}$    &             \\
			APM08279+5255                     & 3.9    & $9.45\times10^{13}$ $^{10}$ &  $3.7\times10^{14}$ $^{20}$  & 7 $^{12}$   & $0.81\pm0.18$ $^{11}$ & $7.3\pm0.8$ $^{29}$   &                       & $<1.1$ $^{29}$        &             \\ 
			PSSJ2322+1944                     & 4.119  & $5.09\times10^{12}$ $^{11}$ &                              & 3.5 $^{12}$ & $0.92\pm0.30$ $^{25}$ &                       &                       & $1.4\pm0.3$ $^{25}$  &             \\
			BR1335-0417                       & 4.407  & $2.76\times10^{13}$ $^{12}$ &                              & 1 $^{12}$   & $0.43\pm0.02$ $^{12}$ &                       & $3.08\pm0.11$ $^{32}$ & $1.04\pm0.1$ $^{12}$ & $26.2\pm4.3$ $^{32}$ \\
			BR1202-0725                       & 4.692  & $3.17\times10^{13}$ $^{13}$ &  $1.3\times10^{14}$ $^{21}$  & 1 $^{12}$   & $0.49\pm0.06$ $^{26}$ &                       & $3.1\pm0.86$ $^{11}$  &               &               \\
			LESSJ033229.4-275619              & 4.755  & $4.20\times10^{12}$ $^{14}$ &  $1.8\times10^{14}$ $^{22}$  & 0 $^{11}$   & $0.09\pm0.02$ $^{14}$ &                       &                       &               &               \\   
			\bottomrule
			\end{tabular}
			\end{adjustbox}
			\end{center} 
			\renewcommand{\arraystretch}{1}
			\begin{adjustwidth}{35pt}{35pt}
			\vspace{-0.2cm}
			\textbf{References: }
			$^1$ \cite{Barvainis_2002},
			$^2$ \cite{2007AJ....133..564W},
			$^3$ \cite{2010ApJ...724..957S},
			$^4$ \cite{2011ApJ...739L..31R},
			$^5$ \cite{2004ApJ...614L..97V},
			$^6$ \cite{2003A&A...409L..41W},
			$^7$ \cite{Downes_2003},
			$^8$ \cite{Carilli_2005},
			$^9$ \cite{2012MNRAS.423.2132S},
			$^{10}$ \cite{2011ApJ...730...18W} ,
			$^{11}$ \cite{2006ApJ...650..604R},
			$^{12}$ \cite{2008ApJ...686L...9R},
			$^{13}$ \cite{2012ApJ...752L..30W},
			$^{14}$ \cite{2010MNRAS.407L.103C},
			$^{15}$ \cite{2021ApJ...923...59V},
			$^{16}$ \cite{2008A&A...491..173A},
			$^{17}$ \cite{1993MNRAS.261..299C},
			$^{18}$ \cite{Lutz_2007},
			$^{19}$ \cite{2006AJ....132.1307P},
			$^{20}$ \cite{2011ApJ...740L..29F},
			$^{21}$ \cite{2013A&A...559A..29C},
			$^{22}$ \cite{Gilli_2011},
			$^{23}$ \cite{Venturini_2003},
			$^{24}$ \cite{1999Sci...286.2493P},
			$^{25}$ \cite{2011ApJ...739L..33C},
			$^{26}$ \cite{2002AJ....123.1838C},
			$^{27}$ \cite{2008AJ....136.1118A},
			$^{28}$ \cite{Bradford_2009},
			$^{29}$ \cite{2007A&A...467..955W},
			$^{30}$ \cite{2004A&A...423..441B},
			$^{31}$ \cite{2012ApJ...753..102W},
			$^{32}$ \cite{2018ApJ...864...38L}
			\end{adjustwidth}
			\end{small}
			\end{sidewaystable}


		\section{Dust SED fitting} 
			\label{sec:dust_sed_fitting}

			\begin{table*} 
			\begin{center}
			\caption{Dust SED fitting results. \label{tab:dust_sed}} 
			\renewcommand{\arraystretch}{1.5}
			\begin{tabular}{@{} >{\raggedright}m{0.25 \columnwidth}>{\centering}m{0.25\columnwidth}>{\centering}m{0.15\columnwidth}>{\centering}m{0.15\columnwidth}>{\centering}m{0.4\columnwidth}>{\centering\arraybackslash}m{0.4\columnwidth}@{}}
			\toprule 
			Name  &  $\log(M_\mathrm{dust} / \mathrm{M_\odot})$ & $T_\mathrm{dust}$ (K) & $\beta$  & $\log(L_\mathrm{FIR, 40-400\mu m} / L_\odot)$ &   $\log(L_\mathrm{FIR, 42.5-122.5\mu m} / L_\odot )$ \\
			\midrule
			P036+03 	&  $8.1_{-0.3}^{+0.5}$	& -										& $1.7_{-0.2}^{+0.3}$		&  $13.21_{-0.26}^{+0.26}$	& $13.14_{-0.26}^{+0.24}$	\\
			J0305-3150 	&  $9.5_{-0.2}^{+0.2}$	& $29_{-\phantom{0}2}^{+\phantom{0}2}$	& $2.1_{-0.1}^{+0.1}$		&  $12.78_{-0.05}^{+0.05}$	& $12.65_{-0.07}^{+0.08}$	\\
			J2348-3054 	&  $8.0_{-0.2}^{+0.3}$	& $61_{-13}^{+18}$						& $1.4_{-0.2}^{+0.2}$		&  $12.93_{-0.13}^{+0.18}$	& $12.87_{-0.13}^{+0.17}$	\\
			\midrule
			J0927+2001  & $8.5_{-0.3}^{+0.5}$	& -	 				 					&	 $2.0_{-0.3}^{+0.4}$	&   $13.60_{-0.33}^{+0.53}$ 	&	$13.52_{-0.32}^{+0.50}$ \\
			J0129-0035  & $8.0_{-0.3}^{+0.5}$	& -	 				 					&	 $1.4_{-0.3}^{+0.4}$	&   $12.94_{-0.25}^{+0.32}$ 	&	$12.87_{-0.25}^{+0.30}$ \\
			J0840+5624	& $8.6_{-0.6}^{+0.5}$	& $40_{-10}^{+25}$	 					&	 $2.0_{-0.6}^{+0.6}$	&   $12.90_{-0.20}^{+0.43}$ 	&	$12.80_{-0.24}^{+0.46}$ \\
			J2310+1855	& $9.3_{-0.1}^{+0.1}$	& $34_{-\phantom{0}1}^{+\phantom{0}1}$	&	 $2.0_{-0.1}^{+0.1}$	&   $13.10_{-0.02}^{+0.02}$ 	&	$13.01_{-0.03}^{+0.03}$ \\
			J2054-0005 	& $8.4_{-0.3}^{+0.3}$	& $45_{-10}^{+13}$	 					& 	 $1.8_{-0.5}^{+0.6}$	&   $12.91_{-0.08}^{+0.08}$ 	&	$12.85_{-0.08}^{+0.08}$ \\
			J1429+5447 	& $8.2_{-0.3}^{+0.5}$	& -	 				 					&	 -						&   $13.03_{-0.34}^{+1.91}$ 	&	$12.97_{-0.34}^{+1.76}$ \\
			J0100+2802 	& $7.7_{-0.3}^{+0.5}$	& -	 									& 	 $0.7_{-0.2}^{+0.5}$	&   $12.52_{-0.30}^{+0.37}$ 	&	$12.45_{-0.32}^{+0.36}$ \\
			J1148+5251 	& $8.4_{-0.4}^{+0.4}$	& -	 									&	 $2.1_{-0.3}^{+0.3}$	&   $13.24_{-0.20}^{+0.36}$ 	&	$13.18_{-0.20}^{+0.33}$ \\
			\bottomrule 
			\end{tabular} 
			\renewcommand{\arraystretch}{1}
			\end{center} 
			\begin{adjustwidth}{25pt}{25pt}
			** In cases where the fit parameters were poorly constrained (essentially returning the priors), we do not quote the results. 
			\end{adjustwidth}
			\end{table*} 

			We fit the dust SEDs with a modified blackbody of the form, 
			\begin{align}
				S\Delta v ^\mathrm{against~CMB} (\nu_\mathrm{rest}) = \dfrac{f_\mathrm{CMB}(1+z)}{d_L^2} M_\mathrm{dust} \kappa_\mathrm{\nu_{rest}} \left(B[T_\mathrm{CMB}] - B[T_\mathrm{dust}] \right)
			\end{align}
			where $d_L$ is the luminosity distance, $B (\nu, T)$ is the blackbody function and the contrast against the CMB is accounted for by $f_\mathrm{CMB} = 1- B_\mathrm{\nu} [T_\mathrm{CMB}(z)] / B_\mathrm{\nu} [T_\mathrm{dust}(z)]$ \citep{2013ApJ...765....9D}. We thereby assume optically thin dust emission at $\lambda>40$ \textmu m and correct for both the heating by and contrast against the CMB. We assumed that the dust opacity scales as $\kappa_{\nu_\mathrm{rest}} = \kappa_{\nu_0} (\nu_\mathrm{rest} /\nu_0)^\beta$ where $\beta$ is the dust spectral emissivity index. Following \cite{2003MNRAS.341..589D} we take $\nu_0 = c/(125$\textmu m) and $\kappa_{\nu_0} = 2.64~\mathrm{m^2 kg^{-1}}$.  We used a Bayesian approach to fit the dust SEDs and recover posterior likelihood distributions for the log of the total dust mass ($\log(M_\mathrm{dust})$), the dust temperature ($T_\mathrm{dust}$) and dust emissivity index ($\beta$). We apply uniform priors of $2.73 (1+z) < T_\mathrm{dust} < 100$, $6 < \log{M_\mathrm{dust}} < 11$ and $1 < \beta < 3$. 

			Because we are assuming thermal dust, we only included observations sampling the dust SED at $\lambda_\mathrm{rest}>70$ \textmu m. At shorter wavelengths, the effects from stochastically heated grains become more dominant \citep[e.g.][]{2021ApJ...919...30D} and contamination by QSO non-thermal and torus emission becomes significant \citep[e.g.][]{Leipski_2014}. We also omitted data at the other end of the SED, in the radio regime where synchrotron and free-free emission dominate ($\nu_\mathrm{rest}\lesssim 100$ GHz). To sample the probability distributions we applied Markov Chain Monte Carlo sampling via the \texttt{emcee} package \citep{2013PASP..125..306F}. The resulting likelihood posteriors as well as the median, 16th and 84th percentile values are shown for our sample and the literature sample in Fig. \ref{fig:dust_seds_our_sample} and \ref{fig:dust_seds_lit_sample}, respectively. The data that were fit are shown as light blue circles. We also show additional continuum data that were not fit, in purple. The derived values are provided in Table~\ref{tab:dust_sed}. 

			\begin{figure*}
				\centering
				\includegraphics[width=0.8\textwidth, trim={0cm 0cm 0cm 0cm},clip]{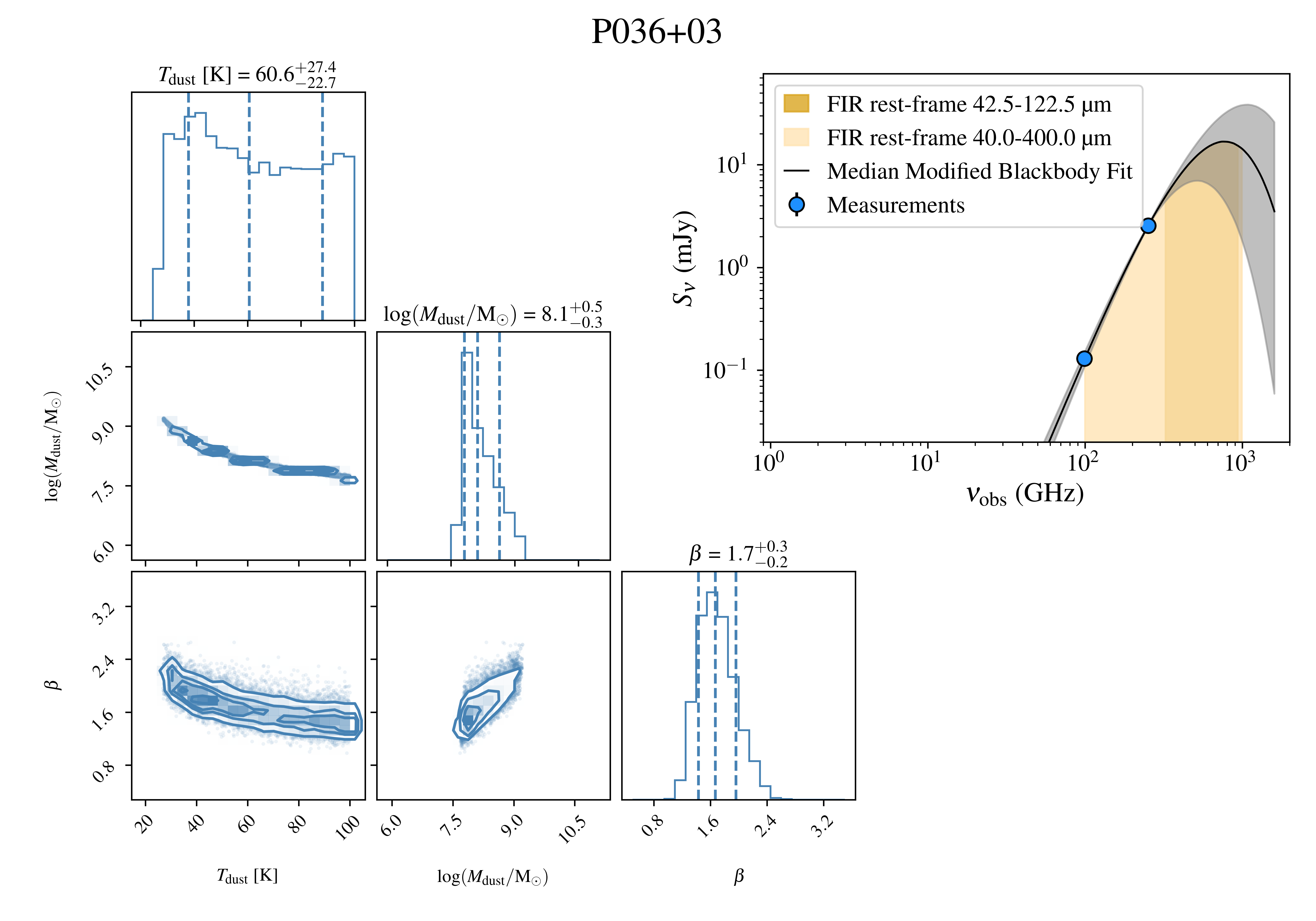}
				\\
				\includegraphics[width=0.8\textwidth, trim={0cm 0cm 0cm 0cm},clip]{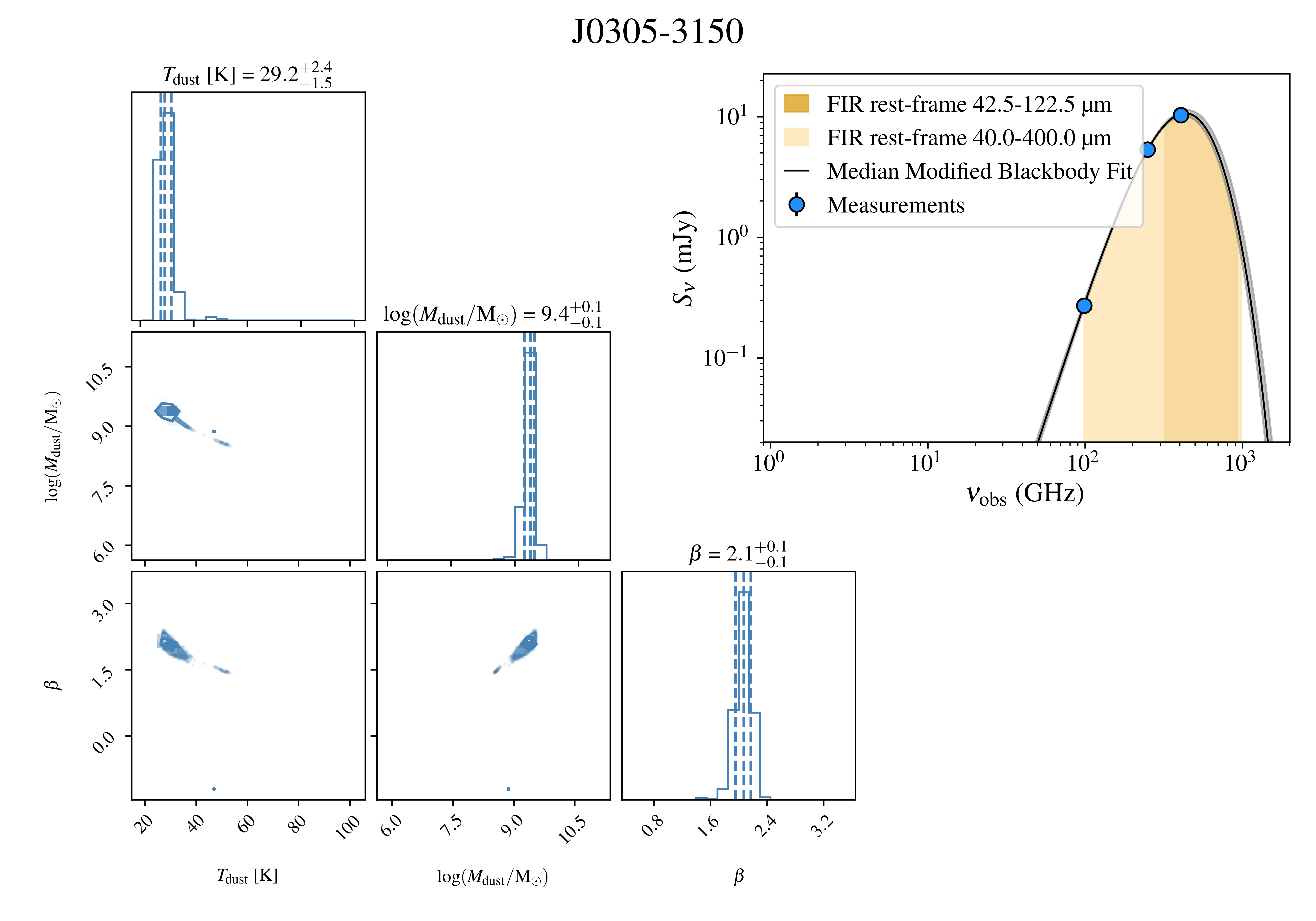}
				\caption{Outputs of the Bayesian dust SED fitting for our three QSO hosts.  For each source (labelled at top), the top right panel shows the best-fit dust models (solid lines) and the observed fluxes (blue and purple points). The data that were included in the fits are shown in light blue, those not included in purple. The remaining panels from top left to bottom right show the marginalised likelihood distributions for the three fit parameters: dust temperature ($T_\mathrm{dust}$), mass ($M_\mathrm{dust}$), and emissivity index ($\beta$). \label{fig:dust_seds_our_sample}}	
			\end{figure*} 

			\begin{figure*}
				\ContinuedFloat
				\centering
				\includegraphics[width=0.8\textwidth, trim={0cm 0cm 0cm 0cm},clip]{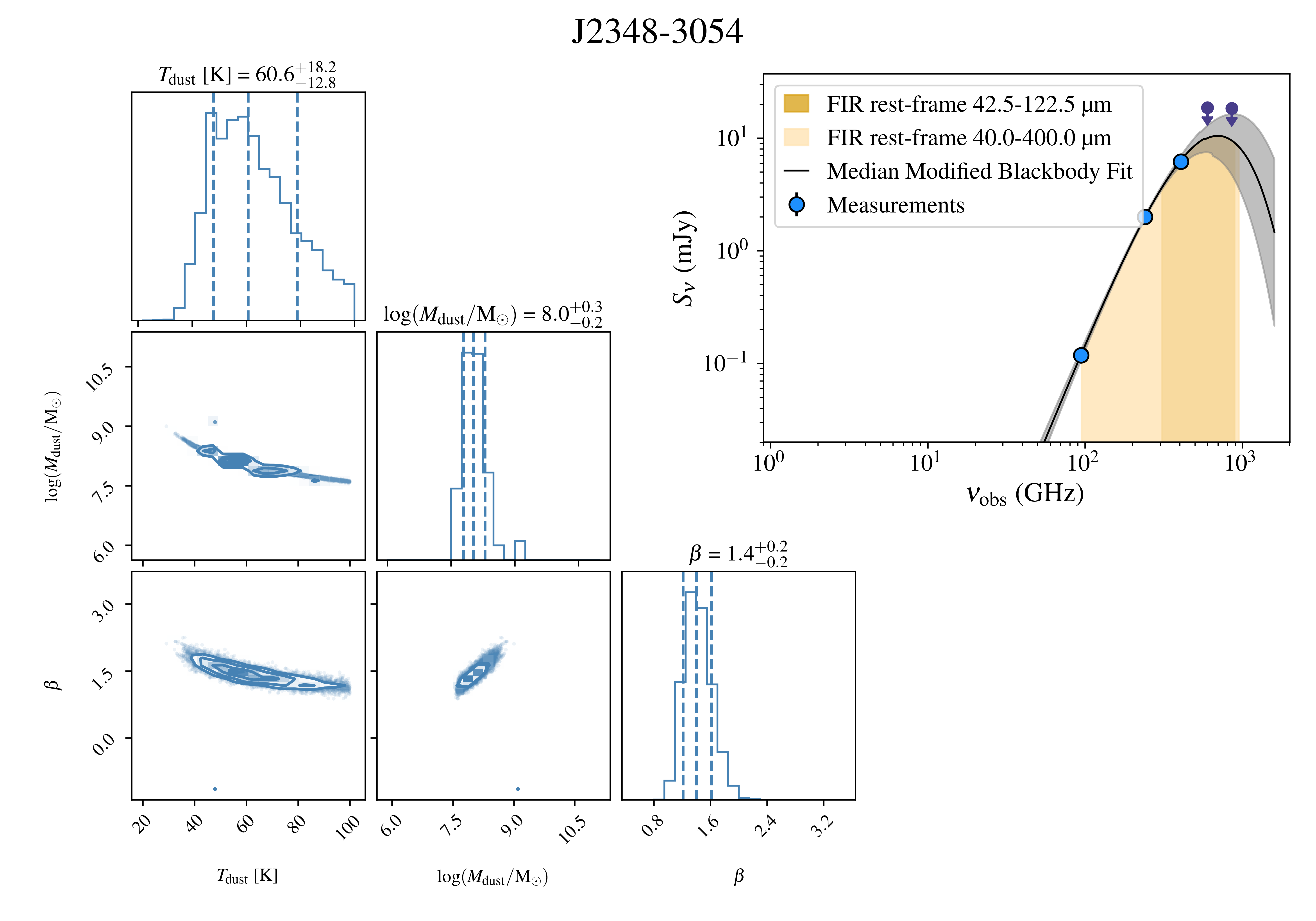}
				\caption{continued}	
			\end{figure*}

			\begin{figure*}
				\centering
				\includegraphics[width=0.8\textwidth, trim={0cm 0cm 0cm 0cm},clip]{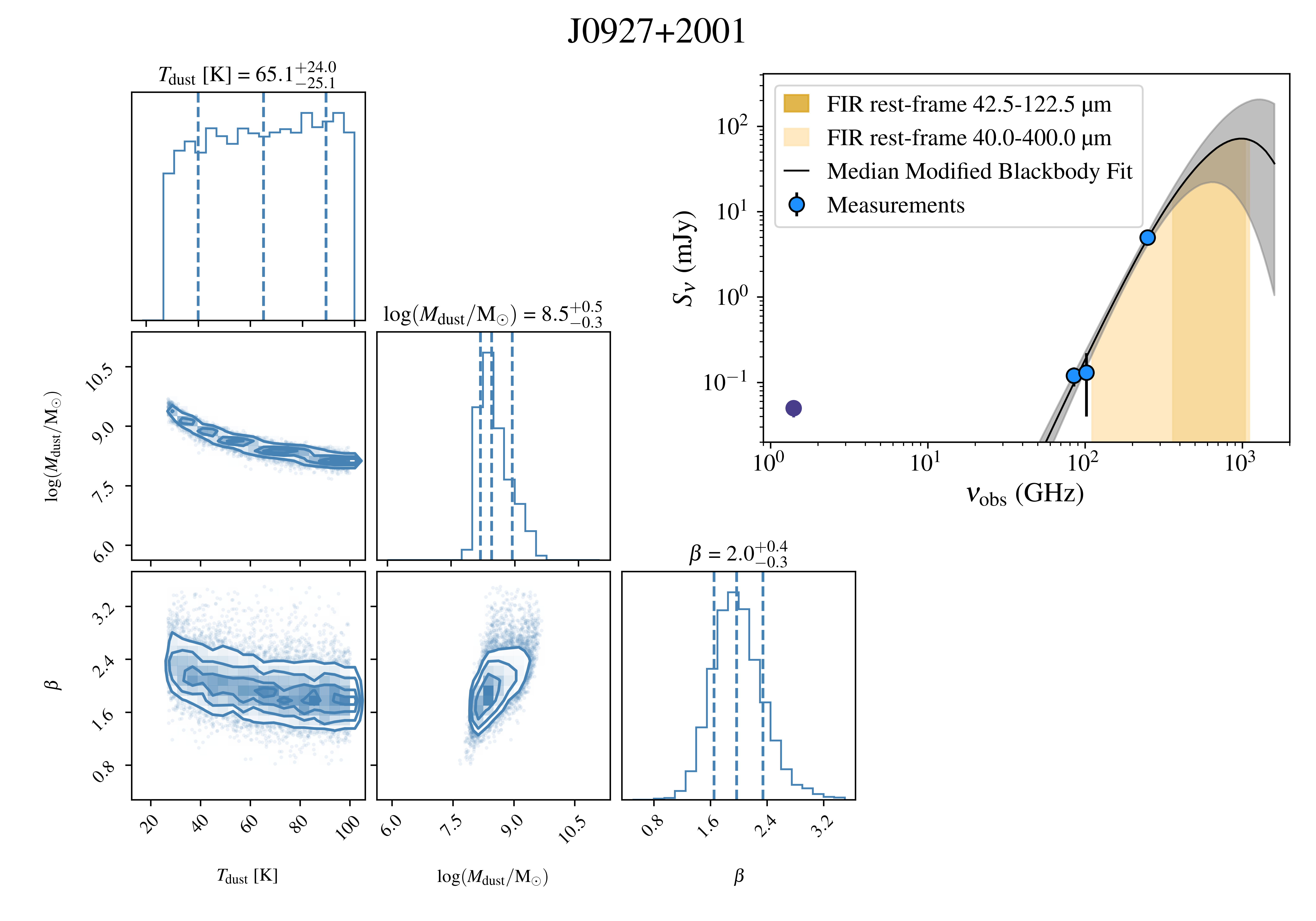}
				\\
				\includegraphics[width=0.8\textwidth, trim={0cm 0cm 0cm 0cm},clip]{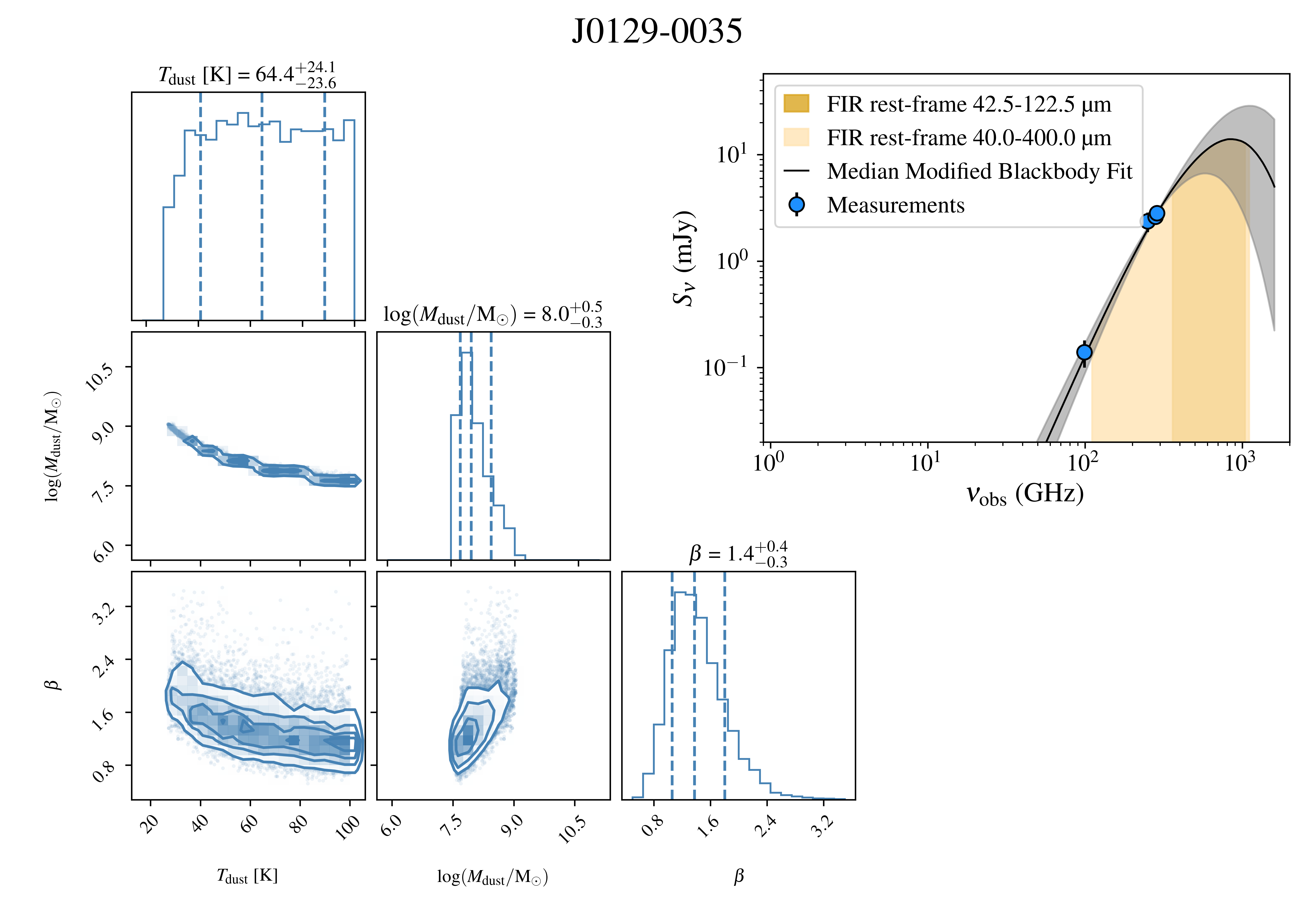}
				\caption{Same as Fig.~\ref{fig:dust_seds_our_sample} but for the literature sample of $z=5.7-6.4$ QSO hosts.  \label{fig:dust_seds_lit_sample}}	
			\end{figure*} 

			\begin{figure*}
				\ContinuedFloat
				\centering
				\includegraphics[width=0.8\textwidth, trim={0cm 0cm 0cm 0cm},clip]{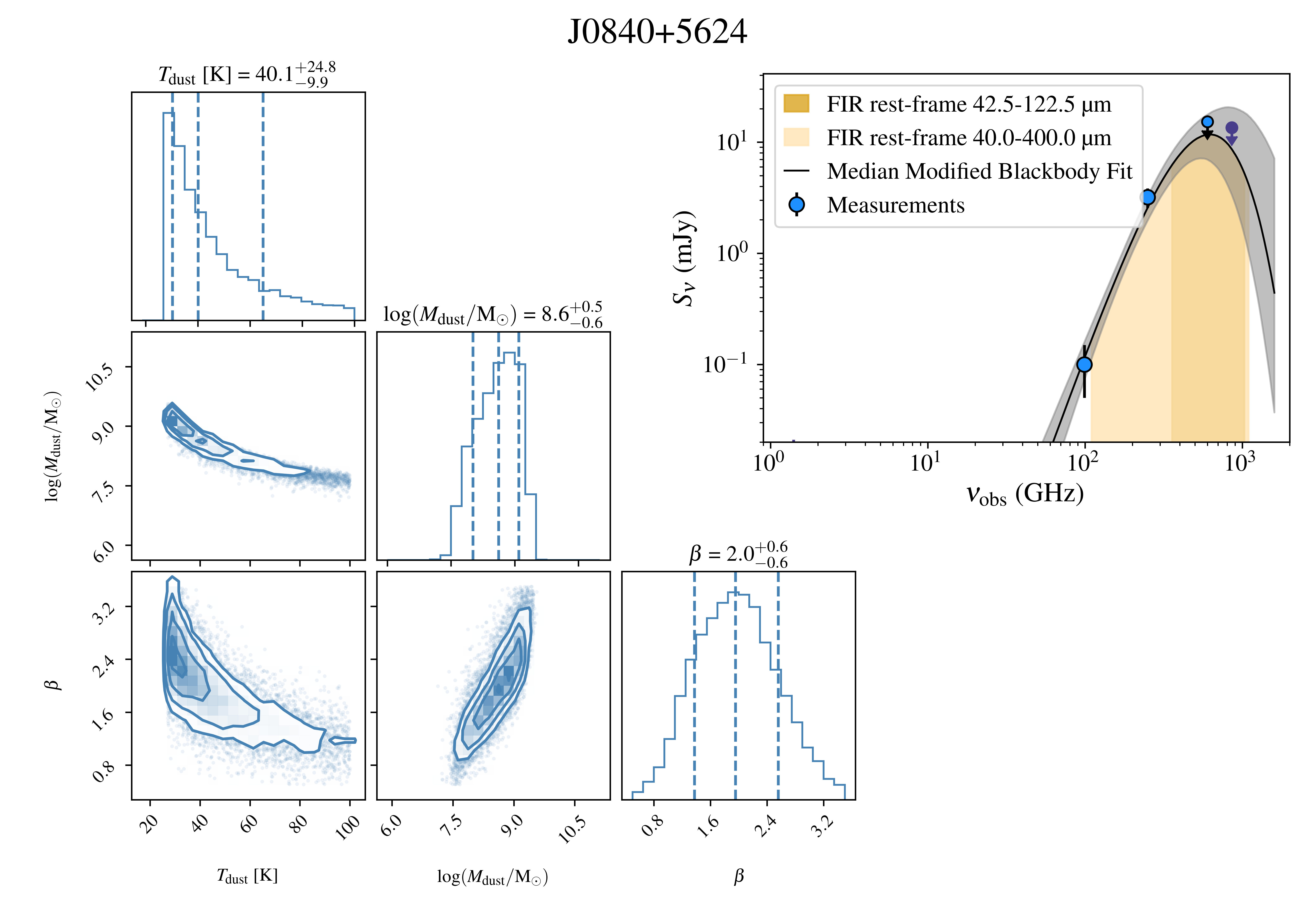}
				\\
				\includegraphics[width=0.8\textwidth, trim={0cm 0cm 0cm 0cm},clip]{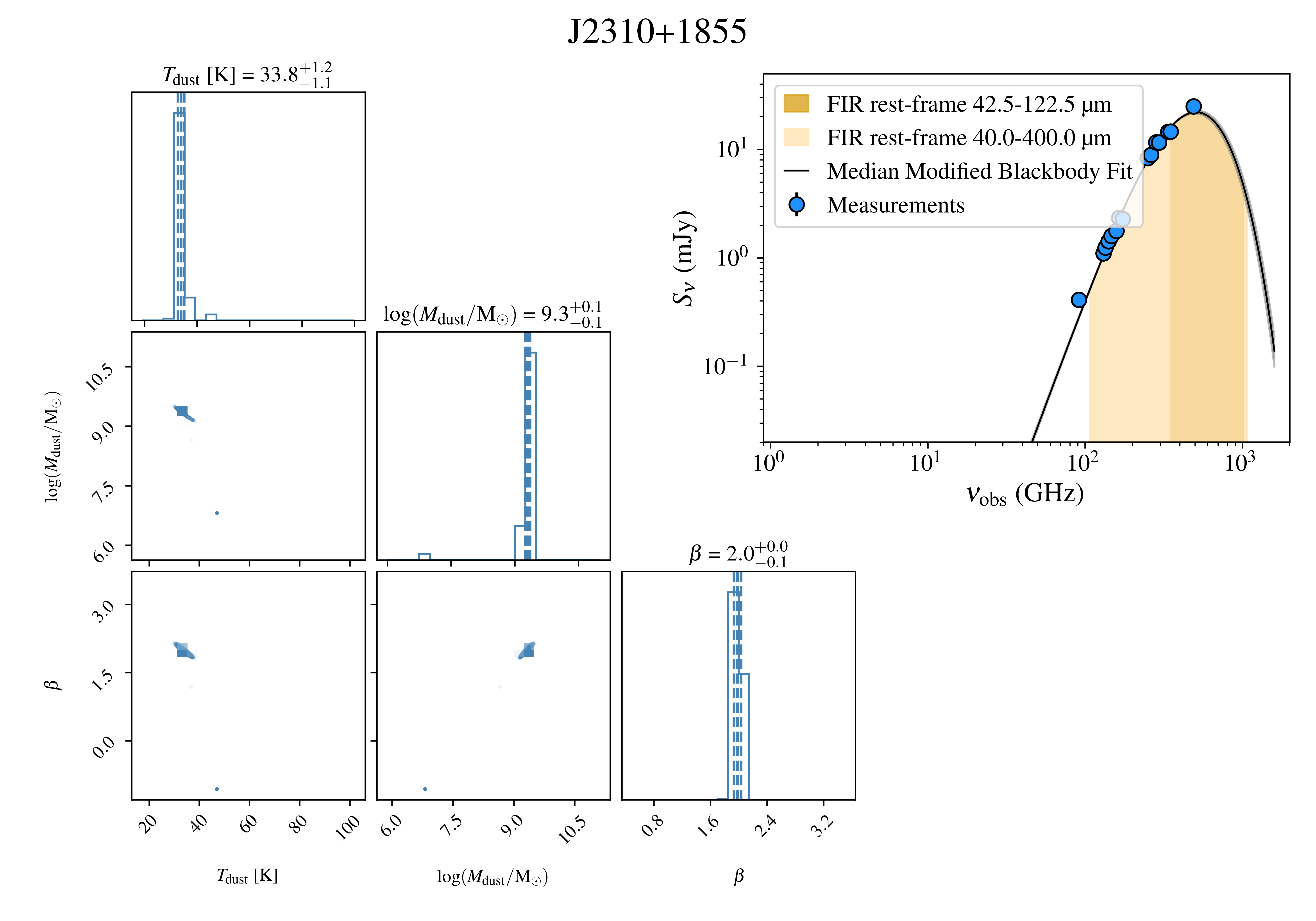}
				\caption{continued}	
			\end{figure*}

			\begin{figure*}
				\ContinuedFloat
				\centering
				\includegraphics[width=0.8\textwidth, trim={0cm 0cm 0cm 0cm},clip]{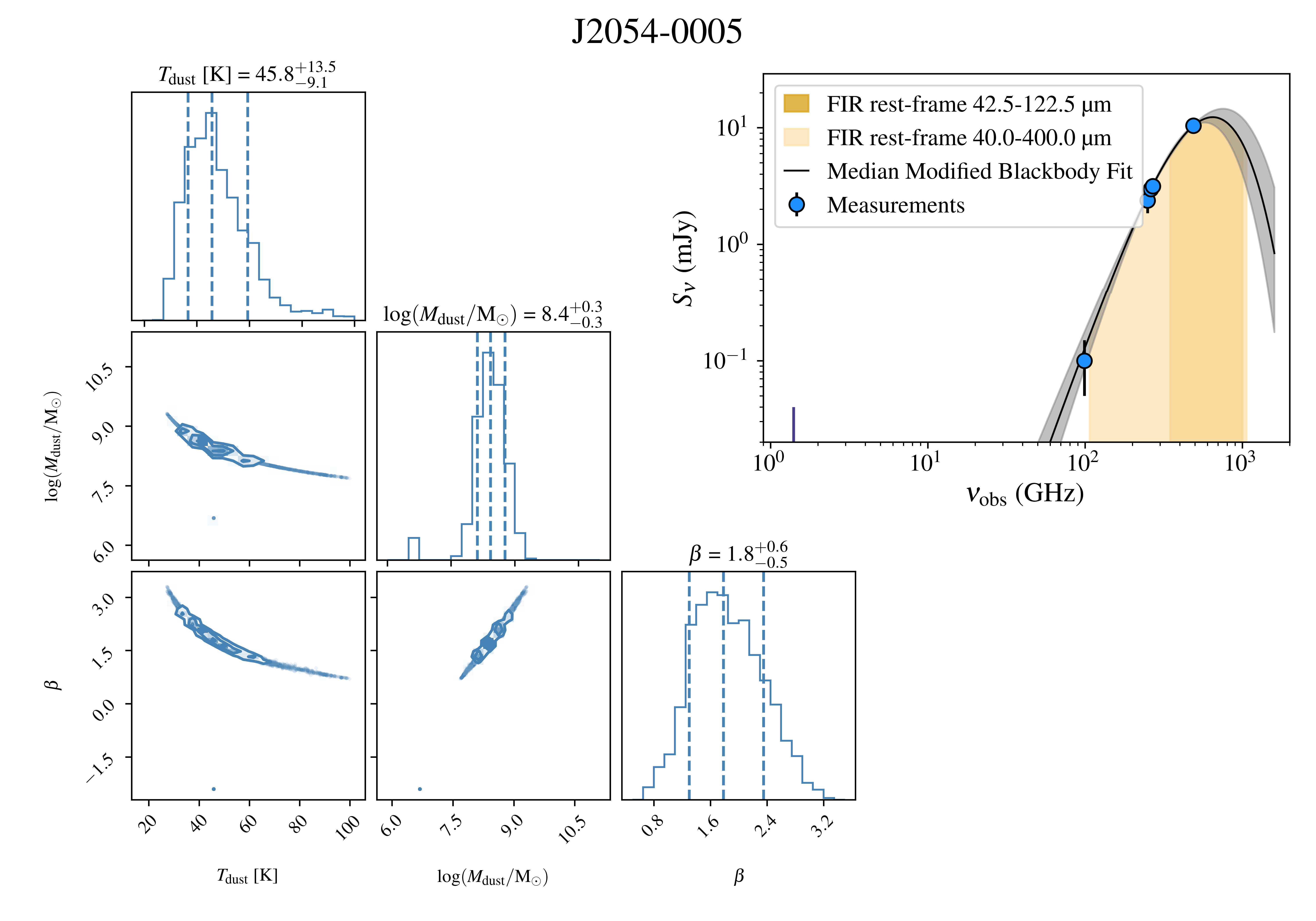}
				\\
				\includegraphics[width=0.8\textwidth, trim={0cm 0cm 0cm 0cm},clip]{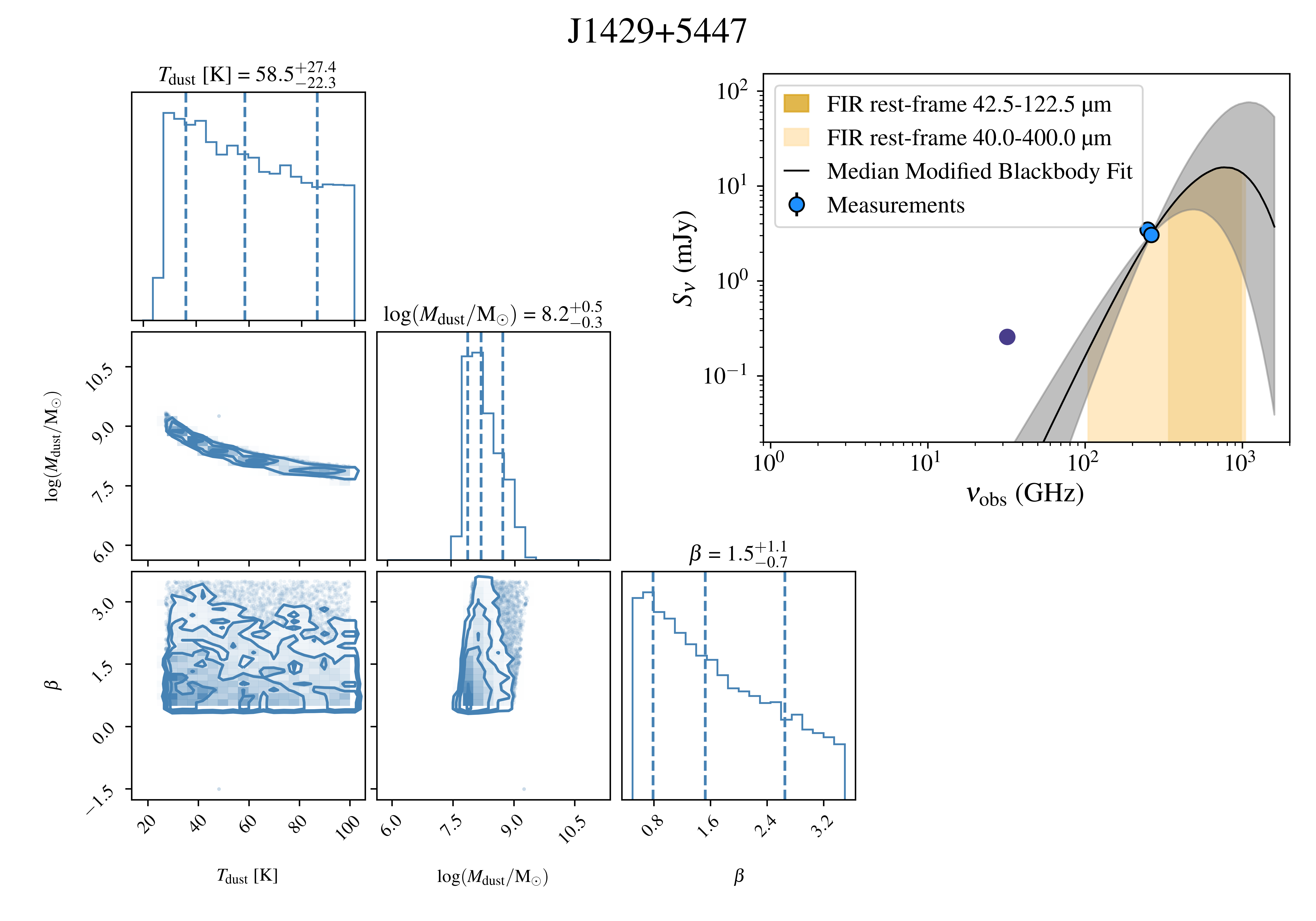}
				\caption{continued}	
			\end{figure*}

			\begin{figure*}
				\ContinuedFloat
				\centering
				\includegraphics[width=0.8\textwidth, trim={0cm 0cm 0cm 0cm},clip]{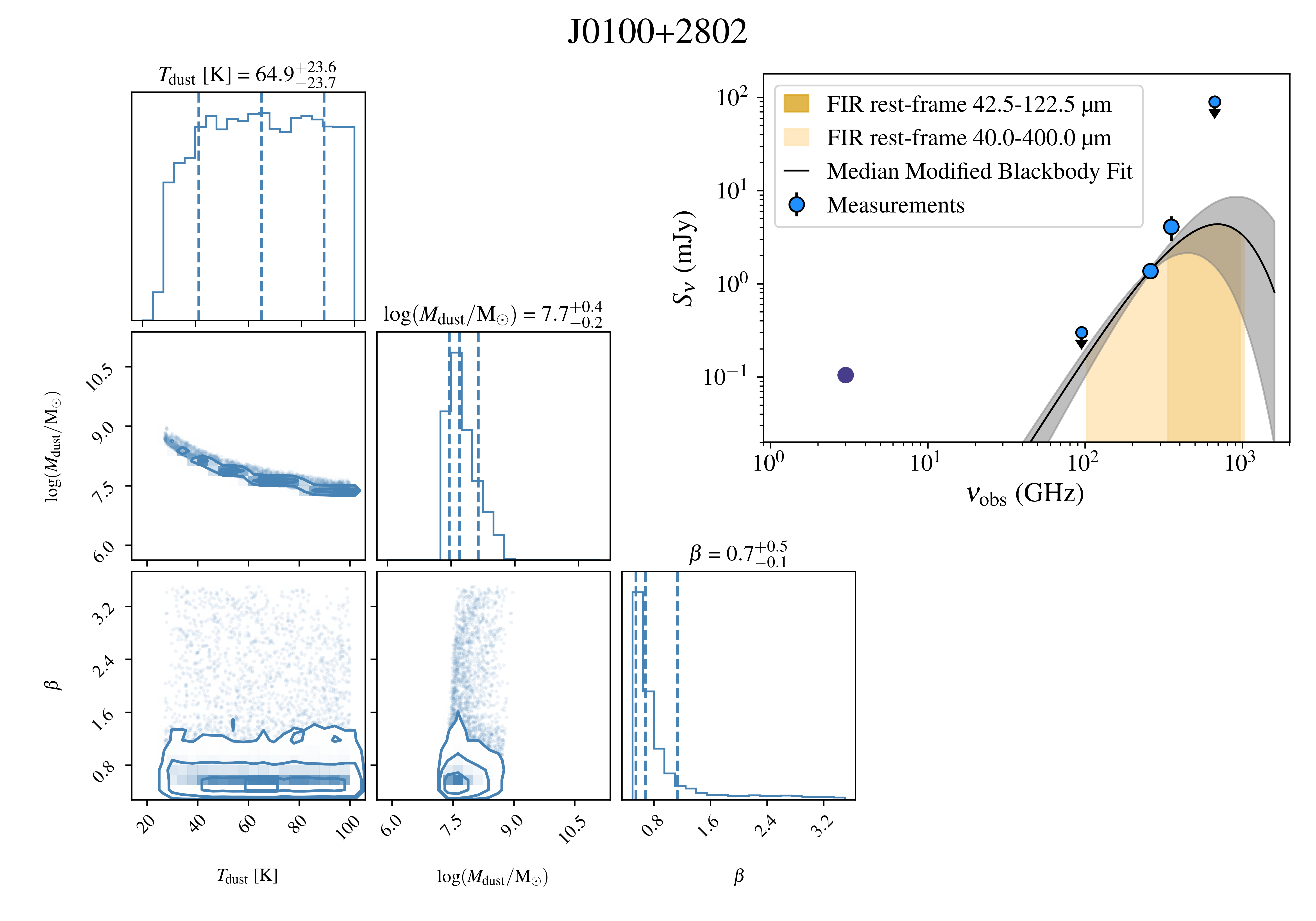}
				\\
				\includegraphics[width=0.8\textwidth, trim={0cm 0cm 0cm 0cm},clip]{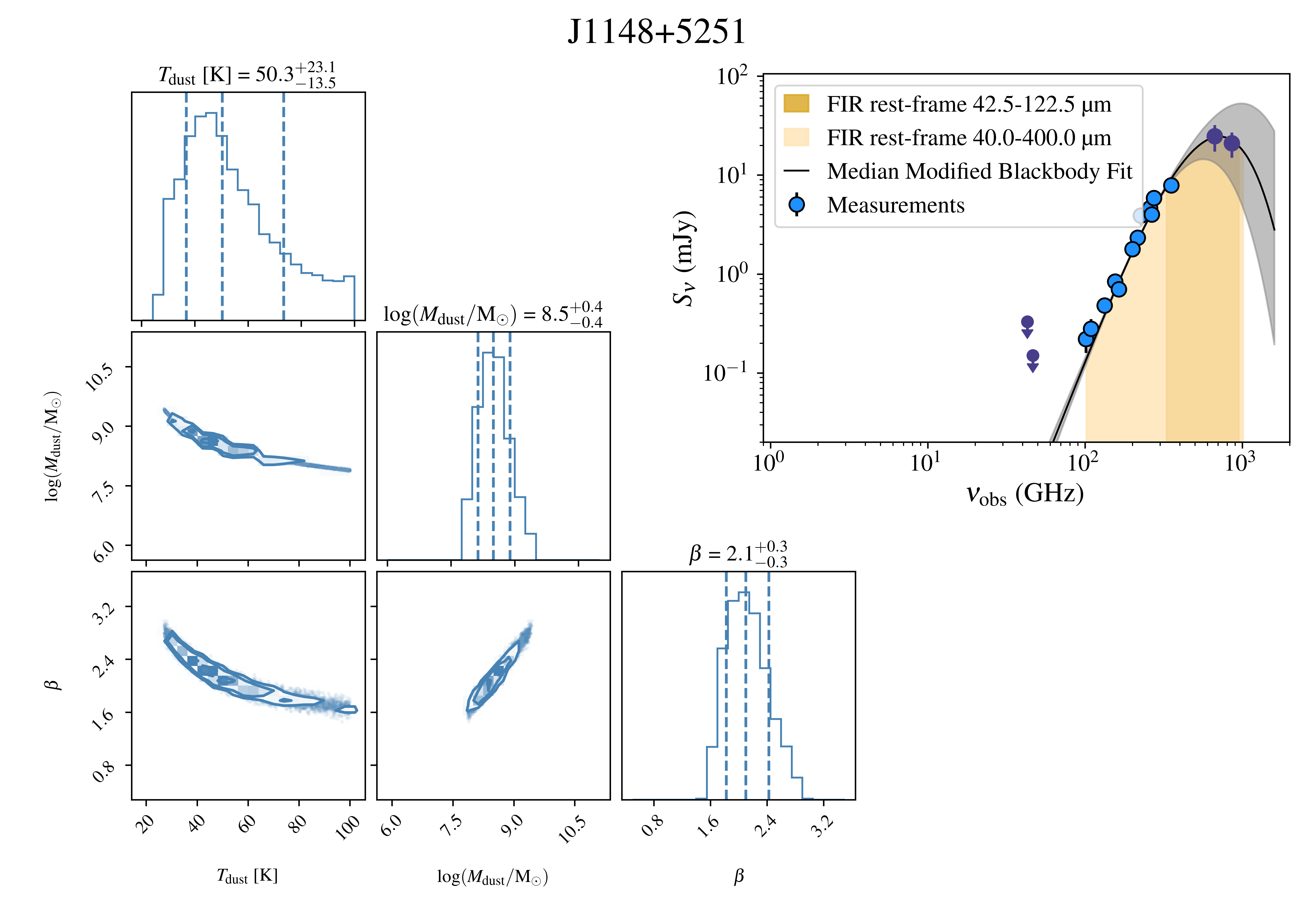}
				\caption{continued}	
			\end{figure*}


	\newpage

	\end{appendix}

\end{document}